\documentclass[12pt,letterpaper,onecolumn,amsmath,amssymb,aps,titlepage,tightenlines] {report}
\usepackage[utf8]{inputenc}
\usepackage{graphicx}
\usepackage{rotating}
\usepackage{amsmath}
\usepackage[margin=1in]{geometry}
\usepackage{threeparttable}
\usepackage{graphicx}
\usepackage{wrapfig}
\usepackage[T1]{fontenc}
\usepackage{ae,aecompl}
\usepackage{dcolumn}
\usepackage{pslatex}
\usepackage{array}
\usepackage{float}
\usepackage{amsmath}
\usepackage{setspace}
\usepackage[font=footnotesize, justification=justified, format=plain, labelfont=bf, textfont=it]{caption}
\usepackage{subcaption}
\usepackage{enumerate}
\usepackage[shortlabels]{enumitem}
\usepackage[square,numbers]{natbib}
\usepackage[most]{tcolorbox}
\usepackage{fancyhdr}
\usepackage{multirow}

\usepackage{pdfpages}
\usepackage[all]{nowidow}

\definecolor{royalblue}{rgb}{0.25, 0.25, 0.88}
\definecolor{aoblue}{rgb}{0.0, 0.0, 1.0}
\definecolor{yaleblue}{rgb}{0.06, 0.3, 0.57}
\definecolor{ultramarine}{rgb}{0.07, 0.04, 0.56}
\definecolor{goldenpoppy}{rgb}{0.82, 0.50, 0.08}
\definecolor{citegreen}{rgb}{0.3,0.7,0}
\definecolor{purple}{rgb}{0.38,0.059,.66}

\usepackage{lmodern}
\usepackage{url}

\DeclareTextFontCommand{\textbfit}{%
  \fontseries\bfdefault % change series without selecting the font yet
  \itshape
}

\usepackage{color}
\usepackage{xcolor}
\usepackage{lipsum}

\usepackage{hyperref}
\hypersetup{
    colorlinks=true, %set true if you want colored links
    linktoc=all,     %set to all if you want both sections and subsections linked
    linkcolor=royalblue,  %choose some color if you want links to stand out
    citecolor=citegreen
}

\newlength{\seplinewidth}
\newlength{\seplinesep}
\colorlet{sepline}{goldenpoppy}
\newcommand*{\sepline}{%
  \par
  \vspace{\dimexpr\seplinesep+.5\parskip}%
  \cleaders\vbox{%
    \begingroup % because of color
      \color{sepline}%
      \hrule width\linewidth height\seplinewidth
    \endgroup
  }\vskip\seplinewidth
  \vspace{\dimexpr\seplinesep-.5\parskip}%
}

\usepackage{titlesec}
\let\oldfootrule\footrule

\fancypagestyle{mystyle}{
  \fancyhf{} % clear all fields
  \fancyhead[RO,LE]{\thepage}
  \fancyfoot[CE,CO]{\leftmark}
  \renewcommand{\headrulewidth}{2pt}
  \renewcommand{\headrule}{\hbox to\headwidth{\color{goldenpoppy}\leaders\hrule height \headrulewidth\hfill}}
  \renewcommand{\footrule}{{\color{goldenpoppy}\oldfootrule}}
  
  }
  
\makeatletter
\renewcommand{\section}{\@startsection{section}{1}{0pt}
	{0.\baselineskip}{0.1\baselineskip}{\Large\textbf}} % was \large
\renewcommand{\subsection}{\@startsection{subsection}{2}{0pt}
	{0.\baselineskip}{0.1\baselineskip}{\large\textbf}}  % was \normalsize
\renewcommand{\subsubsection}{\@startsection{subsubsection}{3}{0pt}
	{0.1\baselineskip}{0.1\baselineskip}{\normalsize\textbf}}
\renewcommand{\paragraph}{\@startsection{paragraph}{4}{0pt}
	{0.1\baselineskip}{-0.5\baselineskip}{\normalsize\textsf}}
\renewcommand{\subparagraph}{\@startsection{subparagraph}{5}{0pt}
	{0.1\baselineskip}{-0.5\baselineskip}{\normalsize\textit}}
\makeatother

\makeatletter
\def\bstctlcite{\@ifnextchar[{\@bstctlcite}{\@bstctlcite[@auxout]}}
\def\@bstctlcite[#1]#2{\@bsphack
  \@for\@citeb:=#2\do{%
    \edef\@citeb{\expandafter\@firstofone\@citeb}%
    \if@filesw\immediate\write\csname #1\endcsname{\string\citation{\@citeb}}\fi}%
  \@esphack}
\makeatother

\begin{document}
\bstctlcite{IEEEexample:BSTcontrol}

\pagestyle{plain}
\begin{titlepage}
\begin{center}
~~\\
\vspace{2cm}
\sepline
\vspace{0.5cm}
\rmfamily{\Huge  \sc A Roadmap}\\
\vspace{5mm}
\rmfamily{\Huge  \sc For Scientific Ballooning }\\
\vspace{0.75cm}
\rmfamily{{\Huge 2020-2030}}\\
\vspace{0.5cm}
\sepline

\vspace{2cm}

\vspace{0.25cm}

\includegraphics[width=\textwidth]{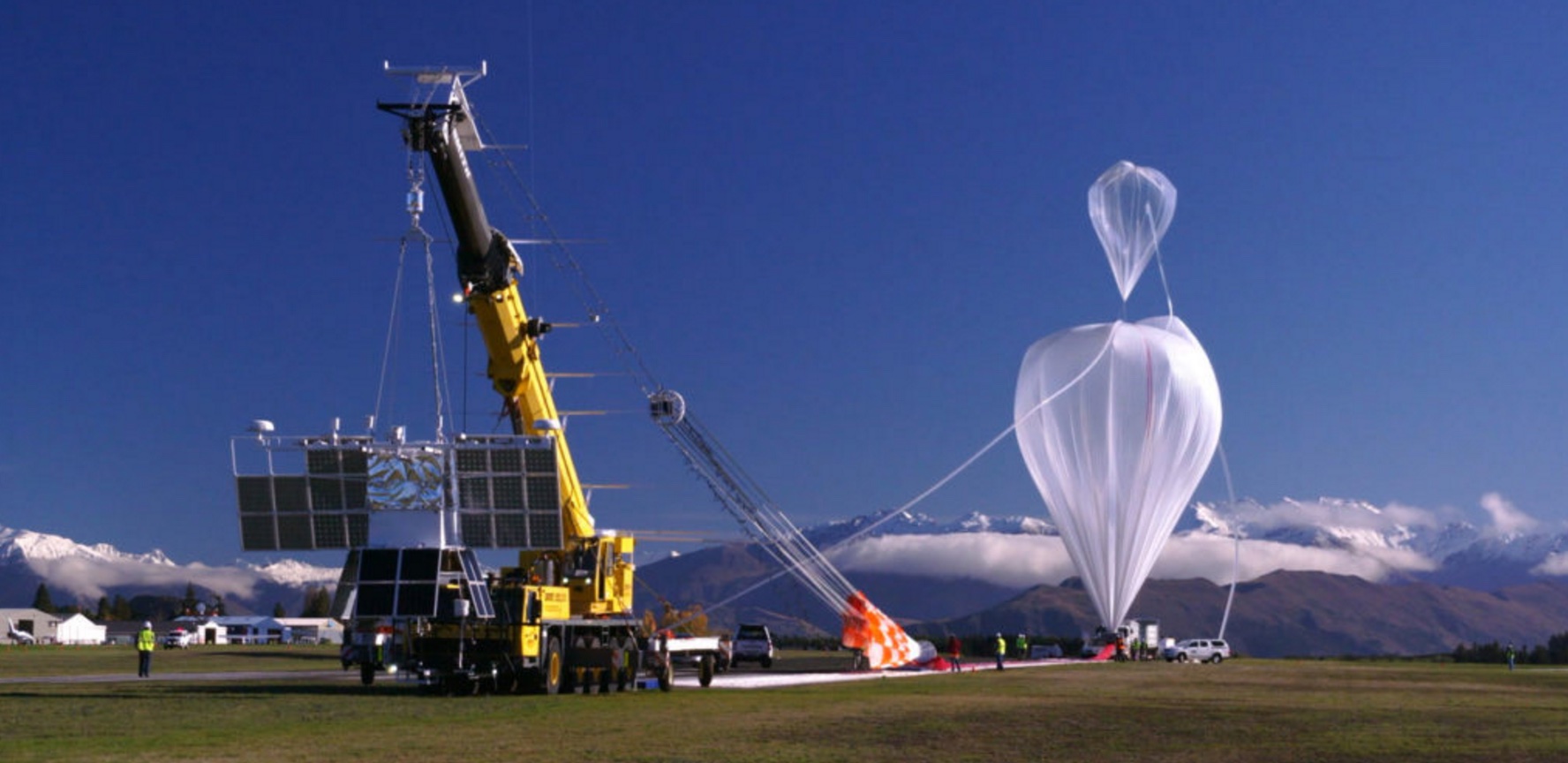}

\vspace{0.25cm}

\rmfamily{{\Large  \sc Prepared for the}}\\
\vspace{.15cm}
\rmfamily{{\Large  \sc National Aeronautics and Space Administration}}\\
\vspace{.75cm}
{\Large  \sc By the}\\
\vspace{0.75cm}
\rmfamily{{\Large \sc NASA Balloon Program Analysis Group}}\\
\vspace{.75cm}

\vspace{.5cm}

\end{center}

\clearpage

%\begin{center}
\pagestyle{mystyle}
\vspace{0.25cm}
\sepline

\vspace{2cm}
\noindent
{\Large The NASA Program Analysis Group Executive Committee: }

\vspace{0.75cm}
\noindent
{\large
\textbf{Peter Gorham, Chair} \\
~~~~~~~University of Hawaii at Manoa

\noindent
\textbf{James Anderson} \\
~~~~~~~Harvard University 

\noindent
\textbf{Pietro Bernasconi} \\
~~~~~~~Johns Hopkins University Applied Physics Laboratory

\noindent
\textbf{Supriya Chakrabarti} \\
~~~~~~~University of Massachusetts Lowell 

\noindent
\textbf{T. Gregory Guzik} \\
~~~~~~~Louisiana State University

\noindent
\textbf{William Jones} \\
~~~~~~~Princeton University

\noindent
\textbf{Carolyn Kierans} \\
~~~~~~~NASA Goddard Space Flight Center

\noindent
\textbf{Robyn Millan} \\
~~~~~~~Dartmouth College 

\noindent
\textbf{Abigail Vieregg} \\
~~~~~~~University of Chicago 

\noindent
\textbf{Christopher Walker} \\
~~~~~~~University of Arizona

\noindent
\textbf{Eliot Young} \\
~~~~~~~Southwest Research Institute 
}
\vspace{1cm}

{\it Cover photo: The Compton Spectrometer and Imager (COSI) payload just prior to 
launch from Wanaka, New Zealand, on a NASA super pressure balloon in May 2016.}

%\end{center}

\clearpage

\begin{small}
% \centerline{\bf Abstract}
% \vspace{10pt}
% The NASA Balloon Program Analysis Group (PAG) has been tasked by NASA to develop scientific and strategic priorities for NASA's scientific ballooning endeavor
% through the next decade. Here we summarize the draft findings to date of the PAG for the purpose of informing the Astrophysics 2020 Decadal survey.
% Scientific Ballooning involves a wide array of disciplines in Astrophysics, Planetary Science, Earth Science, Heliophysics, Technology development,
% and Education/Public Outreach. The program itself is managed under the Astrophysics Division within the NASA Science Mission Directorate,
% and thus we submit these findings to the Astro 2020 Decadal Panel for purposes of indicating recommendations for Balloon Program Support,
% affecting all of the associated scientific disciplines.
\end{small}
%\tableofcontents

\end{titlepage}

\phantomsection
\hypertarget{TOC}{} % create an anchor for the table of contents
\tableofcontents

\pagestyle{mystyle}
\clearpage

\chapter{Executive Summary}

\section{Introduction to Scientific Ballooning}

The exploration of the Earth, the stratosphere, the solar system, and the cosmos beyond it from balloon-borne scientific instruments has a long and illustrious history. 
Scientific ballooning has progressed in a continuous arc from early beginnings in the late 18th century when the first measurements of air temperature versus altitude were made from a balloon payload, to our current era where almost every branch of astrophysics, space science, Earth science, and planetary science garner contributions from balloon missions. 
There is every reason to believe that this rate of 
scientific productivity will continue its remarkable trajectory into and beyond the next decade. 

The near-space conditions offered through balloon flights provide an exceptional science and instrument development opportunity for payloads ranging in size from 5 to over 5000~lbs with observation times lasting from hours to months.
The short time scales for project development, and the relatively low cost compared to satellite missions, allows for innovative technology development and a complete educational experience. The instruments launched with balloons are as varied as the accessible science, but all benefit from the three main advantages of scientific ballooning: 
\begin{itemize}[label={}]
\item {\textcolor{royalblue}{\textbfit{Scientific balloons provide a unique platform for groundbreaking science}} 

Due to the relative low cost and short program timescales, balloon-borne instrument can be purpose-built and optimized for focused, high-impact science.
The high altitudes achieved through balloon flights allow for observations in the electromagnetic spectrum that are not observable from ground or piloted aircraft. 
At these attitudes, the atmosphere is thin enough that there is negligible astronomical seeing and minimal absorption of cosmic particles and high energy photons.
Furthermore, the balloon platform allows for a look back at the Earth as a target for science observations.}
\vspace{2cm}

\item {\textcolor{royalblue}{\textbfit{Balloon-borne missions provide a test bed for future space-flight instruments}} 

The low cost and rapid development cycle of balloon flights, coupled with the near-space environment accessible through the platform, allow for a demonstration of new technologies designed for space-based missions. 
With the almost guaranteed recovery of the payload, improvements can be made on novel technologies and prototypes.}

\item {\textcolor{royalblue}{\textbfit{Balloon projects are a hands-on training ground}} 

A balloon project provides an unparalleled education experience for the next generation of scientists and engineers. 
With project time scales being consistent with the academic schedule, and with a higher acceptable risk for balloons relative to spacecraft, students can contribute to all aspects of a mission, from hardware and software development to leadership.
High altitude balloons, particularly those that can be hand-launched, provide an accessible platform for the education and outreach for younger students.}

\end{itemize}

These three assets of the balloon program, which are a continuous theme throughout this document, are discussed further in the following sections.

\section{Science Highlights from the Last Decade}

Scientific balloons provide a platform for groundbreaking science. 
The science drivers for the balloon program are discussed in detail for each division of NASA's Science Mission Directorate in Chapter~\ref{ch:science}; however, here we highlight a number of science results published in the past decade which exemplify the capabilities of the balloon platform for focused, high-impact science:

\begin{minipage}{0.2\textwidth}
\includegraphics[width=\textwidth]{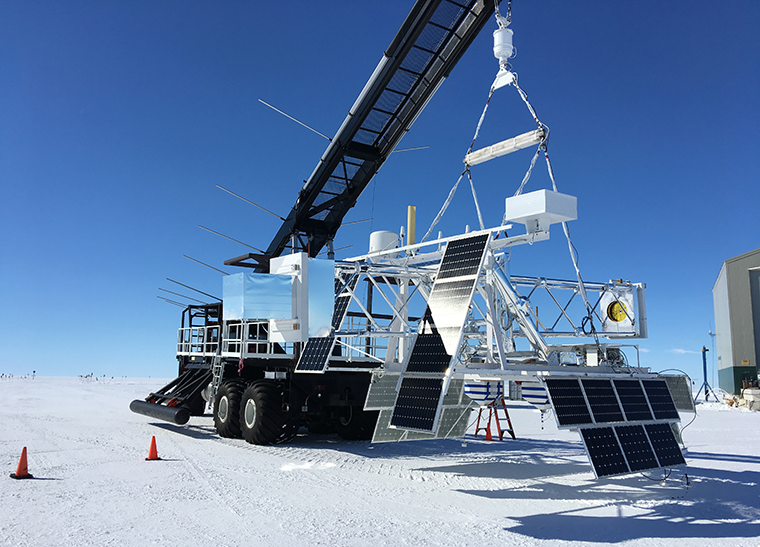}
\end{minipage}%
\begin{minipage}{0.7\textwidth}
\begin{itemize}
    \item  X-Calibur reports observations of X-ray pulsar GX~301-2 with first constraints on linear polarization in the 15-35 keV X-ray regime~\cite{2020ApJ...891...70A}.
\end{itemize}
\end{minipage}

\begin{minipage}{0.2\textwidth}
\includegraphics[width=\textwidth]{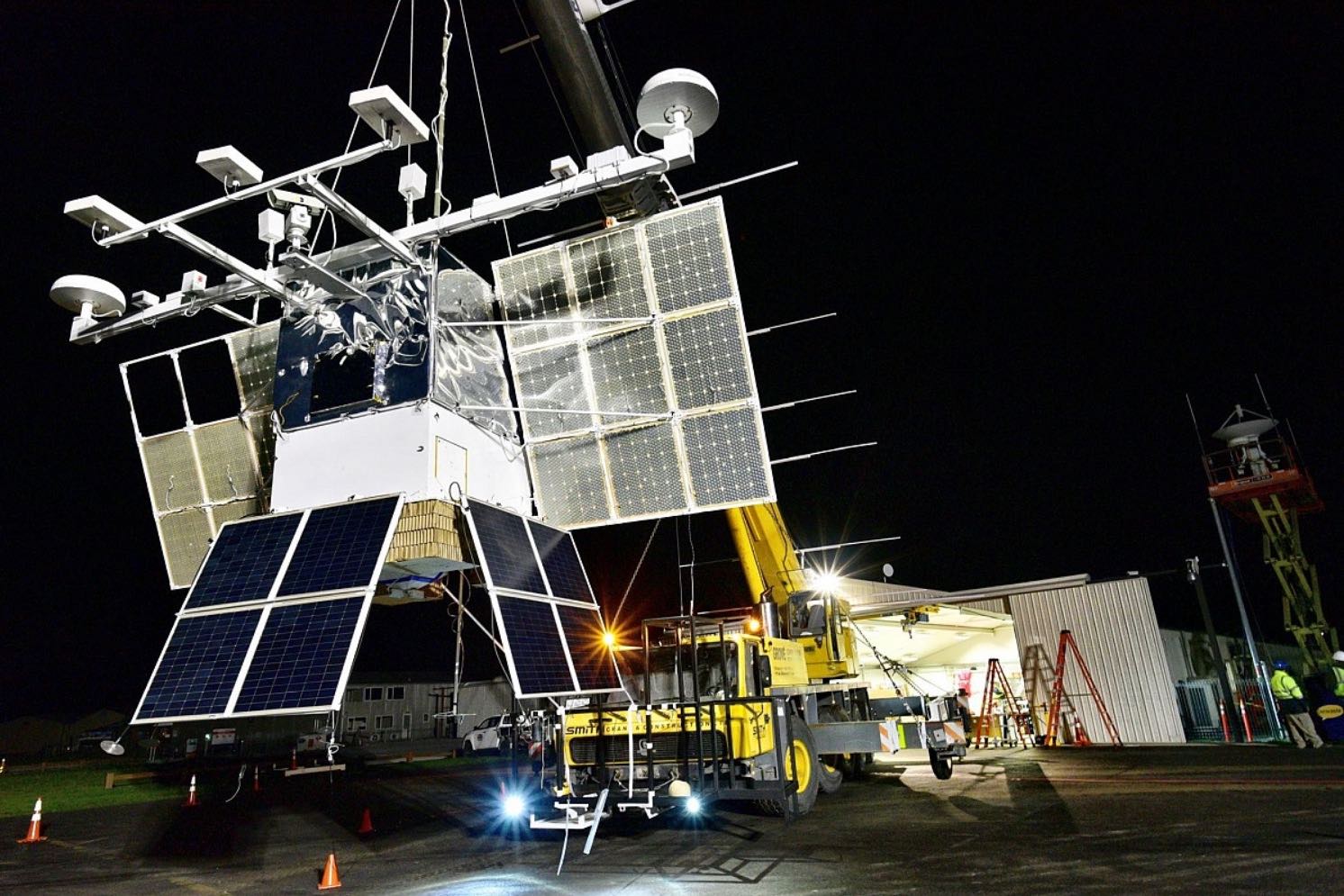}
\end{minipage}%
\begin{minipage}{0.7\textwidth}
\begin{itemize}
    \item COSI reports detection and imaging analysis of the 511~keV Galactic positron annihilation line~\cite{2020ApJ...895...44K, 2020ApJ...897...45S}. COSI set a constraining upper limit on the polarization of GRB160530A of 46\%~\cite{2017ApJ...848..119L}.
\end{itemize}
\end{minipage}

\begin{minipage}{0.2\textwidth}
\includegraphics[width=\textwidth]{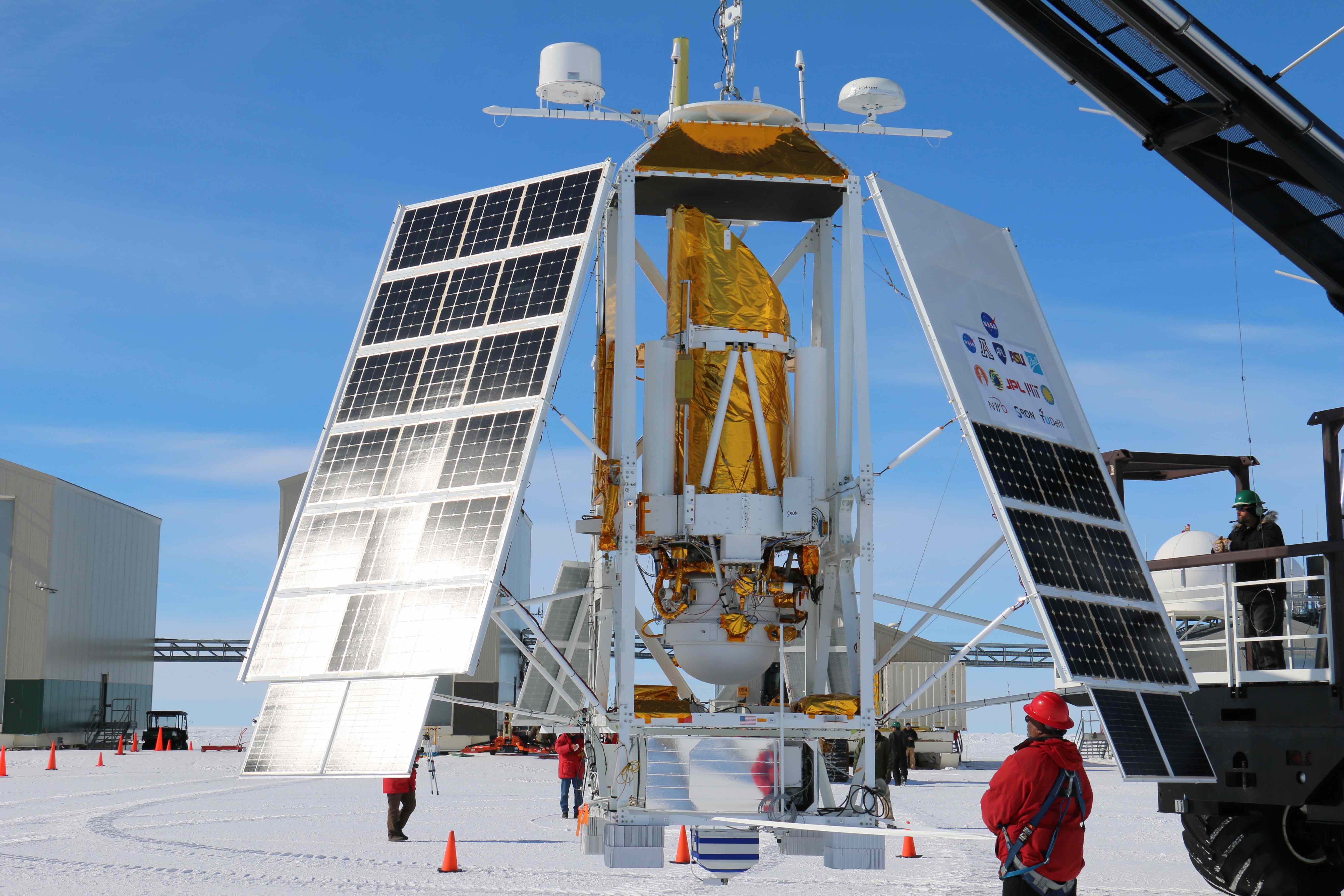}
\end{minipage}%
\begin{minipage}{0.7\textwidth}
\begin{itemize}
    \item STO-2 reports velocity-resolved spectral observations of extreme star formation regions in Carina, via CII~\cite{2019ApJ...878..120S}.
\end{itemize}
\end{minipage}

\begin{minipage}{0.2\textwidth}
\includegraphics[width=\textwidth]{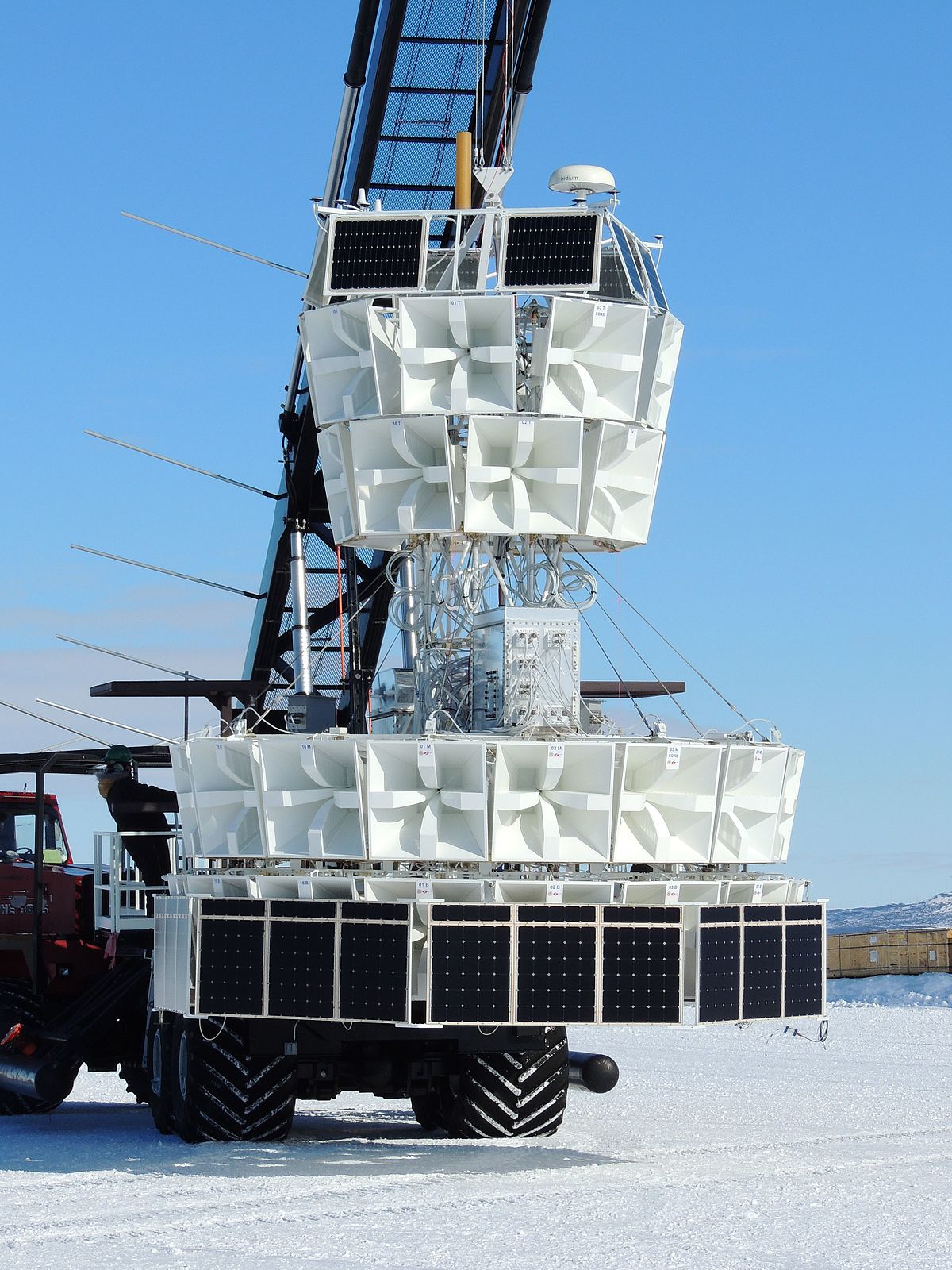}
\end{minipage}%
\begin{minipage}{0.7\textwidth}
\begin{itemize}
    \item ANITA reports strong limits on ultra-high-energy neutrino flux from the 4th flight in 2016~\cite{A4neutrino}. ANITA reports a new anomalous cosmic-ray-like event with possible beyond-Standard Model physics implications~\cite{A3upwardshower}. ANITA reports on Earth-skimming cosmic-ray-like events, including
    four near-horizon anomalous events that could signal tau neutrino-generated air showers~\cite{A4CRs}.
\end{itemize}
\end{minipage}

\begin{minipage}{0.2\textwidth}
\includegraphics[width=\textwidth]{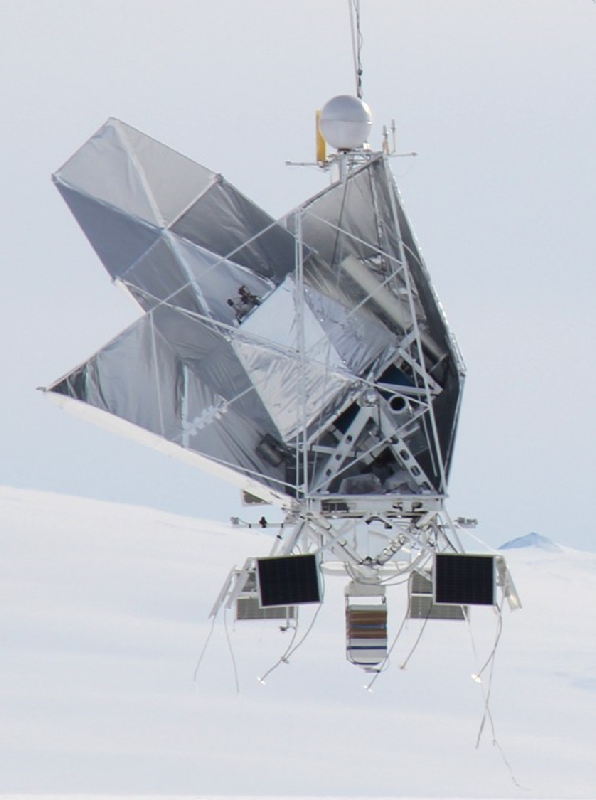}
\end{minipage}%
\begin{minipage}{0.7\textwidth}
\begin{itemize}
    \item BLAST-Pol reports a flat submillimeter polarization spectrum of the Carina Nebula~\cite{Shariff_2019}. BLAST-pol reports first Observation of the Submillimeter Polarization Spectrum in a Translucent Molecular Cloud~\cite{Ashton_2018}. BLAST-Pol finds evidence of relations between density and magnetic field orientation in Vela C molecular complex~\cite{Fissel_2016}.
\end{itemize}
\end{minipage}

\begin{minipage}{0.2\textwidth}
\includegraphics[width=\textwidth]{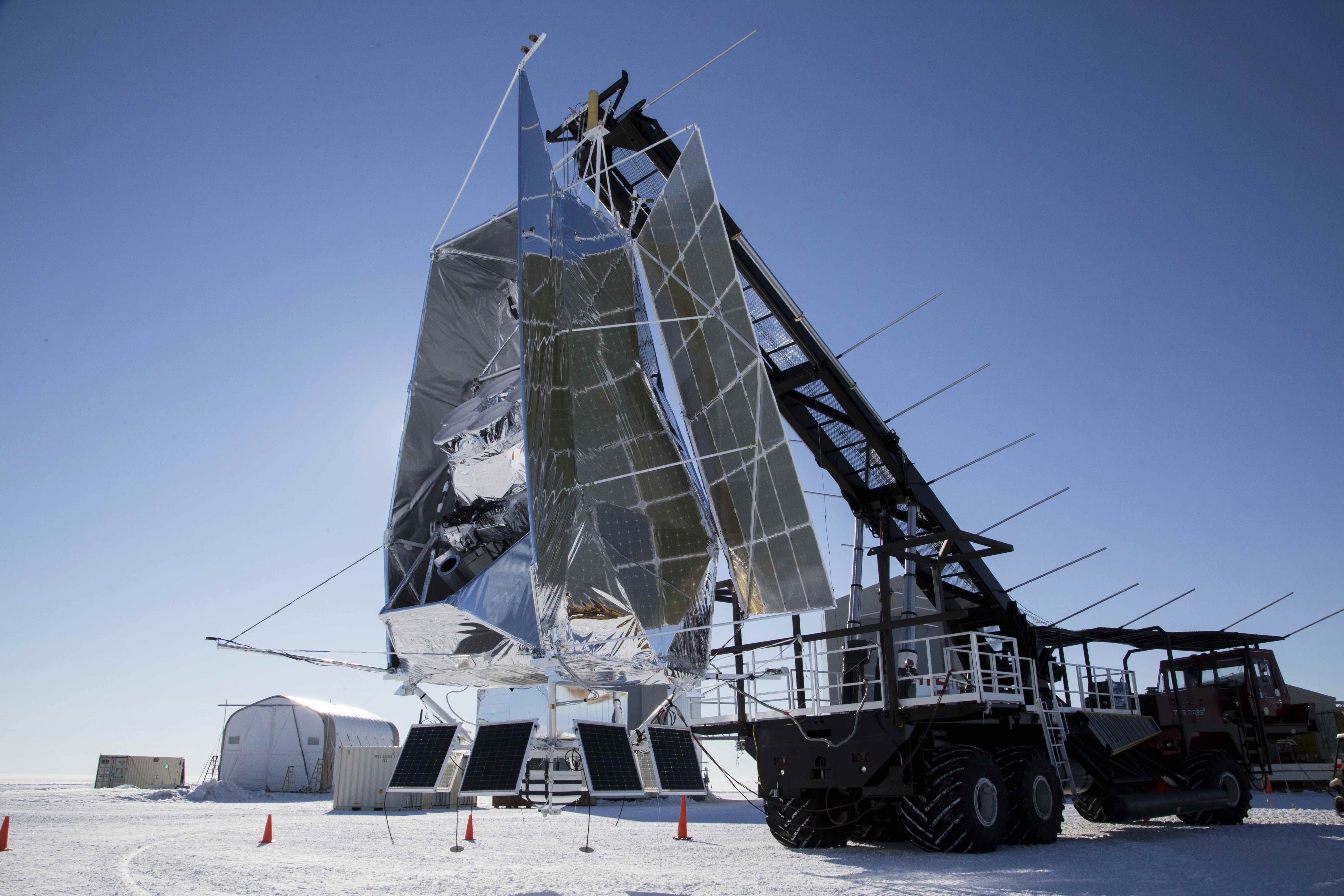}
\end{minipage}%
\begin{minipage}{0.7\textwidth}
\begin{itemize}
    \item SPIDER returns maps of CMB polarization with significantly higher sensitivity than {\it Planck}, and reports a new limit on CMB circular polarization~\cite{Nagy_2017}.
\end{itemize}
\end{minipage}

\begin{minipage}{0.2\textwidth}
\includegraphics[width=\textwidth]{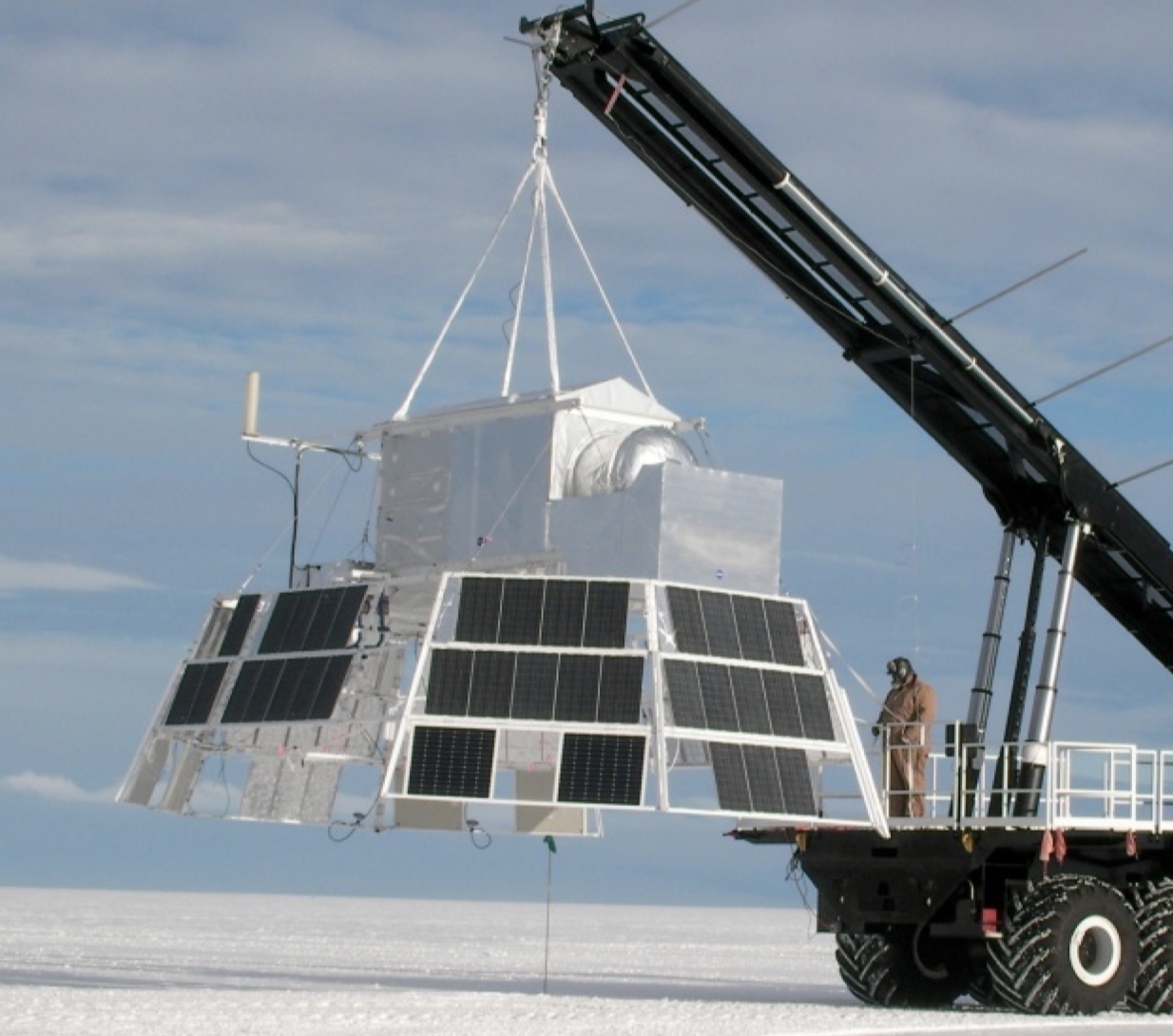}
\end{minipage}%
\begin{minipage}{0.7\textwidth}
\begin{itemize}
    \item BESS-Polar II provided most stringent constraints on the possible abundance of antihelium~\cite{Abe_2016}. BESS-Polar II reports measurements the cosmic-ray antiproton spectrum at solar minimum~\cite{PhysRevLett.108.051102}.
\end{itemize}
\end{minipage}

\begin{minipage}{0.2\textwidth}
\includegraphics[width=\textwidth]{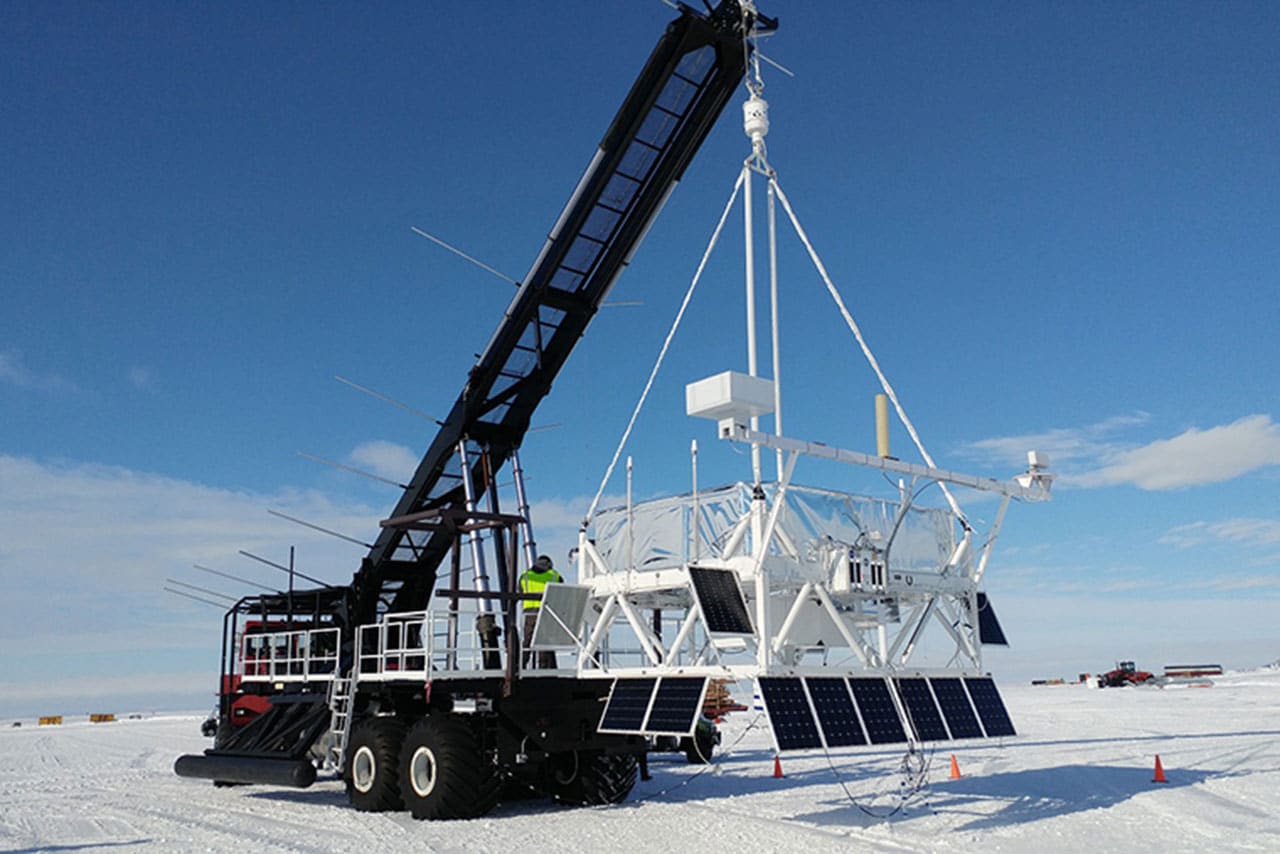}
\end{minipage}%
\begin{minipage}{0.7\textwidth}
\begin{itemize}
    \item SuperTIGER reports that abundances of elements from Iron to Zircon show a 20\% contribution from massive stars~\cite{Murphy_2016}.
\end{itemize}
\end{minipage}

The Astrophysics science groups, which account for approximately one third of all balloon payloads, have published over 150 referred journal articles and 300 conference proceedings within the past decade. There is no doubt that scientific ballooning will continue to have a significant impact in the relevant fields of study in the next decade. 

\section{Space-based Missions Matured through the Balloon Program}

Balloons are a necessary platform for technology development. 
From hand-launches to two-ton payloads, balloons allow for small and large scale tests of promising new technologies in a relevant environment, and advance their Technology Readiness Levels (TRLs).
With the added advantage of payload recovery, the instrument can be prototyped and improved on a fast timescale. %of promising technologies through testing in a relevant environment.
%NASA supports this future mission development through the Astrophysics Research and Analysis (APRA) program and the Heliophysics Flight Opportunities for Research and Technology (H-FORT).

There are numerous examples of NASA space-based missions that matured through the balloon program over the decades. 
The cosmic microwave background (CMB) balloon flights in the late 1980's and 90's laid the critical ground work for the design of the Wilkinson Microwave Anisotropy Probe (WMAP) as well as the focal plane instruments at the heart of the {\it Planck} spacecraft. %; launched 2001). 
The germanium detectors on the Reuven Ramaty High Energy Solar Spectroscopic Imager (RHESSI) %; launched 2002) 
mission were first developed and demonstrated on balloon-borne instruments.
%The Cosmic Ray Isotope Spectrometer scintillating fiber trajectory detector that was flown on the Advanced Composition Explorer (ACE) spacecraft (launched 1997) was demonstrated first in a balloon flight.
%All four instruments on the NASA's second Great Observatory, the Compton Gamma Ray Observatory (CGRO; launched 1991), were developed in balloon flights.
Three balloon flights of the cadmium-zinc-telluride (CZT) detector array produced data needed to design the Swift Burst Alert Telescope instrument, and % (launched 2004). 
balloons supported full engineering prototype flights of the Fermi Large Area Gamma-ray Telescope, which launched in 2008.

%COBE (1989), EOS-Aura (2004), 

\begin{figure}[tb]
\centering
\begin{minipage}{0.45\textwidth}
\centering
\includegraphics[height=2.2in]{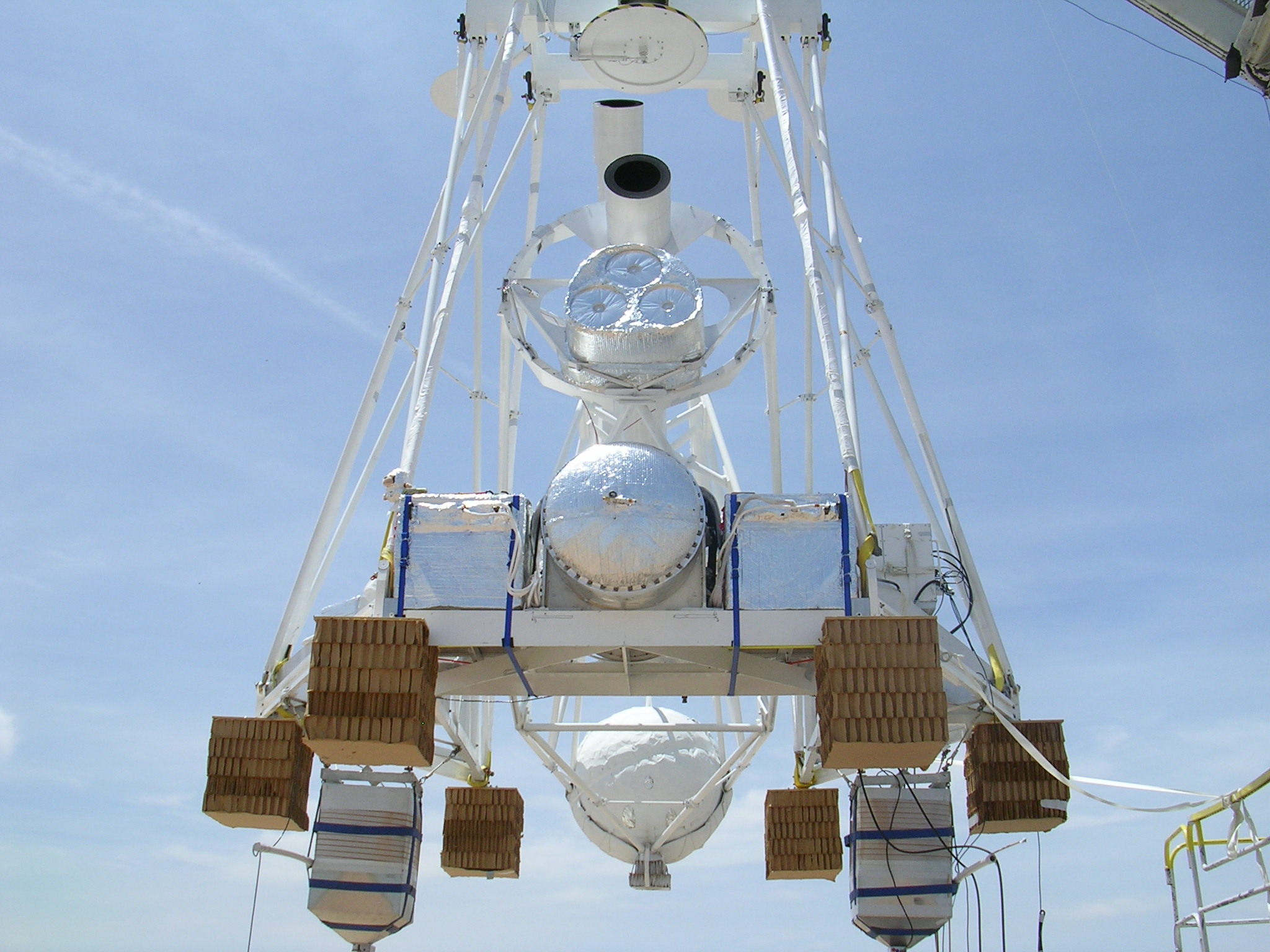}
\end{minipage}%
\hfill
\begin{minipage}{0.55\textwidth}
\centering
\includegraphics[height=2.2in]{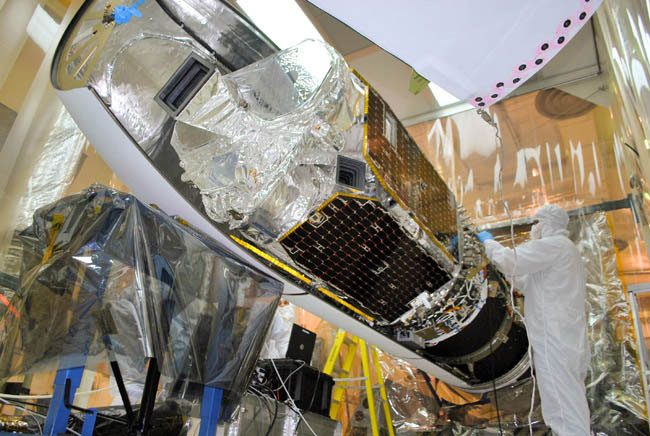}
\end{minipage}
\caption{The HEFT balloon-borne mission (left) matured the optics and detector technology that is currently being flown on NuSTAR (right).}
\label{fig:heft}
\end{figure}

More recently, the High-Energy Focusing Telescope (HEFT)~\cite{2005ExA....20..131H} lead the way for NuSTAR~\ref{fig:heft}. HEFT had its first balloon flight in 2005 and successfully demonstrated the hard X-ray focusing instrument concept with an image of the Crab Nebula.  % (launched 2012).
Grazing incidence mirror X-ray focusing technology, similar to those used in NuSTAR, was first demonstrated in near-space 
on the High-Energy Replicating Optics (HERO) balloon-borne instrument.
HERO-like optics are currently flying in the Astronomical Roentgen Telescope X-ray Concentrator (ART-XC) instrument aboard the Russian-led  Spectrum-X-Gamma mission.

\section{Training Next Generation Scientists and Engineers}

Scientific ballooning has over the years acquired another important role within NASA's broad portfolio of missions: that of a critical training ground for students and early-career investigators at many different levels. 
This is due both to the relatively short duration of the projects, which is well-matched to the several-year tenure of both students and researchers, and to the low relative costs and higher acceptable risk of balloon payload and the associated missions, which enables greater participation of students and early career scientists in project roles that would otherwise be reserved for far more senior personnel. 
Over the past 10 years, the balloon program has trained over 300 undergraduates, 200 graduate students and 100 postdocs.
The diversity of students in science groups and scientific ballooning education programs is encouraging, but needs improvement.
The impact of the balloon program on the workforce development pipeline and the education of the next generation of scientists and engineers is discussed in Chapter~\ref{ch:education}.
%In all of the PAG assessments presented here, we maintain a goal of preserving such access.

%Balloons provide unique and complete education experience: time scales consistent with the academic schedule; students can contribute to hardware with higher acceptable risk relative to spacecraft.

\begin{figure}[t]
\centering
\includegraphics[width=4.5                          in]{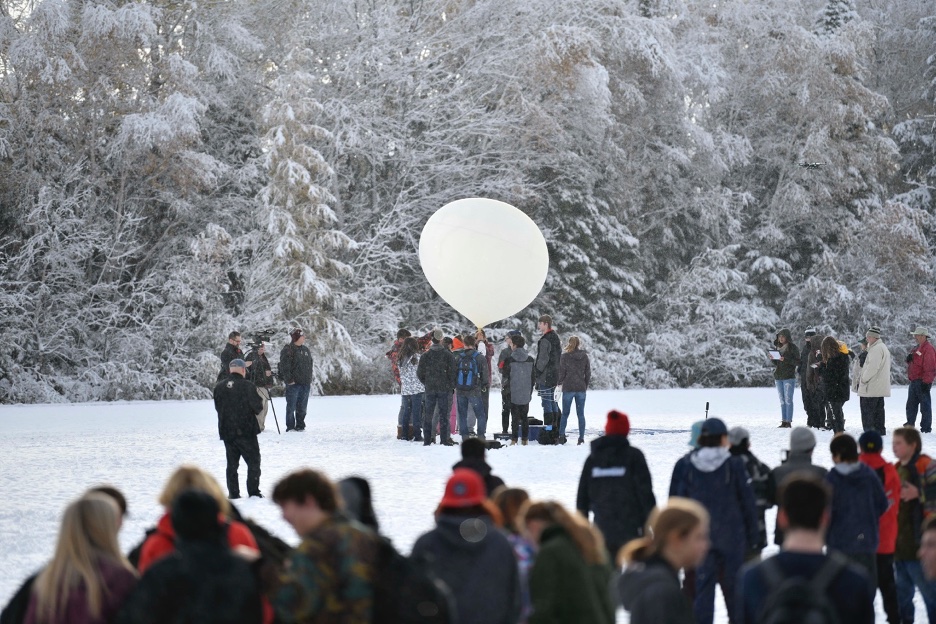}
\caption{Grade 11 students involved in the Underpass program at West Ferris Secondary School have designed and built a hand-launch payload to measure CO$_2$ in the atmosphere.}
\label{fig:underpass}
\end{figure}

\section{Terms of Reference for the Program Analysis Group}

To establish the baselines on which this report is prepared, we quote here a portion of the Terms of Reference which establish the scope of this effort~\cite{PAGletter}:

\begin{quotation}
The Scientific Balloon Roadmap Program Analysis Group (Balloon Roadmap PAG) serves as a community-based, interdisciplinary forum for soliciting and coordinating community analysis and input in support of the NASA Scientific Balloon Program and its capability planning, development and activity prioritization. 
It provides findings of analyses to the NASA Astrophysics Division Director. 
The material will also be made available to the upcoming 2020 Astrophysics Decadal Survey.

The objective of NASA's Balloon Program is to provide high-altitude scientific balloon platforms for NASA's scientific and technological investigations. 
These investigations include fundamental scientific discoveries that contribute to our understanding of the Earth, the solar system, and the universe. 
Scientific balloons also provide a platform for the demonstration of promising new instrument and spacecraft technologies that enable or enhance the objectives for the Science Mission Directorate Strategic Plan, as well as training the next generation of scientists and engineers to carry out NASA's future missions.

The Balloon Roadmap PAG enables direct regular communication between NASA and the community, and within the community, through public meetings that give the community opportunities to provide its scientific and programmatic input. 
Structurally, the Balloon Roadmap PAG Chair and the Balloon Roadmap PAG Executive Committee (EC) are appointed members whose responsibilities include organizing meetings and collecting and summarizing community input with subsequent reporting to the Astrophysics Division Director.
The full Balloon Roadmap PAG consists of all members of the community who participate in these open meetings.

\vspace{0.75cm}

\textbf{The Balloon Roadmap PAG is tasked to carry out the following:}
\begin{itemize}
\item Articulate and prioritize the key scientific drivers and needed capabilities for NASA's Balloon Program \textcolor{royalblue}{[see Chapter~\ref{ch:recommendations}]};
\item Evaluate the expected capabilities of potential balloon-borne missions for achieving the science goals and maturing important and strategic technologies of SMD \textcolor{royalblue}{[see Chapter~\ref{ch:science}]};
\item Evaluate Balloon Program goals, objectives, investigations, and required measurements on the basis of the widest possible community outreach \textcolor{royalblue}{[see Chapter~\ref{ch:education}]};
\item Articulate and prioritize focus areas for needed balloon mission technologies \textcolor{royalblue}{[see Chapter~\ref{ch:tech} and \ref{ch:recommendations}]}; and
\item Summarize and assess balloon launch opportunities and mission capabilities provided by emerging commercial providers  \textcolor{royalblue}{[see Section~\ref{sec:commercial}]}.
\end{itemize}

\end{quotation}

\noindent {\large \textcolor{royalblue}{\textbfit{Balloon Program Analysis Group and Community Input}}}

Through over six different townhall-style public meetings conducted in 2018-2019, the Executive Committee (EC) and the PAG have coordinated and compiled input from the wider community in response to these tasks. This includes sessions at the January 2018 and 2019 American Astronomical Society meetings, at the summer 2018 COSPAR meeting, and in 2019 at the December American Geophysical Union meeting and the April American Physical Society meeting. In each of these meetings
we encouraged and received numerous presentations, and
we also encouraged members of the scientific community to submit white papers describing their ideas 
for the scientific directions of ballooning in the next decade. All of these contributions were considered in
the execution of this report.

Our goal in this report is thus to distill the scientific context and directions from a diverse and creative community of investigators, educators, and technologists, in a way that will help to prepare us to meet the scientific needs of the next decade and the scientific needs of a new generation of investigators who are engendering new avenues of research.

\section{Overview of Document}

In the following chapters, we identify high priority scientific drivers in all of the major disciplines relevant to the NASA balloon program, including Astrophysics (Sec.~\ref{sec:astro}), Earth Science (Sec.~\ref{sec:earth}), Planetary Science (Sec.~\ref{sec:planetary}), and  Heliophysics (Sec.~\ref{sec:helio}). We give an overview of the current Balloon Program capabilities and technologies (Chapter~\ref{ch:tech}) as well as the Workforce Development Pipeline and Education through the program (Chapter~\ref{ch:education}).
Chapter~\ref{ch:recommendations} concludes this document with a prioritized set of findings and recommendations to the NASA Balloon Program that are needed to achieve the science goals and strategic technologies of the SMD.
%Here we list the top priority recommendations to the Balloon Program that are discussed in more detail in .
%overall (Ch.~\ref{sec:recommendations}). 
%Many additional recommendations and findings are captured in the final chapter of the report.

\noindent {\large \textcolor{royalblue}{\textbfit{Priority Recommendations}}}

The recommendations from the PAG fall into four general categories: Balloon Capabilities; Launch Site \& Facilities; Funding Opportunities; and Workforce Development and Education \& Outreach. The EC has prioritized the recommendations in each of these four categories based on the input from the PAG, and we list the top recommendation from each category here. 

\noindent\textbf{The complete set of  of findings and recommendations is presented in Chapter~\ref{ch:recommendations}.}

\begin{itemize}

{\item  \textbf{\textcolor{royalblue}{Balloon Capabilities:} \\ Super-pressure balloons.} 
The PAG recommends that the NASA super-pressure balloon program continue to pursue the goal of 100-day flights including at mid-latitudes. 
NASA should strive to advance the lift capability and float altitude to the point where SPBs are commensurate with current zero-pressure balloon capabilities.}

{\item \textbf{\textcolor{royalblue}{Launch Sites \& Facilities:} \\ Multiple-Payload building in Wanaka.}
The PAG commends the NASA Astrophysics Division and Balloon Program for developing and maturing the 18 Mcf super-pressure balloon, and the new launch facility in Wanaka, New Zealand, which supports SPB launches.  
These developments will enable new science investigations from Wanaka with science returns comparable to significantly more costly space flight missions, and complementing to NASA flagship missions. 
The PAG recommends continued support for the growth of this facility, including a new payload integration building that could accommodate multiple payloads. 
}

{\item \textbf{\textcolor{royalblue}{Funding Opportunities:} \\  Earth \& Planetary Science.} 
The PAG notes that currently NASA scientific ballooning offers no funding opportunities in Planetary science, and very limited opportunities in Earth Science.
As we have detailed in previous chapters, there are many different scientifically compelling investigations in both Earth and Planetary Sciences.
The PAG strongly recommends that NASA to consider ways to allow balloon payloads to compete for Earth and Planetary Science funding opportunities. }

{\item {\bf \textcolor{royalblue}{Workforce Development, Education \& Outreach:} \\ Workforce Development.} 
The PAG recommends that the community and Balloon Program Office work to foster high altitude ballooning as a key element in NASA’s workforce development pipeline from pre-college to new scientists. 

The PAG recommends specifically that NASA:
\begin{enumerate}[(i)]
\item Engage with nationwide entities that are already supporting transdisciplinary learning. 
\item  Improve accessibility to flight options for groups involved with experiential projects. 
\item Develop safety and performance standards for balloon projects at a level significantly below spacecraft standards.
\item Support technical workshops for entry level scientists to provide a venue for instrument design sharing, lessons learned, and best practices.
\end{enumerate}
}
\end{itemize}

\hyperlink{TOC}{Return to table of contents.}

\chapter{Science Drivers for the Balloon Program}
\label{ch:science}

\begin{wrapfigure}{r}{0.3\textwidth}
\centerline{~~\includegraphics[trim=0 0 0 0, clip, width=1.8in]{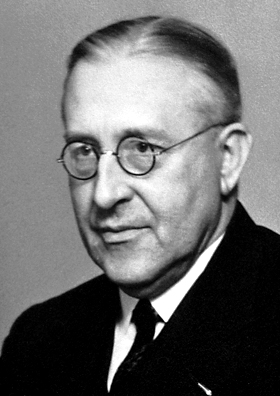}}
\caption{ \footnotesize \it   Victor Hess, discoverer of cosmic rays via a balloon flight in 1912,
for which he received the Nobel Prize in 1936.
\label{Hess}
}
\end{wrapfigure}
It has been just over a century since Austrian physicist Victor Hess flew three electroscopes up to an altitude of nearly 5~km in a manned 1912 balloon flight~\cite{Hess28} that Bruno Rossi called ``the beginning of one of the most extraordinary adventures in the history of science~\cite{Rossi64}.'' 
His efforts represent the dawn of what is now known as {\it particle astrophysics}, although it took several more decades for physicists such as Robert Millikan and Arthur Compton to coin the term ``cosmic rays,'' and begin to understand their true nature. 
Hess was awarded the 1936 Nobel Prize in physics for the discovery of cosmic rays, shared with Carl D. Anderson for the discovery of the positron. 
Hess' prize was the first Nobel with direct connections to phenomena in astrophysics, but it would not be the last in which balloon-borne instruments played an important role.

During the middle of the twentieth century, balloons made numerous ground-breaking science discoveries. 
These include solar measurements by Charles Abbot, director of the Smithsonian Astrophysical Observatory, establishing for the first time the much larger flux of solar energy at altitudes of up to 32 km~\cite{Abbot44}. 
In the early 1950's, balloons were used to assist sounding rocket launches, a hybrid system called a ``Rockoon,'' which enabled James Van Allen to gather the initial evidence for trapped particle radiation belts, now known as the Van Allen belts, in the Earth's exosphere. 
In a pattern that has continued to the present, this balloon-assisted mission provided critical information used to develop the Explorer I satellite which conclusively identified the belts~\cite{vanAllen}.
Several years later, the first balloon-borne astronomical telescope, with a 12" primary mirror and special closed-circuit camera guiding  system, was launched as part of the Stratoscope project, and made photographs of sunspots a higher resolution than any instrument had ever done~\cite{Stratoscope}. 

The follow-up mission, Stratoscope II, increased the scientific scope dramatically with a functional observatory as a balloon payload, with a 91 cm primary mirror, and a pointing system which allowed the instruments to measure infrared spectra from the Moon, Mars, Venus, and other stellar targets. 
The data acquired through this program had significant impact on both planetary and stellar astrophysics of the time, and was among the first scientific balloon payloads managed by NASA, under Nancy Grace Roman, who later became NASA's first chief astronomer.

The 1960's and 1970's saw a rapid rise in scientific ballooning in astrophysics, as well as atmospheric and planetary physics, in parallel with the development of both manned and unmanned spacecraft. 
The favorable economics of balloons and their large payload capacity ensured that scientific ballooning remained a vital and active part of the new National Aeronautics and Space Administration. 
In the 1960's the National Scientific Balloon Facility (NSBF) was established under the National Science Foundation, initially in Colorado, but soon moved to its current home in Palestine, Texas. 
In the early 1980's NSBF was transferred to NASA oversight, where it has remained under the purview of the Astrophysics Division since that time. In the following two decades, experiments developed for the balloon platform revolutionised the field of observational cosmology through observations of the Cosmic Microwave Background.  These missions, including Maxima and Boomerang, returned the first resolved, high signal-to-noise images of the CMB anisotropy and established the $\Lambda$CDM paradigm as the standard model of cosmology \cite{b98_nature,b98_lange,Lee_2001}.

In the following decade technologies developed for, and demonstrated by the balloon platform went on to enable pioneering orbital missions, such as {\it Planck} and {\it Herschel}~\cite{psb_methods}.
In 2003, the loss of the Columbia Space Shuttle over West Texas during reentry prompted a name change to the Columbia Scientific Balloon Facility (CSBF).

%Launched in 1989, the COsmic Background Explorer (COBE) NASA spacecraft found evidence for the blackbody spectrum and anisotropy of the relic Big-Bang cosmic background radiation. 
%Two balloon-borne payloads, MAXIMA and BOOMERANG, along with a second space mission, the Wilkinson Microwave Anisotropy Probe (WMAP), later confirmed and extended these results. 
%The COBE investigators John Mather and George Smoot were later awarded the Nobel Prize in 2006 as a result.

As the 21st century now enters its third decade, the CSBF and the NASA Balloon Program Office (BPO) that oversees it have demonstrated a new level of professionalism within scientific ballooning that has established this discipline as one of the key launch and science missions providers within NASA, with a host of payloads that have rivaled spacecraft payloads in their science return, and a new community of investigators who have discoveries in astrophysics and many disciplines beyond. 
We begin within the Astrophysics realm as an acknowledgement in part of the roots of scientific ballooning in the 20th century, but as the subsequent sections will show, scientific ballooning's contributions extend far beyond, and we anticipate that growth will not abate in the next decade.

\section{Astrophysics}
\label{sec:astro}

\begin{tcolorbox}[colback=royalblue!8!white,colframe=royalblue,fonttitle=\bfseries,title=Relevance to SMD Science Goals]
Balloon-borne astrophysics addresses all of the questions posed in Enduring Quests Daring Visions: \textit{Are We Alone? How Did We Get Here?} and \textit{How Does Our Universe Work?}

\vspace{0.75cm}

\textbf{Particle and High-Energy Astrophysics} (Sec.~\ref{sec:astroparticle}) studies specifically address the Stellar Life Cycles and the Evolution of the Elements, as stated in Section~3.1 of NASA's 2013 Astrophysics Roadmap: 
\begin{quote}
The study of cosmic ray particle acceleration is an important physics question in its own right, but this high-energy particle component of any actively star-forming galaxy is also responsible for the creation of some crucial elements (lithium, beryllium, and boron). It bathes the galaxy in a diffuse glow of high- energy gamma rays, which represents a significant fraction of its total energy output; the presence of the cosmic ray component can even play a substantial role in the self-regulation of the star-formation rate.
\end{quote}
The studies of the high-energy universe achievable from a balloon platform also address the science goals outlined in Section 4.2 of the Roadmap, \textit{Revealing The Extremes of Nature}, through x-ray and gamma-ray observations of nature’s most powerful engines.

\vspace{0.75cm}

Studies of \textbf{Exoplanets} (Sec.~\ref{sec:exoplanets}) directly address the question \textit{Are We Alone?} posed in Enduring Quests Daring Visions, while balloon-borne \textbf{Stellar Astrophysics} answers \textit{How Did We Get Here?}

\vspace{0.75cm}

The field of \textbf{Observational Cosmology}  (Sec.~\ref{sec:cosmology}) uses empirical data to better understand the evolution, contents, and growth of structure within the Universe in the context of the general theory of relativity and standard model physics.  The observed dominance of dark matter-energy represents a unique opportunity to use the Universe as a laboratory for beyond-standard model physics.

\end{tcolorbox}

% These are files for individual Astro subdisciplines, in separate sections of this chapter
\subsection{Particle and High-Energy Astrophysics}
\label{sec:astroparticle}

The study of high-energy particles and photons entering our atmosphere from space was the first astrophysics discipline to rely extensively on balloon-borne instruments. 
In initial efforts, altitude played the role of a proxy for the degree of absorption of cosmic particles, but as payload altitudes increased, it became possible to achieve atmospheric mass overburdens that were a fraction a percent of the sea-level pressure allowing for groundbreaking observations of the high-energy universe.

\vspace{5mm}
\begin{figure}[!ht]
\begin{center}
\centerline{\includegraphics[width=6in]{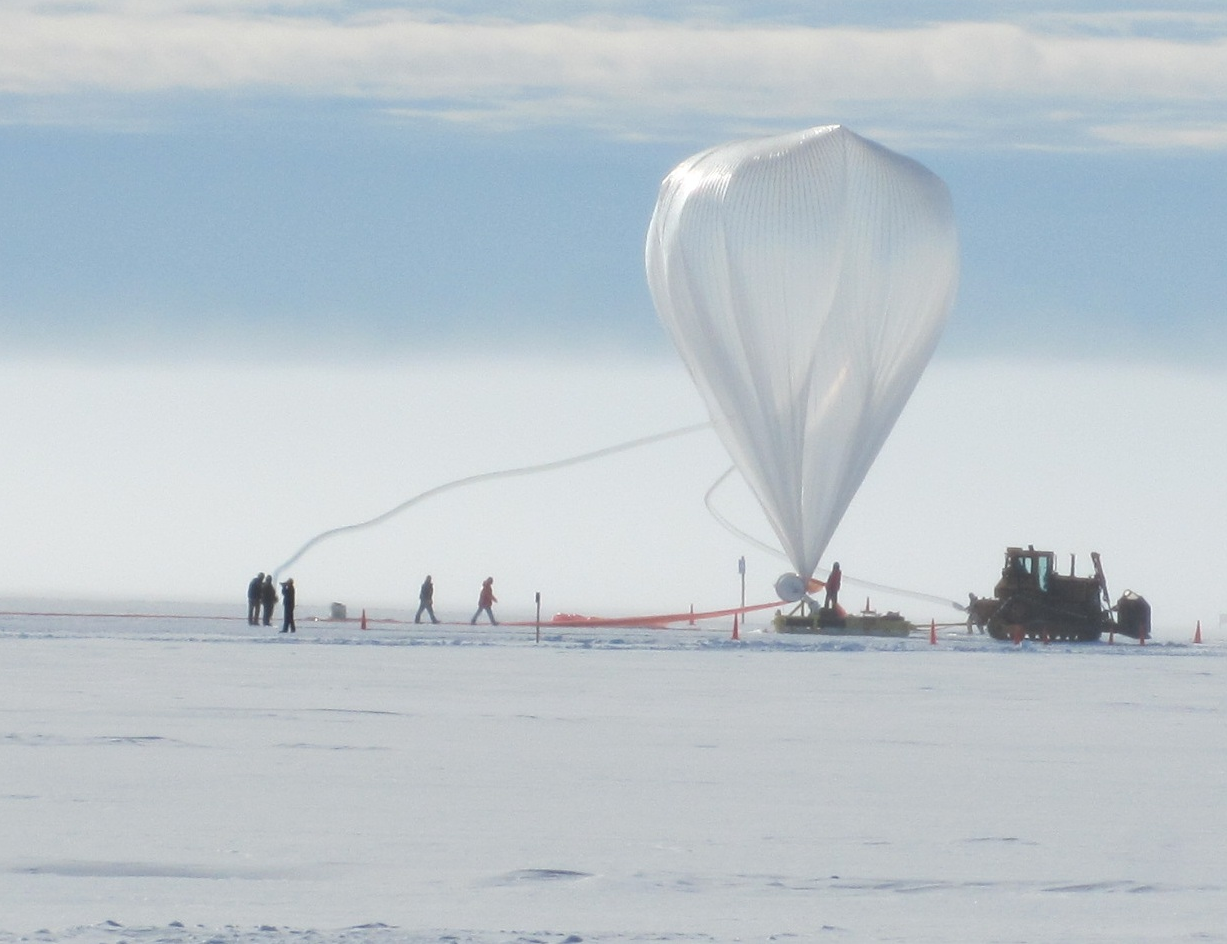}}
\caption{Balloon inflation for the LDB Super-Trans-Iron-Galactic Element Recorder (SuperTIGER) launch in 2013.
\label{STlaunch}}
\end{center}
\end{figure}

In recent efforts, this has allowed the direct study of the composition and flux of the primary cosmic rays up to single-particle energies approaching $10^6$~GeV (1~PeV), where the population of particles includes many different atomic species up to and beyond iron, with implied center-of-momentum energies that rival those of the CERN Large Hadron Collider. 
The relatively low cost of lofting large masses has enabled direct collecting areas of several square meters, an important parameter in making statistically significant measurements at the very low fluxes encountered at the highest particle and photon energies.

The stratospheric access to near-space with a very low atmospheric overburden also allows investigation into potential exotic constituents of the primary cosmic ray flux, specifically antimatter, which is otherwise unobservable.
Some such constituents, such as anti-helium, if detected, could signal entirely new physics.

In fact, beyond the PeV energy scale, direct detection of cosmic particles via an onboard detector target becomes impractical.
However, from a stratospheric vantage point, a balloon payload can also observe other potential particle interaction targets, including Earth's atmosphere, and homogeneous land and sea, or even ice masses, such as the Antarctic ice sheets. 
These opportunities have served to open up a new window in ultra-high energy particle astrophysics of both cosmic-rays and neutrinos.

High-energy astrophysics, namely gamma-ray and x-ray astronomy, has benefited from the balloon platform with large loft capabilities and high-altitude observations.
NASA's high-energy missions \textit{Swift}, \textit{Fermi}, and the Compton Gamma-Ray Observatory (CGRO) all had significant instrument development through balloon flights prior to the successful satellite missions. 
Scientific balloons continue to provide a unique and invaluable platform for high-energy astrophysics instrument development, in addition to novel scientific pursuits.

\noindent {{\textbfit{\textcolor{goldenpoppy}{Cosmic Ray composition and spectrum below the ``knee''}}}}

Measurements of the composition and flux of cosmic rays in the GeV to PeV ($10^{9-15}$~eV) energy range provides unique visibility into important high-energy astrophysics processes in the galaxy and interstellar medium. 
Galactic cosmic rays (GCR) begin to dominate the solar-system particle flux at GeV energies, and such measurements  must be done above the terrestrial atmosphere. 
The GCR dominate for another 5 decades in energy up to the PeV scale, when their gyroradii in the galactic magnetic fields become large enough to escape to intergalactic space. 
This causes a break in the energy spectrum known as the ``knee.'' 
Prior to this, the fluxes of a wide range of elements and their isotopes can be used to test a wide variety of theory about GCR acceleration and sources.

This topic was in fact one of the central drivers for iconic spacecraft-borne payloads such as the Pioneers, Voyagers, and Ulysses, and cosmic-ray missions have remained a part of NASA's astrophysics mission portfolio into the new century.
In 1997, NASA launched the Advanced Composition Explorer (ACE) to the L1 Lagrange point, and the spacecraft is still producing data as of April 2020, with propellant enough to maintain its orbit until $\sim 2024$. 

Important spacecraft instrumentation has also been derived from balloon-borne cosmic-ray instruments.  
As one example, scintillating-optical-fiber hodoscopes suitable for spacecraft were originally used and proven in cosmic-ray instruments on balloons.  
They were subsequently incorporated into the Cosmic Ray Isotope Spectrometer (CRIS), which is still performing flawlessly on the ACE spacecraft after 23 years in space.  
This same technology is being used since 2015 in a hodoscope of the CALorimetric Electron Telescope (CALET) on the International Space Station, and as a component of the Fermi Gamma-Ray Telescope’s anti-coincidence shield.

The Payload for Antimatter Matter Exploration and Light-nuclei Astrophysics (PAMELA) was launched in 2006 with a goal of  precise measurements of cosmic antiparticles, electrons, and light nuclei in the energy range from 0.1 to 1000~GeV, and flew until 2016, producing numerous results on GCR. 
Like ACE, PAMELA benefited from prior balloon flights to validate some of the technologies flown.
More recently, the second Alpha Magnetic Spectrometer payload (AMS02) was launched and deployed to the International Space Station (ISS) in 2011 and continues operation.

Despite the dominance of space-based over the lower range of the GCR spectrum, balloon payloads have and still do play a critical role in the exploration of particles over the wide energy regime, since they are able to deliver large mass, large area to altitudes high enough for primary composition measurements at low cost. 
As important as spacecraft observations are for study of cosmic rays, balloon-borne investigations are underway, or planned for the near term, that are not yet possible with instruments on spacecraft.

\noindent\textcolor{blue}{\textbfit{Notable recent and planned cosmic-ray payloads}}

\begin{wrapfigure}{r}{0.45\textwidth}
\centerline{~~\includegraphics[trim=0 0 0 0, clip, width=2.8in]{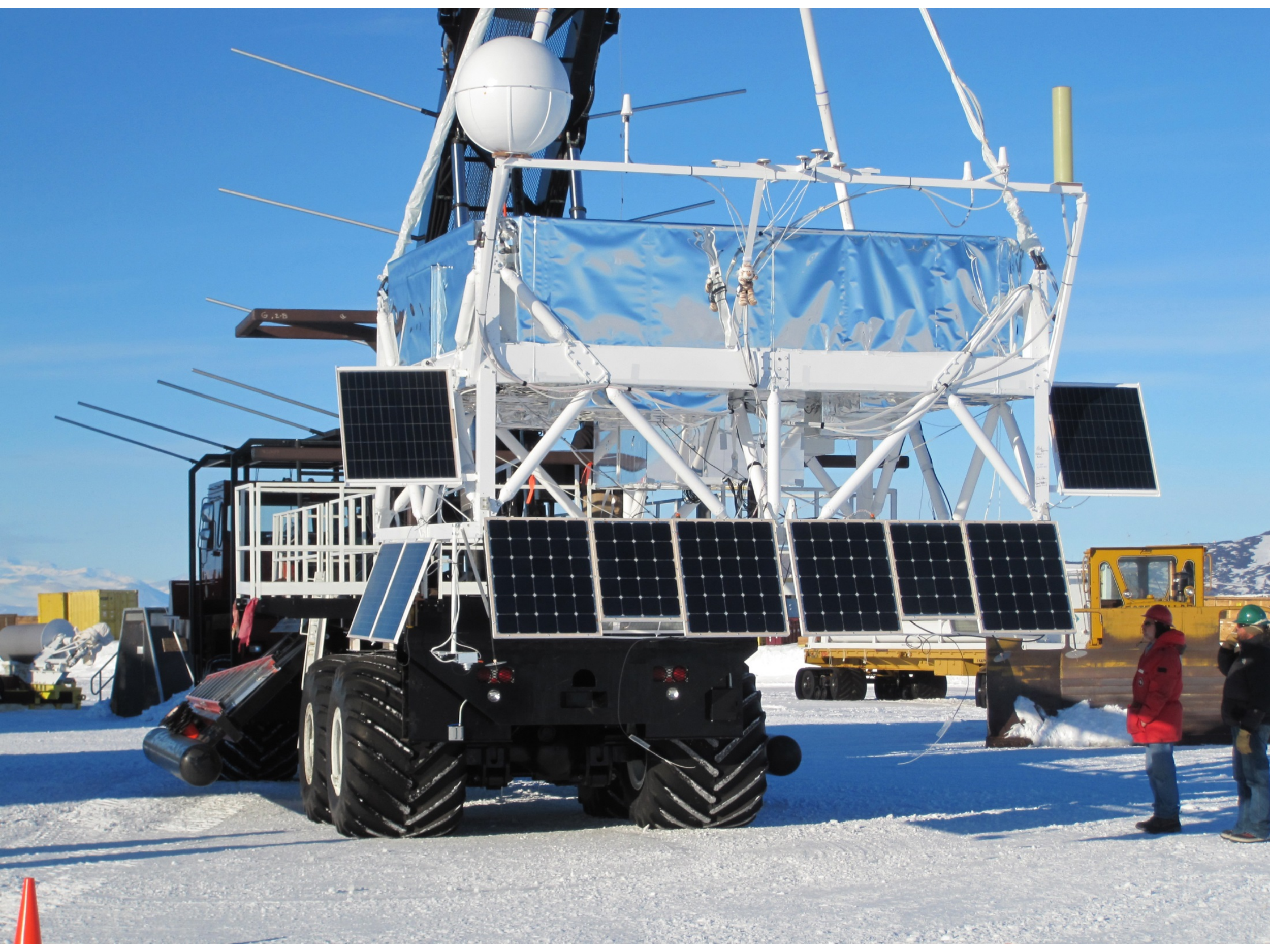}}
\caption{SuperTIGER prepares for launch in Antarctica in 2019.
\label{supertiger}
}
\end{wrapfigure}

The two primary thrusts in galactic cosmic-ray astrophysics are the determination of the composition, and the absolute fluxes for
the various elemental constituents. Theoretical model builders develop predictions for these observables for various
source classes as a function of composition and energy. Two recent payloads that have addressed these complementary
aspects of the study of GCR are the Cosmic-Ray Energetics and Mass (CREAM) payload, which achieved a total of 169 days
of flight exposure over five LDB flights in 2004-2012, and the Trans-Iron Galactic Element Recorder (TIGER, and Super-TIGER)
which has achieved similar exposure and remains an active payload.

The SuperTIGER instrument (Fig.~\ref{supertiger}), which most recently had a month-long flight over Antarctica in December 2019, takes advantage of balloon capability of handling very large area ($\sim 10$~m$^2$), heavy (2 ton) instruments that are necessary for determining the elemental composition of the rare ultra-heavy (UH) elements with atomic number Z > 30.  Data from this 2019 flight, along with data from a flight in December-January of the 2012-2013 austral summer have demonstrated that ST has the charge resolution necessary to determine relative abundances of individual elements at least up to Z = 56.  With increased statistics from an additional ST flight we expect to begin distinguishing between Supernovae and Binary Neutron Star Mergers as sites where the r-process elements originate.  Ideally these measurements would be made on spacecraft outside the earth’s atmosphere, and ST is a proof-of-principle for a new instrument, TIGERISS, which is being proposed for the International Space Station.

The High Energy Light Isotope eXperiment (HELIX) is a magnetic spectrometer that will be ready for balloon flight in the next year or two.  It is designed to measure light cosmic-ray isotopes from 0.2 to 10 GeV/nucleon, particularly optimized for determining the ratio of radioactive 10Be to stable 9Be.  This ratio has already been measured by CRIS below 0.5 GeV/nucleon, thereby putting serious constraints on the origin and galactic propagation of low-energy cosmic rays.  Anomalies in the high-energy positron fraction have been variously interpreted as intrinsic astrophysical source effects or as effects of galactic propagation or even exotic physics such as annihilating dark matter.  Extending the Be isotope measurements into the range of several GeV/nucleon will constrain the galactic propagation models and either limit or support exotic possibilities.  Although AMS-02 is a magnetic spectrometer on the International Space Station, it lacks the mass resolution to address the Be isotopes that HELIX will study.

\begin{wrapfigure}{r}{0.33\textwidth}
\centerline{~~\includegraphics[trim=0 0 0 0, clip, width=2in]{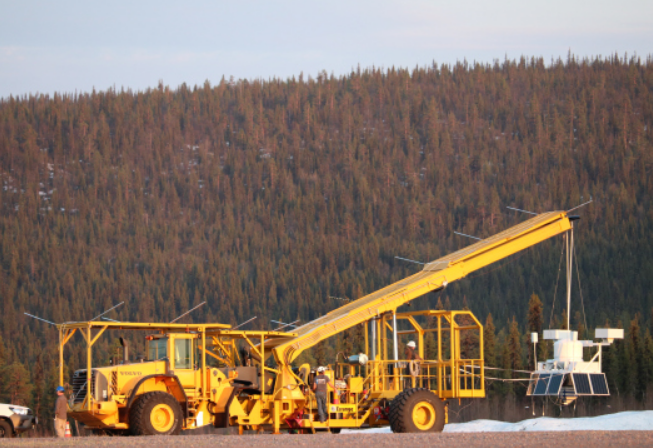}}
\caption{Aesop-lite prepares for launch in Sweden in 2018.
\label{aesoplite}
}
\end{wrapfigure}

The Anti-Electron Sub-Orbital Payload (AESOP-Lite, Fig.~\ref{aesoplite}) instrument is a balloon-borne compact magnetic spectrometer designed to measure low energy (20 – 300 MeV) primary electrons and positrons, with a primary objective of exploring the origin of the negative spectral index in the cosmic-ray electron spectrum below 100 MeV.  Also, measurements of spectra of electrons and positrons separately at low energy are essential for understanding the solar modulation of galactic cosmic rays, which is different in alternate eleven-year periods of the solar cycle, and for understanding the charge-sign dependence in geomagnetic cutoff diurnal variations.  This instrument has had one successful high-latitude balloon flight from Sweden to Canada in 2018, on a 40-million cubic foot balloon.  It would be able to reach lower-energy electrons and positrons if the 60-million cubic foot balloon under development were operational.  We note that this measurement is only possible on balloons flying in polar regions where the Earth’s geomagnetic cutoff is close to zero.  Any satellite in low-earth orbit can have at most a small fraction of its time in such regions of low cutoff.

%\subsubsection*{Cosmic Antiparticles}
\noindent {{\textbfit{\textcolor{goldenpoppy}{Cosmic Antiparticles}}}}

At tens to hundreds of GeV, cosmic antiparticles are observed by both spacecraft and balloon-borne payloads. The spectral energy distribution of positrons and antiprotons may indicate the presence of nearby pulsars in the solar neighborhood, but may also arise from dark matter decay scenarios. Detection of certain particle species, such as anti-helium nuclei or anti-deuterons, could provide compelling evidence for particle dark matter. 

\textcolor{blue}{\textbfit{Notable recent and planned cosmic antiparticle payloads}}

The General AntiParticle Spectrometer (GAPS) is planning an Antarctic balloon flight in the 2021 or 2022 Austral summer to measure the flux of low-energy (< 250 MeV/nucleon) cosmic-ray antiprotons and anti-deuterons as well as anti-helium.  GAPS will be the first experiment dedicated to detecting anti-deuterons, a signal of which can provide a smoking-gun signature of dark matter annihilation.  GAPS identifies antiparticles through a unique technique in which an antiparticle stopping in a ten-layer Si(Li) tracker forms an exotic atom, which quickly decays to produce X-rays of specific energies and a star of pions and protons from the annihilation event.  Measurements of the X-ray energies and the trajectories in the tracker of the charged particles from the star discriminate against possible backgrounds and give GAPS complementary capabilities to magnetic spectrometers.  Antimatter at these energies can be detected near Earth only in polar regions, where there is essentially no geomagnetic cutoff.  With its large aperture, efficient detection technique, and lack of geomagnetic cutoff, GAPS in 100 days will achieve an improved anti-deuteron sensitivity compared to AMS-02 in 10 years, in a lower kinetic energy band.

\noindent {{\textbfit{\textcolor{goldenpoppy}{Ultra-high energy cosmic rays and neutrinos}}}}

The mystery of the sources of the highest energy cosmic rays, at Exavolt ($10^{18}$~eV) to Zetavolt ($10^{21}$~eV) energies, remains a compelling problem in the field of particle astrophysics. Detectors using Antarctic ice have now begun to open a new observing window for neutrinos in this area, and payload sensitivities to EeV neutrinos have grown steadily, providing important and unique astrophysical constraints, and possible evidence of unexpected signals. Balloon-born payloads have a unique physics reach in the ultra-high energy range, with float altitudes above Antarctica providing a very large observable target volume, but at a proximity not available from space, thus giving an important advantage in a low particle energy threshold.

The basis for these efforts arose in the 1960's with predictions by Armenian-born Russian physicist Gurgen Askaryan, who
proposed that high energy electromagnetic particle cascades,then a source of active research in the area of high energy
particle physics, would lead to significant observable coherent radio emission for cascades induced by 
the highest energy cosmic particles, whether charged nuclei, gamma-rays, or even neutrinos. This proposal led to
a burst of effort to observe this effect in high energy cosmic-ray air showers, and in fact radio signatures of these
atmospheric particle cascades were observed, but were found to be due to another related effect, a form of 
geomagnetically-induced coherent synchrotron radiation. 

The original Askaryan effect, as it became later known, was forgotten for several decades until  renewed interest in
detection of cosmogenic high-energy neutrinos -- those arising from cosmic interactions of the highest energy cosmic rays with the
3K microwave background radiation -- caused a renaissance of efforts to exploit Askaryan's prediction. The effect itself
was finally directly measured in 2001, and since then has become the mainstay process for a host of both ground-, balloon-,
and even space-based efforts to exploit the coherent radio impulses for neutrino detection.

Other terrestrial target volumes such as the atmosphere can provide visible-light signals to photonics-based detectors via nitrogen fluorescence (N2Fl) and optical Cherenkov emission when an ultra-high energy cosmic ray or neutrino interacts in the atmosphere. 
The development of large ground-based cosmic-ray arrays in the latter half of the twentieth century high precision measurements of ultra-high energy cosmic rays via this emission. 
Because the N2Fl process allows observation of cosmic rays at EeV energies where the fluxes are more plentiful, the techniques could be more highly refined, and although the requirement of moonless, very clear nights severely limited the duty cycle, the technique can be extended to a balloon-borne or space-based instrument, where the large area exposure overcomes the ground-based limitations.
While a space-based vantage for observing N2Fl emission will likely be necessary, balloon payloads can offer an excellent proving ground for the technology needed to achieve maximum sensitivity in a spacecraft-based instrument.

\textcolor{blue}{\textbfit{Notable recent and planned ultra-high energy payloads}}

\begin{wrapfigure}{r}{0.4\textwidth}
\centerline{~~\includegraphics[trim=0 0 0 0, clip, width=2.5in]{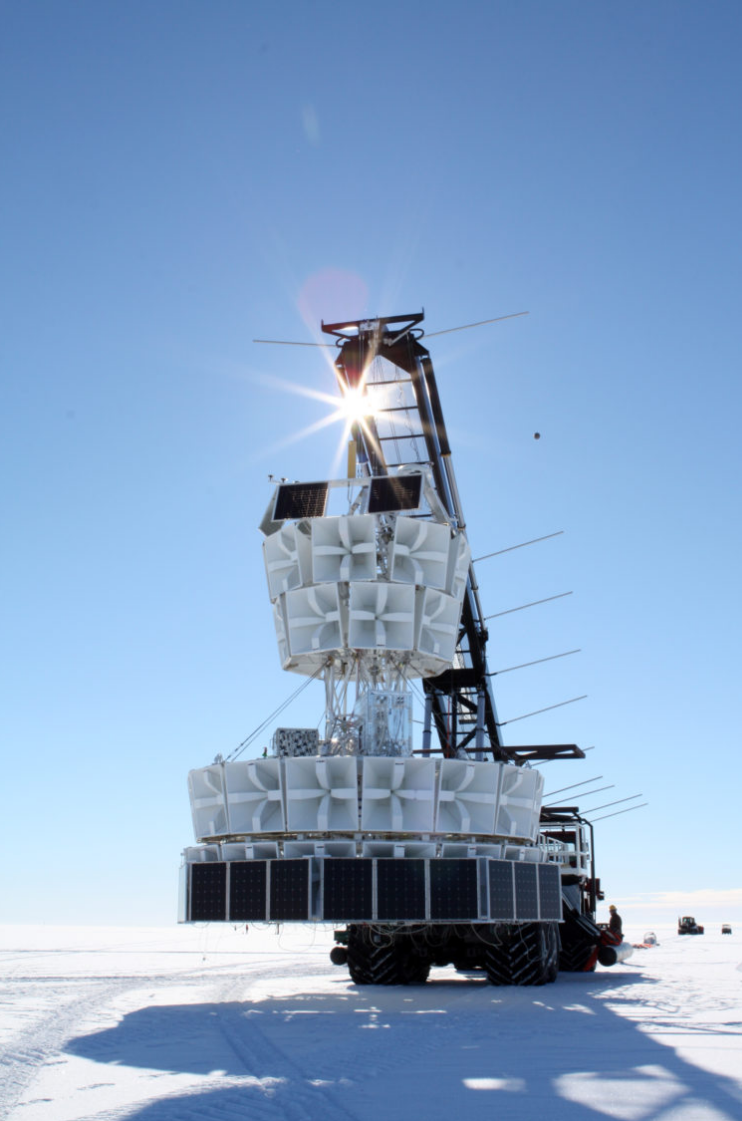}}
\caption{The 8-meter high ANITA-III payload suspended from the LDB launch vehicle during launch preparations in 2013.
\label{ANITA3}
}
\end{wrapfigure}

To date only a single payload has been flown to search for ultra-high energy neutrinos from a suborbital
vantage point, the ANtarctic Impulsive Transient Antenna (ANITA, Fig.~\ref{ANITA3}). ANITA was the first NASA payload of
any kind to pursue efforts in the nascent field of neutrino astronomy. This discipline has been under development
for decades, but the first evidence for cosmic extragalactic neutrinos appeared first within the last decade
as the IceCube detector at the South Pole found evidence for a flux of TeV to PeV neutrinos, some of which
may be tied to active blazars~\cite{IC79}. ANITA's goals were specifically the cosmogenic neutrino flux, which is
tied directly to the UHECR source question, and still remains open. ANITA's four flights searching for
the unique Askaryan signals have led to
among the best world constraints on the UHE neutrino flux~\cite{A4neutrino}, which appears unrelated to the IceCube detections,
and remains an elusive and problematic astrophysical question. Certainly the field of neutrino astronomy is
among the youngest and most exciting in high energy astrophysics.

The forward-beamed geomagnetically-induced radio signals of an ultra-high energy cosmic ray air shower are detectable
at even hundreds of km distance due to coherence -- the tendency of the cascade of charged particles to appear
as a single high-$Z$ charge at the long wavelengths of interest. Coherence of the emission 
allows for a huge additional amplitude to overcome the otherwise
dominant thermal noise floor. The resulting forward-beamed radio impulse has been observed by ANITA, extending measurements from the 1960's and 70's to higher frequencies
than were previously observed. While most UHECR radio signals are seen in reflection off the relatively radio-smooth
ice sheets, ANITA also observed for the first time ultra-high energy cosmic-ray air
showers that transited entirely within the stratosphere, at 20~km altitude or more, developing tangentially
through the atmosphere, and pointing directly at the payload~\cite{ANITA_CR}.

ANITA's sensitivity to UHECR through their geomagnetic radio emission was novel in itself, but has in fact
led to even more surprising results, in the form of anomalous events that appear to arise from air showers
emerging from the ice, rather than reflected off of it~\cite{Upshowers, A3upwardshower}. 
ANITA's precision polarization and time-domain waveform
measurements allow clear separation of reflected and non-reflected signals due to the 180$^{\circ}$ phase-inversion
that a reflection causes. ANITA has now observed a total of six such anomalous events in 4 flights, a result
that may signal the first direct measurements of cosmic $\tau$-neutrinos, or even possible beyond-Standard-Model 
physics~\cite{A4CRs}. Further work on these exciting possibilities is clearly needed.

The proposed successor to ANITA is the Payload for Ultra-high Energy Observations (PUEO), which is projected to have
a sensitivity at least an order of magnitude greater than that of ANITA. This is accomplished by extending the
design of ANITA to a much larger number of antennas, 120 instead of 48, and utilizing state-of-the-art
radio-frequency digital beamforming techniques. These methods allow coherent phasing of the signals 
of up to two dozen antennas at once,
allowing the detection of radio impulses much closer to the thermal noise level, which sets the
detection limit for such experiments. These state-of-the-art digital radio methods will allow 
the PUEO detection sensitivity to be tailored to multiple physics channels, allowing
for independent triggering on both cosmic-ray-like events, with the potential for
tau neutrino sensitivity to transient point sources; and for Askaryan-based in-ice
events that are expected to arise from the diffuse cosmogenic neutrino flux.

% PUEO is also a candidate, as are several other payloads discussed in this report, for the new
% Astrophysics Pioneers program, as the size and complexity of the planned payload will significantly exceed
% the resources normally available in NASA's standard balloon payload offerings. Pioneers are
% budgeted for costs up to \$20M, more than twice the most expensive ROSES/APRA payloads. 
% In PUEO's case, the added complexity and large increase in the number of signal channels significantly
% elevates both the science return and the costs, and combination well-suited to the Pioneers mandate.

\begin{figure}[htb!]
\vspace{3mm}
\begin{center}
\centerline{\includegraphics[width=6.5in]{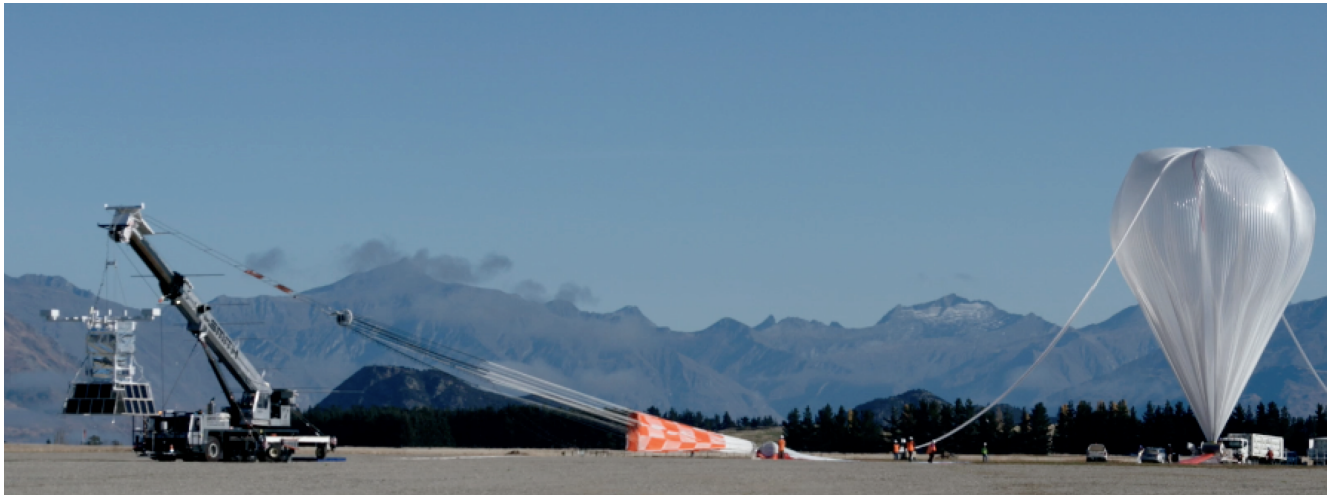}}
\caption{
EUSO-SPB-2 as it prepares for a super-pressure balloon launch from Wanaka, New Zealand, in 2017.
\label{albedo}}
\end{center}
\end{figure}

Another payload seeking to use stratospheric altitudes for observation of UHE events is the 
The Extreme Universe Space Observatory on a Super Pressure Balloon-2 (EUSO-SPB2), which will observe
N2Fl and optical Cherenkov events from UHECR, as a pathfinder mission toward a larger space-based effort.
By observing N2Fl which can be detected from UHECR air showers over a very wide solid angle, EUSO-SPB and its successors
can compensate the dark-sky exposure limitations, and their suborbital-to-orbital vantage points then
allow orders of magnitude more exposure for UHECR than ground-based cosmic-ray observatories. 

EUSO-SPB-2 differs from ANITA in its need for mid-latitudes to provide for night observations, an option
that is not possible in Antarctic flights, which are limited to the austral summer when the stable
circumpolar winds of 
polar vortex are active. In addition, the desire for maximal exposure for science reasons leads to the
proposed use of the Super-pressure balloon (SPB)  technology that NASA's Balloon Program Office has
brought to maturity with the advent of the now-certified 18.9 Mcft SPB midlatitude capability. These
launches take place out of an airport at Wanaka, New Zealand.

\noindent {{\textbfit{\textcolor{goldenpoppy}{High Energy Astrophysics}}}}

%High-energy astrophysics spans over ten decades of photon energies, from x-rays at tens of keV to ultra-high-energy gamma rays above 100's of TeV, and the balloon platform has proven to be 
Scientific ballooning has provided invaluable opportunities for high-energy astrophysics science and technology development, particularly for instruments which are sensitivity to photon energies from tens of keV to tens of MeV. 
Due to the absorption in the atmosphere, lower energy x-ray telescopes need the much higher altitudes achieved by sounding rockets or satellites, and the highest energy gamma-rays can be detected with ground-based telescopes via Cherenkov radiation in the atmosphere. 
However, hard x-ray and soft- to medium-energy gamma rays from the cosmos are accessible via the balloon platform.

%These energetic photons require large, massive detector volumes, and therefore many of these instruments have similar lift requirements as in the cosmic ray field. % and an observation altitudes above 110 kft, all capabilities that are achievable with NASA's current LDB and superpressure balloon programs.
%Long duration balloon flights are generally required for high-energy polarization measurements and transient source detection to allow for sufficient exposure of this relatively weak emission, and at the same time give a higher probability of catching a flaring source. 
%These capabilities are achievable with NASA's current LDB and superpressure balloon programs.

There is a long history of technology demonstration and advancement with balloon-borne high-energy instruments.
All four instruments on NASA's second Great Observatory, the Compton Gamma-ray Observatory (CGRO) that was launched in 1991, were developed through balloon flights. 
This trend continues through today where NASA's three current high-energy missions, the Neil Gehrels Swift Observatory, the Fermi Gamma-ray Large Area Telescope, and NuSTAR also had significant instrument development through balloons. 

One particularly challenging field in high-energy astrophysics is the soft- to medium-energy (MeV) gamma-ray regime, where Compton scattering is the predominate photon-matter interaction. 
There has been no imaging mission dedicated to the MeV regime since CGRO in the 1990s, and many interesting science topics remain from gamma-ray line measurements of Galactic nucleosynthesis and positrons annihilation to continuum studies of high energy events and sources such as pulsars, high red-shift blazers, and gamma-ray bursts. 
Not only has the balloon platform enabled the necessary technology development in this challenging range over the past decade, but with the realization of NASA’s new superpressure balloon, ultra-long duration flights will soon allow for sufficient exposure to accomplish competitive science goals at the relative low cost of a balloon mission. 
Additionally, the newly available southern hemisphere launch location in Wanaka, New Zealand, allows for exposure of the Galactic Center region, which is an important MeV science target, especially for studies of Galactic positron annihilation~\cite{2020ApJ...895...44K}. 

Polarimetry in the hard x-ray and gamma-ray regime is another field of study with balloon-borne instruments at the forefront of developing this novel technique. 
These measurements give unique insight into high energy phenomena with a diagnostic of the emission mechanisms and source geometries for gamma-ray bursts~\cite{2017NewAR..76....1M}, galactic black holes~\cite{2019BAAS...51c.150K}, and active galactic nuclei~\cite{2019BAAS...51c.348R}. 
Recent measurements of hard x-ray and gamma-ray polarization from balloon-borne instruments have demonstrated this technology~\cite{2017ApJ...848..119L,2020ApJ...891...70A,2017NIMPA.859..125C}, and the timing is optimal to complement the upcoming Imaging X-ray Polarimetry Explorer (IXPE) satellite mission~\cite{IXPE}, which will perform polarization measurements at lower energies. 
%The Wallops Arc Second Pointer (WASP) has enabled much of this science in the hard x-ray regime. 

\noindent {\textcolor{royalblue}{\textbfit{Notable recent and planned payloads in High Energy Astrophysics}}}

%\begin{figure}[htb]
%\begin{center}
%\centerline{\includegraphics[width=5in]{figures/COSIpayload.jpg}}
%\caption{\small \it The COSI payload in preparation for launch in Wanaka, New Zealand, in 2016.
%\label{cosi1}}
%\end{center}
%\end{figure}

\begin{wrapfigure}[14]{r}{0.43\textwidth}
\vspace{-4mm}
\centerline{~~\includegraphics[trim=0 0 0 0, clip, width=2.5in]{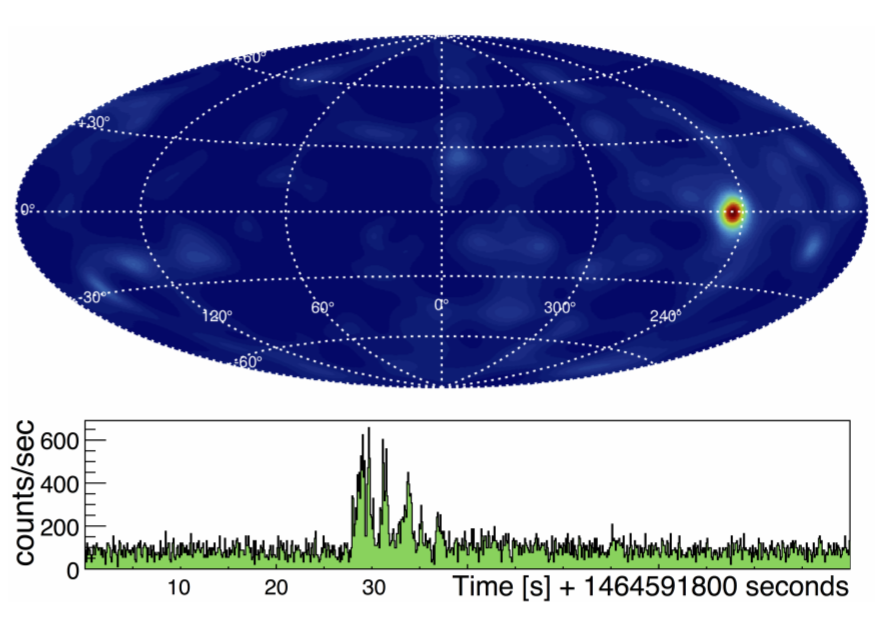}}
\caption{   
2016 COSI observations of GRB160530A was the first GRB reported to the Gamma-ray Coordination Network (GCN) from a balloon payload~\cite{cosigcn}. 
Top: All-sky image showing localization of GRB.
Bottom: time series of count rates. 
\label{cosigrb}
}
\end{wrapfigure}
\begin{figure}[b]
    \centering
    \includegraphics[width = 5in]{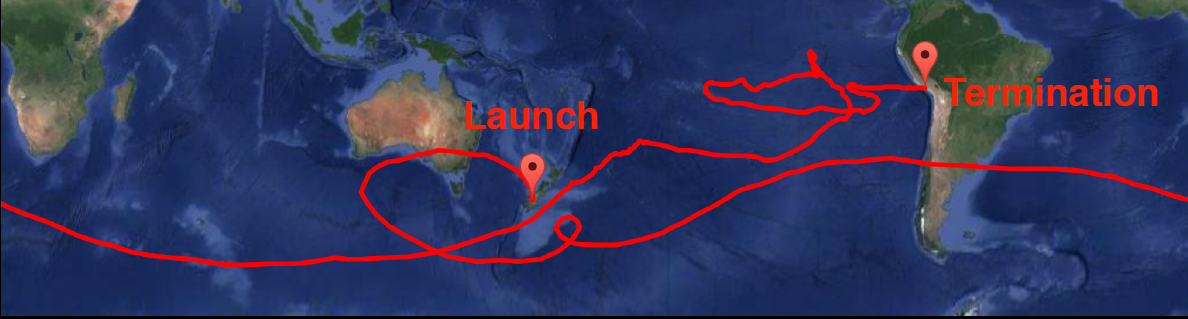}
    \caption{The COSI balloon-born gamma-ray telescope set a mid-latitude flight duration record in 2016. Launched from Wanaka, New Zealand onboard a superpressure balloon on May 17, 2016, the payload landed in the desert north of Arequipa, Peru 46 days later. The instrument was recovered, repaired, and is slated to fly again in 2021.}
    \label{fig:cosi_flight_path}
\end{figure}

In 2016, the Compton Spectrometer and Imager (COSI; see title page) flew for 46 days after a launch from the new NASA flight facility in Wanaka, New Zealand, setting a record duration for a mid-latitude balloon flight. 
%COSI observes the sky in the soft-gamma-ray regime, 0.2-5 MeV, with excellent $\sim 0.3$\% spectral resolution and good polarization sensitivity. 
COSI's 2016 flight resulted in a variety of scientific measurements, including imaging and spectra of a number of compact gamma-ray sources~\cite{kierans2016}, the detection and polarization analysis of the gamma-ray burst GRB160530A~\cite{2017ApJ...848..119L} (Fig.~\ref{cosigrb}), and a measure of the spatial extent of the positron annihilation emission in the Galactic center region~\cite{2020ApJ...895...44K}.
%COSI's technology has since been in continuing development and improvement, and the payload is scheduled to fly again once the launch scheduling can resume its normal cadence.
COSI represents one of the first strong success stories for the burgeoning NASA superpressure balloon program, which has enabled  major improvements in the mid-latitude science capabilities of balloon missions~\ref{fig:cosi_flight_path}.

%The Advanced Particle Telescope (APT) is being developed, initially as a balloon-flight instrument for proof-of principle and subsequently a probe-class mission.  
%As a gamma-ray instrument it combines a pair tracker and Compton telescope in a monolithic design using scintillating fibers for the tracker and wavelength-shifting fibers to readout CsI detectors.  
%It would achieve an order of magnitude improvement in sensitivity compared with Fermi at GeV energies and several orders of magnitude improvement in MeV gamma-ray sensitivity compared to existing instruments. 
%At the same time it would detect rare ultra-heavy cosmic rays, making this a multi-purpose astroparticle-physics observatory.  
%In the short term, a concept for a small 0.45m x 0.45m prototype balloon instrument is being developed.  
%On a high altitude Big60 Antarctic balloon flight this instrument could advance the TRL for APT while providing scientific results through prompt degree-scale localization and polarization constraints on several gamma-ray burst (GRB) events.

\begin{figure}[htb]
    \centering
    \includegraphics[width=4.5in]{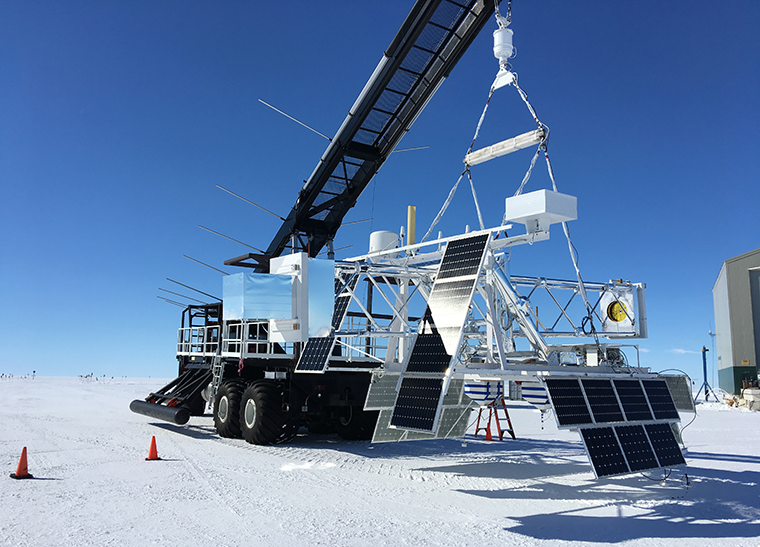}
    \caption{The hard x-ray polarimeter X-Calibur prior to launch in McMurdo Station, Antarctica in 2018 (Photo Cred: Dana Braun). X-Calibur combines BPO's WASP system for arc-seconding pointing with an x-ray telescope and scattering polarimeter detector to be sensitive to the polarized emission from x-ray pulsars, neutron stars and black holes.}
    \label{fig:xcalibur}
\end{figure}

\begin{figure}[tb]
\centering
\begin{minipage}{0.45\textwidth}
    \centering
    \includegraphics[height=3in]{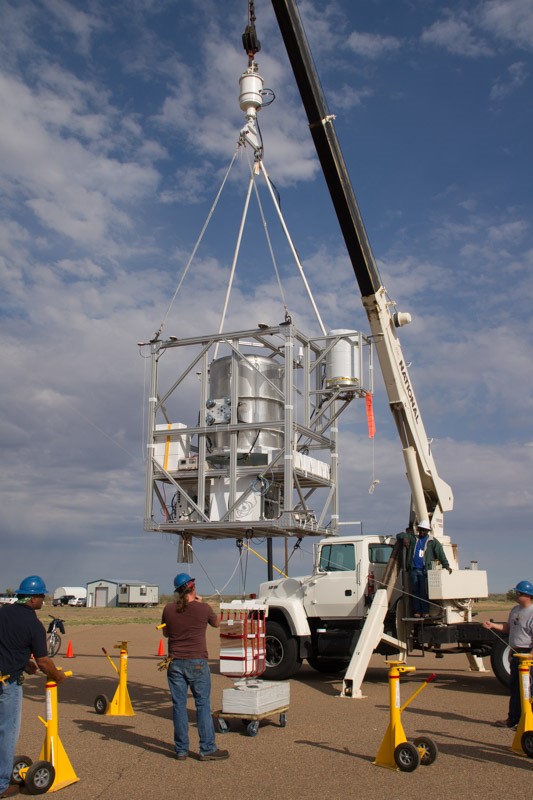}
    \caption{The GRAPE instrument prior to its launch from Fort Sumner in 2011. GRAPE aims to measure polarization of gamma-ray busts.}
    \label{fig:grape}
\end{minipage}%
\hspace{0.5cm}
\begin{minipage}{0.45\textwidth}
    \centering
    \includegraphics[height=3in]{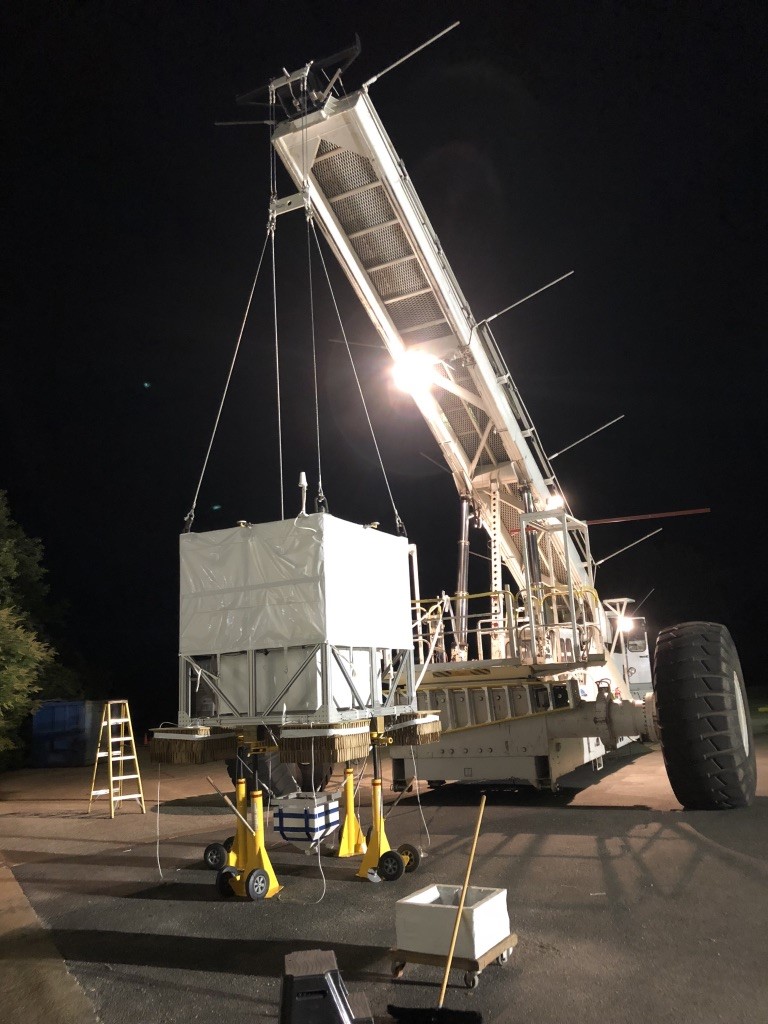}
    \caption{The Advanced Scintillator Compton Telescope (ASCOT) instrument had a test flight in 2018 from Fort Sumner.}
    \label{fig:ascot}
\end{minipage}
\end{figure}

The emerging field of hard x-ray and gamma-ray polarization has been predominately propelled with balloon borne instruments and numerous polarimeters have been developed and flown within the past decade. 
X-Calibur, which combines an 8m long x-ray telescope with the Wallops Arc Second Pointer (WASP) and a scattering polarimeter detector, had its first test flight in September 2016~\cite{10.1117/1.JATIS.4.1.011004} and was launched from McMurdo Station in 2018, see Fig.~\ref{fig:xcalibur}. During the 2018 science flight, X-Calibur had 8~hours of observations of x-ray pulsar GX301-2 that were contiguous with \textit{NICER}, \textit{Swift} and \textit{Fermi}-GBM, and obtained phase-resolved spectro-polarimetric measurements of the emission~\cite{2020ApJ...891...70A}. At higher energies, the Gamma-Ray Polarimeter Experiment (GRAPE)~\cite{2014SPIE.9144E..3PM}, which flew from Fort Sumner in 2011 and 2014 (Fig.~\ref{fig:grape}), consists of traditional scintillation technologies and aims to use its wide field of view to study the polarization of GRBs. 

%and the Swedish instrument the Polarized Gamma-ray Observer (PoGO+)~\cite{2018Galax...6...30F}, which was flown from Esrange Space Center in Sweden in 2016.

The wide-field hard X-ray telescope ProtoEXIST~\cite{Hong_2016} uses a coded-aperture mask and an array of CZT detectors, with Application Specific Integrated Circuits (ASICs) designed for NuSTAR, to enable a wide range of time domain astrophysics. ProtoEXIST had test flight in 2009 and an upgraded instrument, which has better angular resolution than \textit{Swift}-BAT, was flown in 2012.

There are a number of notable projects which are currently being developed for balloon flights that aim to be used in the next-generation space-based MeV mission.
These range from Probe-class mission concepts submitted to the 2020 Decadal Survey, such as the All-Sky Medium Energy Gamma-ray Observatory (AMEGO)~\cite{2019BAAS...51g.245M} and the Advanced Particle-astrophysics Telescope (APT)~\cite{2019BAAS...51g..78B}, to smaller instruments. 
The Advanced Scintillator Compton Telescope (ASCOT)~\cite{2018SPIE10699E..5XB}, is based on high-performance scintillators, such as Cerium Bromide (CeBr3) and p-terphenyl, and aims to follow on the success of the COMPTEL instrument on CGRO by using time-of-flight background rejection. It had its first test flight in 2018, see Fig.~\ref{fig:ascot}.

Recent success of high-energy technology development on the balloon platform was seen in the selection of the 2019 Small Explorers (SMEX) call. 
An upgraded version of the COSI mission~\cite{2019BAAS...51g..98T} was one of two SMEX missions selected for a Phase A study, and the LargE Area burst Polarimeter (LEAP~\cite{2020AAS...23537308M}), which is based on the success of GRAPE balloon project, was selected as a Missions of Opportunity through the same call.

\hyperlink{TOC}{Return to table of contents.}

%\subsubsection{Particle Astrophysics Findings and Recommendations}

%\begin{itemize}

%\item The PAG recommends that NASA continue strong support for missions which can elucidate GCR fluxes and composition, with particular emphasis on improved sensitivities  approaching the spectral knee at cosmic-ray total energies around 1 PeV, enabled by the cost-effectiveness of balloon payloads for large effective areas and target masses necessary in this energy range.

%\item The PAG finds that the astrophysics case for continuing the search for antimatter probes of dark matter is compelling. 
%PAG recommends that NASA continue priority support for cosmic antiparticle searches within the balloon program, specifically where new capabilities for the detection of antihelium and antideuterium are enabled.

%\item The PAG commends NASA's willingness to support exploratory payloads seeking evidence for and constraints on the sources of the highest energy particles in the universe.
%The PAG recommends a robust continuing investment by NASA in payloads that seek to measure these particles, in both the cosmic-ray and neutrino sector. 

%\item The PAG recommends support of technology development for MeV telescopes, in addition to NASA’s Super Pressure Balloon platform to enable ultra-long duration balloon flights.

%\item The PAG recommends support of polarization measurements in the hard x-ray and soft gamma-ray regime and the continued support of WASP.

%\end{itemize}

\subsection{Exoplanets \& Stellar Astrophysics}
\label{sec:exoplanets}

High altitude balloons, especially with ultra-long-duration flights, are real alternatives to large satellite missions to progress exoplanet and stellar astrophysics.
%They demonstrate cutting-edge technologies, are recovered, refurbished and reused to address important scientific questions. 
%In certain disciplines, they have been targeted as the ideal platform for moving the field forward.
%Exoplanetary studies is one example of such usage.
With the demonstration of cutting-edge technologies, exoplanetary and stellar astrophysics studies have targeted the balloon platform as ideal for moving the field forward.

\noindent {{\textbfit{\textcolor{goldenpoppy}{Exoplanetary direct imaging}}}}

An area of exoplanetary studies that balloons excel in is that of exozodiacal debris. 
The bright zodiacal light in our solar system is produced by the scattering of sunlight off of a population of 1-200 $\mu$m dust grains that is continually replenished by outgassing comets and colliding asteroids~\cite{bryden2006}. 
Exozodiacal dust (or exozodi) is the exoanalog of this population and is the inner stellar system complement to the more easily observed debris disks~\cite{eiroa2013}, which can extend out to hundreds of AU from the host star~\cite{greaves2005, kalas2007, stapelfeldt2011}.
Exozodiacal dust is the dominant astrophysical background that needs to be characterized and quantified to image exoplanets. 
Structures such as rings and clumps that form in the exozodiacal cloud can both confound and benefit exoplanet observations. 
Furthermore, coronagraphic studies of debris disk morphology are important in answering fundamental questions such as how circumstellar disks evolve and form planetary systems.

Due to low debris disk-to- star contrasts ($>10^{-6}$), they are accessible to balloon-borne measurements employing moderate sized telescopes (< 1m diameter). 
Given their relatively high brightness, a representative number of debris disks are accessible to even conventional zero-pressure balloon-borne missions flown from northern and southern hemisphere locations.
Even though these direct imaging studies focus on visible light that allows probing regions closer to the parent stars, the bright signals from debris disks can be observed during the day time, thus even a circumpolar flight with suitable instruments can make significant contributions to these studies.

\begin{figure}[htb!]
\begin{center}
\centerline{\includegraphics[width=5in]{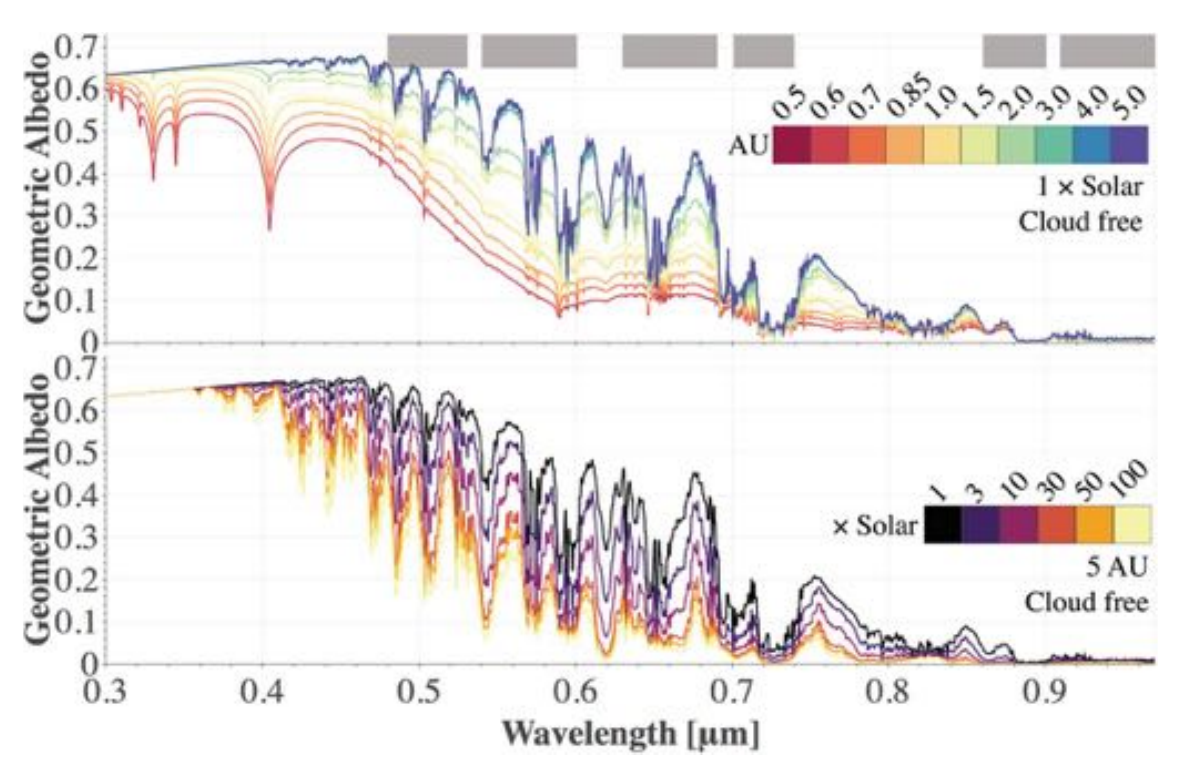}}
\caption{
Model albedo spectra for a range distances of cloud free giant planets characterized characterized by planetary parent-star distances and metallicities. 
Top: Spectra for a range of planetary masses for a fixed (1 $\times$ solar metallicity) atmosphere. 
Bottom: for a system with a planet-star separation of 5 AU~\cite{Batalha2018}.
\label{planetaryalbedo}}
\end{center}
\end{figure}

Theoretical computation of the albedo spectra in the 400 – 1,000 nm range from model atmospheres of giant planets with a broad range of masses and atmospheric compositions located at 2 AU from a G0V type star (Figure~\ref{planetaryalbedo}). 
Model spectra like these will allow for the identification of chemical and cloud signatures that could be discerned with WFIRST-CGI exoplanet coronagraphy. 
With a 1-m class telescope, it should be possible to detect strong methane and ammonia absorption features and test some of these model predictions from balloons. 
Recent studies that have focused on what can be gleaned from reflected light observations of exo-Jupiters~\cite{lupu2016, nayak2017, gao2017, mayorga2016} have shown that a broad range of planetary parameters shape the planet’s spectrum. 
Note that $R\sim 50$ provides sufficient spectral resolution to provide meaningful constraints on atmospheric composition and cloud structure.

Direct imaging of selected exo-Jupiters may also be possible with balloon-borne coronagraphic experiments.
A handful of exo-Jupiters could be observed with 2-6 hours of night time observations. These path-finding measurements could provide much needed experiment design, observation strategies and validation of data analysis algorithms necessary for larger, probe-class missions.

\vspace{1cm}

\noindent {{\textbfit{\textcolor{goldenpoppy}{Exoplanetary Transits}}}}

The phase curve of an exoplanet describes changes of its brightness during an orbital period. 
Wavelength dependent phase curve (or spectroscopic phase curve) observations can be used to determine the inhomogeneous chemical composition and cloud coverage~\cite{Parmentier2019}. 
The value of spectroscopic observing of phase curves for transiting plates to determine their atmospheric composition, thermal structure and dynamics of exoplanets has been  demonstrated with ground and space-based observations.

\begin{figure}[htb!]
\begin{center}
\centerline{\includegraphics[width=5in]{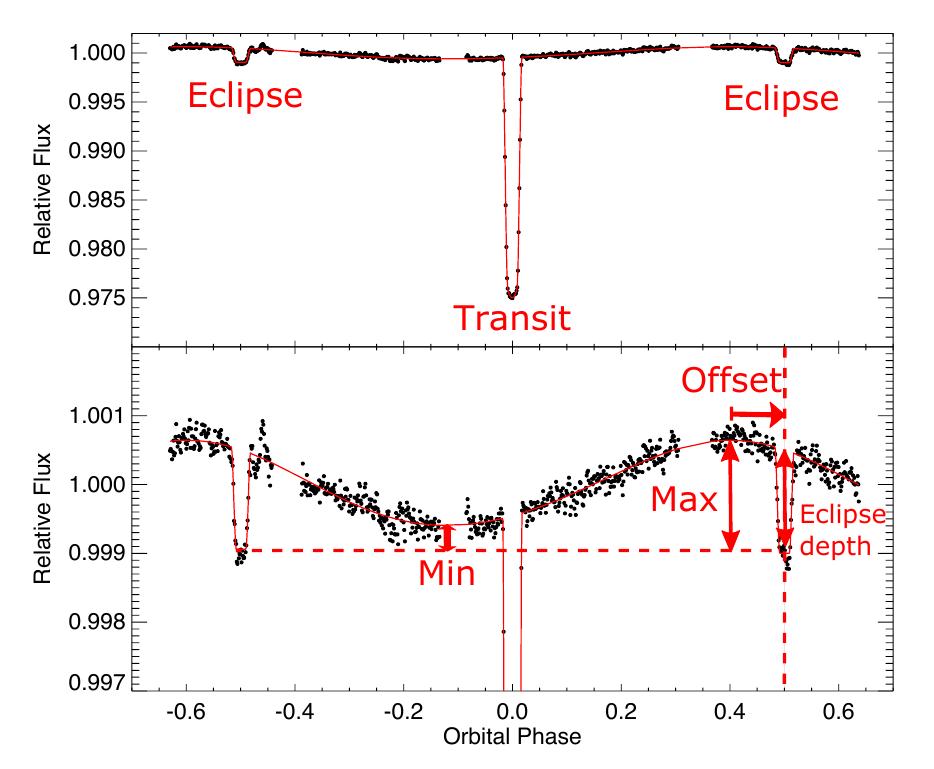}}
\caption{
 Phase curve of HD189733b observed at 3.6$\mu$m with the Spitzer Space Telescope~\cite{knutson2012}. The bottom panel is a zoomed-in version of the top panel, with the measurable quantities of interest annotated~\cite{Parmentier2019}.
\label{phasecurve}}
\end{center}
\end{figure}

Figure~\ref{phasecurve} shows an example phase curve of HD189733b observed by the Spitzer Space Telescope.
The key measurables – amplitude, phase of the peak compared to the secondary eclipse (the offset from the substellar point) and the depth of the secondary eclipse can be used to infer important parameters of exoplanetary atmospheres, such as, the radiative and dynamical time scales, winds, etc.

Even though considerable progress has been made both with atmospheric modeling as well as in the observational fronts, further progress requires spectroscopic observations of complete phase curves. 
Such spectroscopic phase curves could be obtained during an entire orbital period from long-duration balloons for short period transiting planets (see for example, \url{https://arxiv.org/pdf/1903.09718.pdf}).
These measurements are not possible from Low Earth Orbit (LEO) satellite platforms.

\noindent {{\textbfit{\textcolor{goldenpoppy}{Stellar Astrophysics}}}}

The major impediment to high precision IR photometry is water vapor; thus stratospheric balloons, which float above the tropopause and its associated cold trap, are  ideal platforms for near-infrared  photometric measurements.

The high sensitivity of most modern infrared surveys means that even moderately bright stars are saturated, leading to large uncertainties in their photometry. 
Consequently, the brightest ~12,000 stars in the near infrared have photometry no better than 20\%, a level of precision that is an order of magnitude larger than what is achievable and necessary for an accurate understanding of the physics of these stars.

Such measurements are critical in tying together theory and modelling in a number of disparate fields. Physics-based models of the atmospheres of exoplanets, cool stars, and evolved stars,  are tuned to fit high resolution data from interferometers, coronagraphs, and high resolution spectrometers. 
High resolution necessarily implies lower sensitivity and thus these instruments can observe only bright stars. It is precisely these stars for which we have very limited absolute photometry. 
Such stars are also those most easily studied from a high-altitude balloon.

\noindent {{\textbfit{\textcolor{goldenpoppy}{Studies of the Interstellar Medium}}}}

\begin{figure}[htb!]
\begin{center}
\centerline{\includegraphics[width=4in]{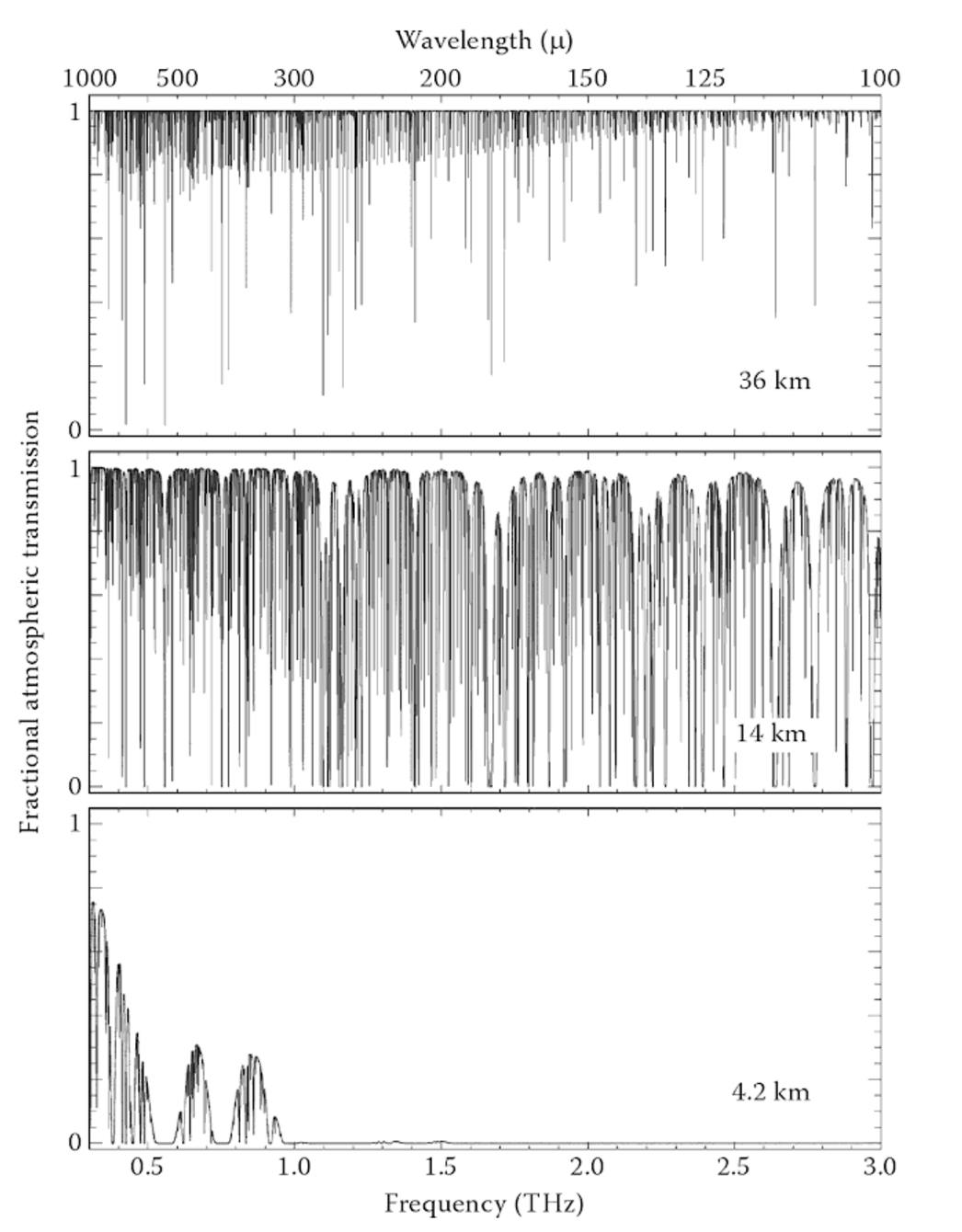}}
\caption{THz atmospheric transmission at mountaintop (bottom), aircraft (middle), and balloon-borne (top) altitudes~\cite{Walker-book}.
Atmospheric transmission is seen to increase exponentially with altitude. Likewise, \textbfit{scientific return also increases exponentially with altitude}, thereby motivating the development of larger balloons capable of taking payloads closer to the edge of space
\label{THz}}
\end{center}
\end{figure}

The TeraHertz (THz) portion of the electromagnetic spectrum (1 to 10 THz) provides us with a powerful window into cosmic evolution. 
THz photons arriving at Earth provide valuable insights into everything from the birth and death of stars to the cataclysmic events associated with the origin of galaxies and the Universe itself. 
Many of the THz photons we observe are emitted by the gas and dust between the stars, that is, the interstellar medium (ISM). 
At THz frequencies we can observe photons associated with the ISM of our own galaxy, the Milky Way, as well as from the ISM’s of distant galaxies.

Today the ISM accounts for only ~1\% of the total gravitational mass of the Milky Way, with the balance being in stars (~9\%) and dark matter (~90\%). 
However, in the beginning, most of the Galaxy’s baryonic mass (everything made from protons and/or neutrons) was in ISM. 
It was, and continues to be, the mass and energetics of the ISM that is a principal driver of galactic evolution. 
Indeed, we owe our very existence to violent processes occurring in the local ISM ~4.7 billion years ago. 
It was these events that led to the formation of our solar system. Before then every atom that makes up the Earth and us (!) was part of the ISM, floating in space, the ash of ~1000 generations of stellar birth and death.

Energy sources within the ISM include gravity and electromagnetic (EM) radiation. 
Sources of EM radiation include the cosmic background (i.e. Big Bang, peaking at microwave/GHz frequencies), photospheres of stars (mostly at UV/optical/IR wavelengths), thermal dust emission (peaking in the far-IR/THz frequencies), emission lines of atoms and molecules (found across the entire spectrum), and free-free and synchrotron radiation (arising mostly at millimeter/microwave/radio frequencies). 

The evolution of THz astronomy has been driven largely by two factors; 1) atmospheric absorption of THz light and 2) the availability of detector technology. 
Water vapor in the Earth’s atmosphere is a very efficient absorber of THz photons (see Figure~\ref{THz}). 
Therefore, THz observations are best conducted from high altitude balloon-borne telescopes or space-based telescopes. 
Utilizing recent breakthroughs in detector technologies, such observatories have or will serve as powerful probes of the life cycle of the ISM, both in our own Milky Way and beyond. 

Due to their ability to take heavy payloads to the edge of space, high altitude balloons allow the possibility of realizing large aperture telescopes ( >5 meters) with sophisticated cryogenic instruments at a fraction of the cost and time associated with achieving similar capabilities on an orbital mission. 
With such observatories it will be possible to probe the intricacies of star and planet formation, as well as the origins of galaxies.

\noindent {\textcolor{royalblue}{\textbfit{Notable recent and planned payloads for Exoplanets and Stellar Astrophysics}}}

\begin{figure}[t]
\centering
\includegraphics[width = 4in]{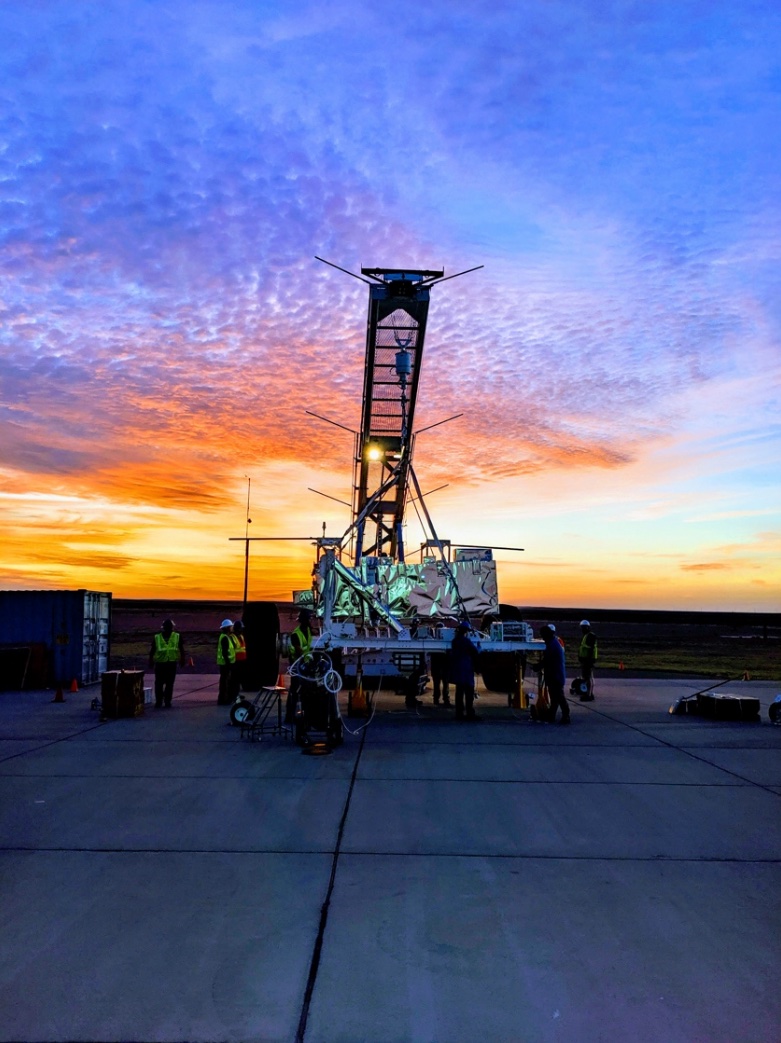}
\caption{PICTURE-C integrated with the WASP gondola awaiting launch for its engineering flight in September 2019.}
\label{picC}
\end{figure}

%%BLAST-TNG
%%BETTII?

High altitude balloon platforms are ideal for the demonstration and technology maturation and risk reduction efforts for exoplanetary studies. 
One such effort is the Planetary Imaging Concept Testbed Using a Recoverable Experiment - Coronagraph (PICTURE – C) program~\cite{picturec}, which saw its engineering flight on September 28, 2019 (Figure~\ref{picC}). 
This experiment demonstrated several key technologies – active wavefront control, use of a high actuator count deformable mirror, a state-of-the-art coronagraph, a 5 milliarcsecond pointing system and an unobstructed telescope. 
These technologies have already been baselined for two future space flight missions, HabEx~\cite{habex} and LUVOIR~\cite{luvoir}, two major astrophysics missions that have undergone detailed studies for the Astro2020 Decadal Survey.
%While HabEx is primarily focused on exoplanetary studies, LUVOIR has a larger science goals.
The science flight of PICTURE-C has been delayed due to COVID-19 and other safety concerns. 
When it is allowed to fly PICTURE-C will characterize one or more debris disks with a contrast of $10^{-7}$.

The Astrophysics Stratospheric Telescope for High Spectral Resolution Observations at Submillimeter wavelengths (ASTHROS) mission, with a planned launch around December 2023, will loft a far-infrared-to-submillimeter, 2.5 m-aperture telescope, one of the largest ever flown on a balloon mission. ASTHROS will observe velocity-resolved motions of gas in galactic stellar 
\begin{wrapfigure}{r}{0.3\textwidth}
\centerline{~~\includegraphics[trim=0 0 0 0, clip, width=2in]{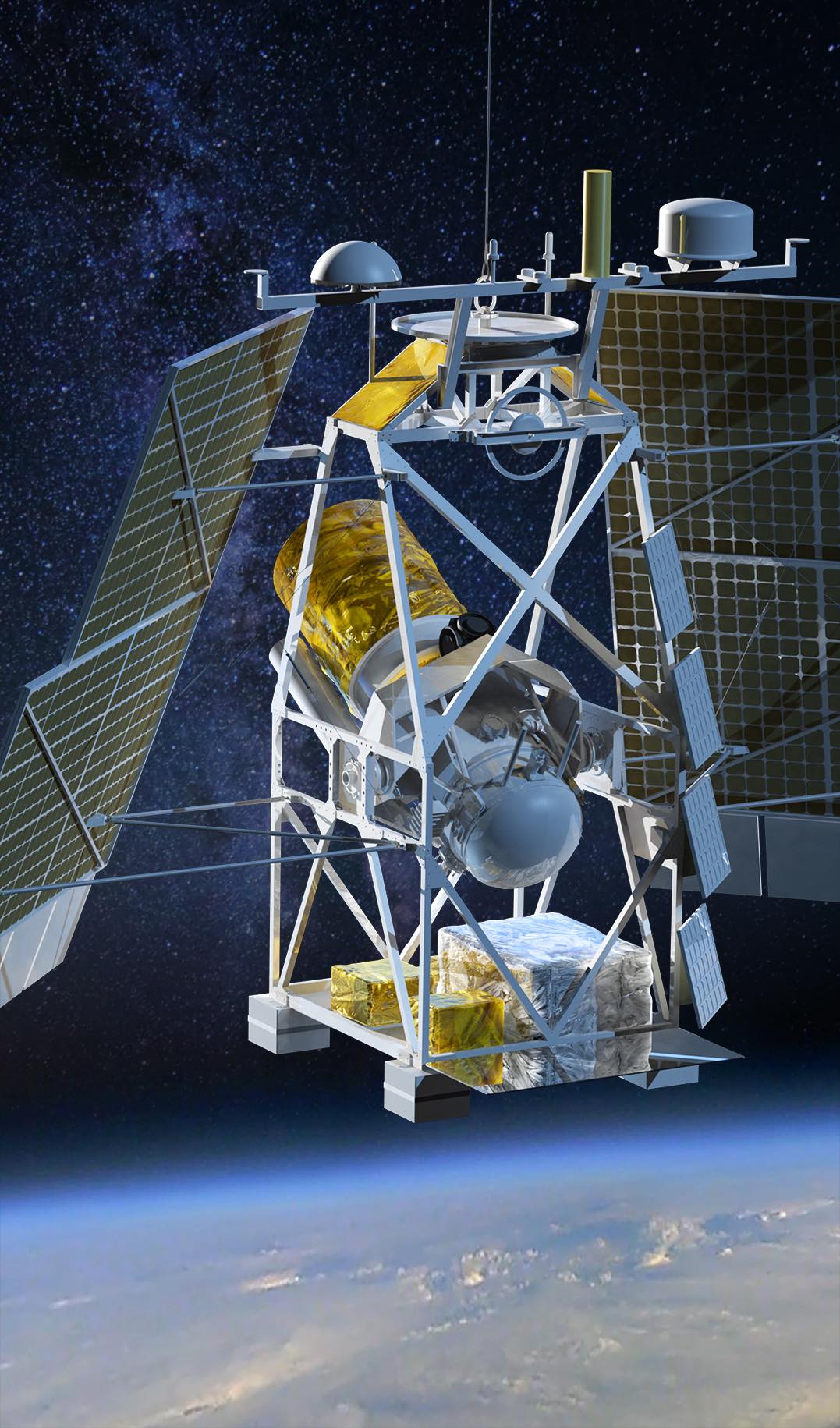}}
\caption{Conceptual image of the GUSTO payload on orbit. GUSTO is set to fly from McMurdo Station in 2021 (credit APL/Johns Hopkins).
\label{gusto}
}
\end{wrapfigure}
systems, including TW Hydrae, a young star system where planet formation is hypothesized to be underway. ASTHROS is envisioned to become a multiply-flown facility with access by the astrophysics commmunity, complementing the SOFIA airborne observatory by providing southern hemisphere observations, and access to longer wavelengths.

%The primary objective of HabEx is to search for and directly image Earth-like exoplanets and spectroscopically characterize their atmospheres for evidence of biosignatures. 
%It will also be able address a large number of other exoplanetary questions as well as fundamental questions on the formation and evolution of the universe. 
%Laboratory tests have validated several enabling technologies and their technology readiness levels have been increased through balloon programs such as PICTURE-C.

%As envisioned, LUVOIR is a general purpose mission, much like the highly successful Hubble Space Telescope. 
%Exoplanetary science, specifically searching for signs of life in tens of exoplanets, is one of its focus areas that include equally fundamental questions on comparative planetology of the solar system objects, cosmological quests as well as stars, galaxies and their evolution. 
%Currently two implementation have been considered. They differ in the size of the primary mirror – one uses a 8m dia while the other a 15m dia one. 
%As such, LUVOIR can be considered the astronomical successor to upcoming James Webb Space Telescope.

The Gal/Xgal Ultra-Long Duration Balloon-borne Spectroscopic THz Observatory (GUSTO, Fig.~\ref{gusto}) is a \$40M Astrophysics Explorer Mission of Opportunity scheduled to fly on a superpressure balloon from Antarctica in 2021~\cite{gusto}. 
GUSTO is a successor to the Stratospheric THz Observatory (STO \& STO-2) which flew successfully in 2011 and 2016, providing compelling heritage that elevated GUSTO to an Explorer-class mission.

GUSTO's science goals center on improving our understanding of the Universe by THz measurements of interstellar gas, determining its life-cycle and constituents throughout the Milky Way and Large Magellanic Cloud (LMC). 
GUSTO will fly a 0.9 m Cassegrain telescope and cryogenic detection system. 
During the planned 100 day SPB flight, GUSTO plans to survey 124 square degrees of the Milky Way and all of the LMC in three important interstellar lines: [CII], [OI], and [NII] at 158, 63, and 205 $\mu$m, respectively.

\hyperlink{TOC}{Return to table of contents.}

%\subsubsection{Findings and recommendations in Exoplanets and Stellar Astrophysics.}

%\begin{itemize}
%\item The PAG recommends that direct-imaging investigations of exoplanets using balloon-borne coronographs be considered as a priority for technology development and risk reduction for larger probe-class direct-imaging missions.

%\item The PAG recommends that priority be given to a survey-class balloon mission to perform precision infrared photometry on a large bright-star sample. 

%\item In order to realize the full potential of THz astronomy within the balloon program, the PAG recommends support for technology development of large aperture balloon-borne THz telescopes, which have the potential for targeted stellar astrophysics missions with sensitivty that is competitive with spacecraft-borne payloads, at a small fraction of the cost.
 
%\end{itemize}

\clearpage

\begin{figure}[ht]
    \centering
    \includegraphics[width = 5in]{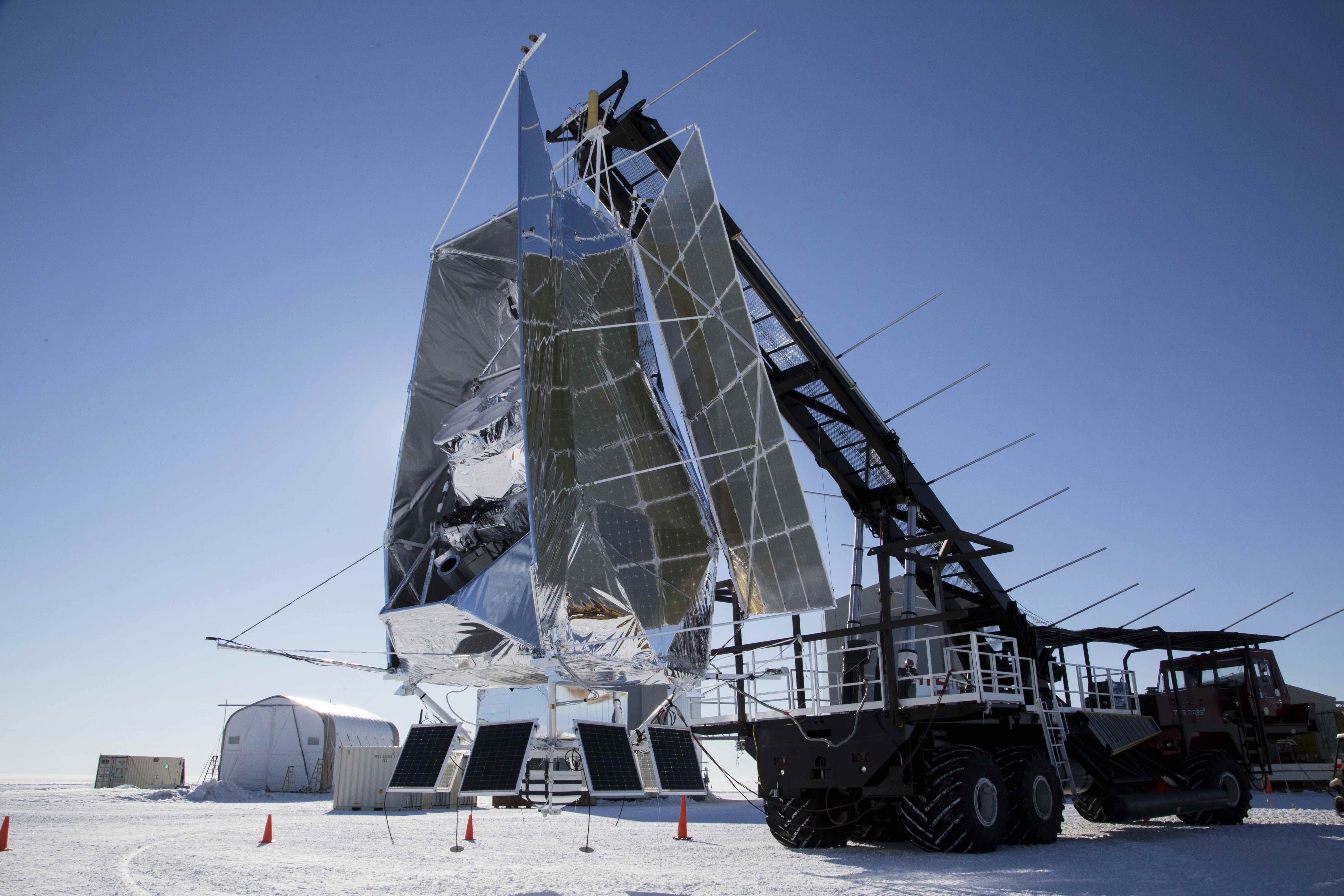}
    \caption{The Spider experiment prior to the Antarctic Long Duration Balloon flight in 2014/15. This flight demonstrated the first operation of fully autonomous, time-domain multiplexed arrays of superconducting transition edge bolometers in a near-space environment.  The compact wide-format focal plane array technology enables massive scaling of the number of cryogenic mm-wave detectors that can be deployed in a resource-constrained platform, such as a future orbital mission.}
    \label{fig:spider}
\end{figure}

\subsection{Observational Cosmology}
\label{sec:cosmology}

\begin{wrapfigure}{r}{0.45\textwidth}
\vspace{-3mm}
\begin{minipage}{0.45\textwidth}
\centering
\includegraphics[height = 2in]{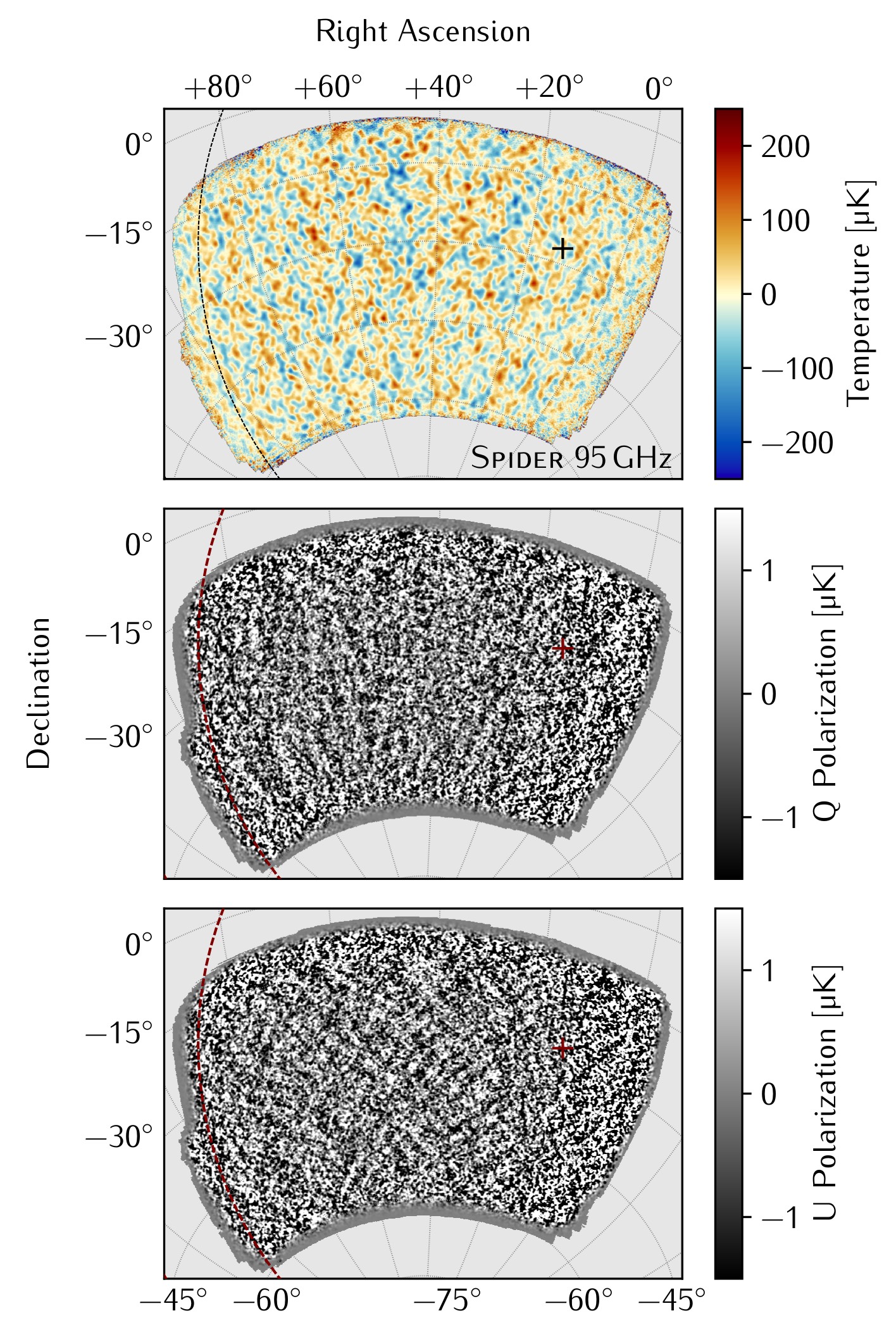}
\includegraphics[height = 2in]{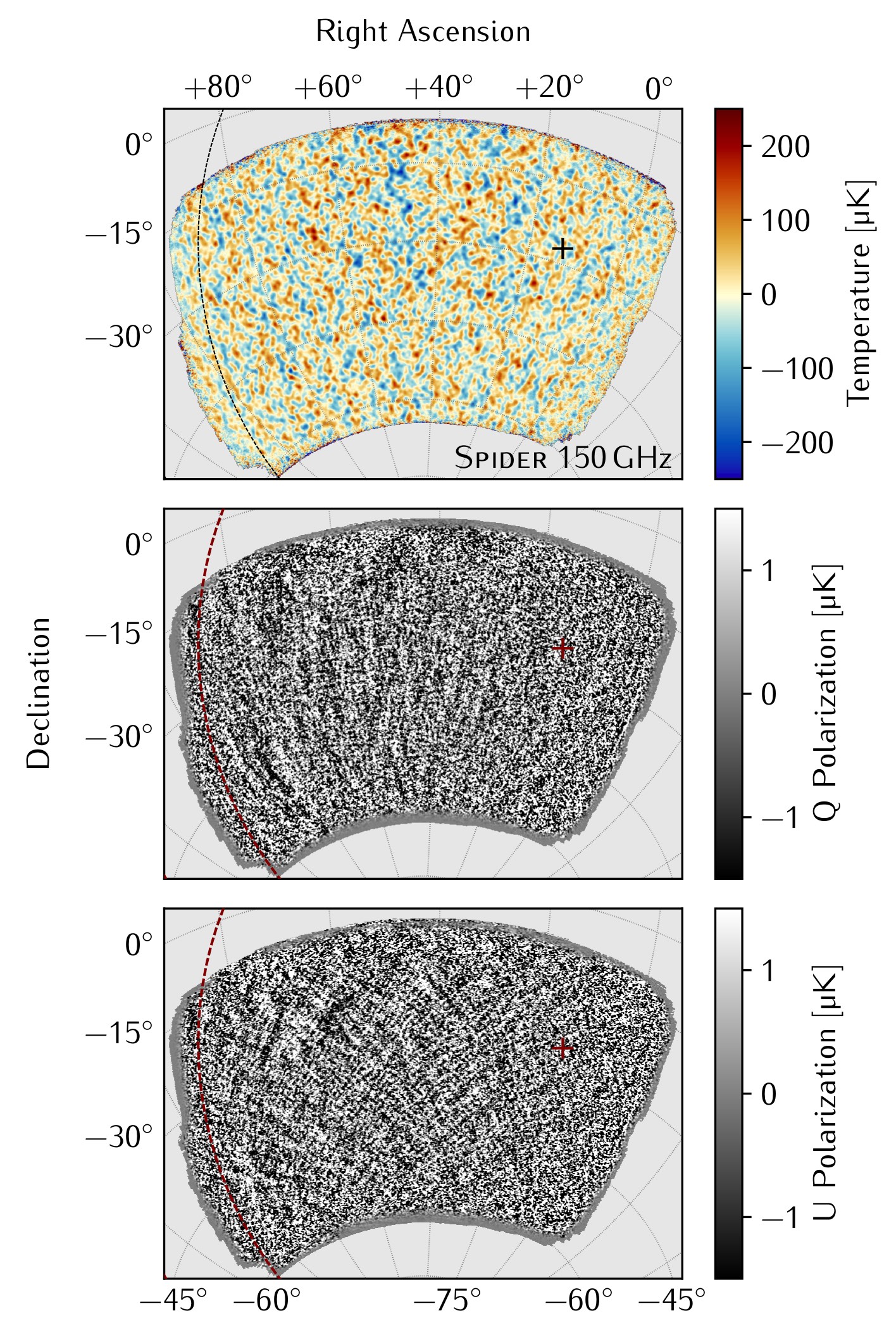}
\caption{The data from Spider produced the most sensitive maps of the CMB to date in a region covering more than 10\% of the southern Galactic hemisphere.  The left and right columns show the total intensity (Stokes I parameter), and linear polarization (Stokes Q and U parameters) at 95 and 150 GHz, respectively.  These data represent a significant advance over the sensitivity of the {\it Planck} legacy survey.}
\label{fig:spiderdata}
\end{minipage}%
\end{wrapfigure}
Over the past several decades, qualitatively new technologies and techniques have enabled cosmologists to use the Universe as a laboratory for physics that cannot be explored in terrestrial laboratories.  NASA's investment in these endeavors, including the women and men who have developed and enabled them, has been transformational.   

The balloon program is central to NASAs leadership in the field; many of the most productive orbital missions derive directly from NASA's investment in sub-orbital payloads and the balloon program.  These include groundbreaking missions such as COBE, CGRO, Spitzer, WISE, WMAP, Planck, Herschel, NuSTAR, NGRST, and JWST.  
The future holds opportunities for the balloon program to extend this heritage to groundbreaking discoveries in fundamental physics. 

\vspace{1cm}

\noindent {{\textbfit{\textcolor{goldenpoppy}{The Cosmic Microwave Background}}}}

In the post-{\it Planck} era, and in anticipation of the ambitious next generation of ground-based CMB projects, the scientific ballooning platform represents a unique and enabling scientific opportunity.  Broadly speaking, the measurements enabled fall in two classes;  at frequencies above 220 GHz, near the peak of the CMB intensity and where Galactic polarized dust emission dominates, and at large angular scales.  

\begin{figure}[htb]
\centering
\includegraphics[width = 5in]{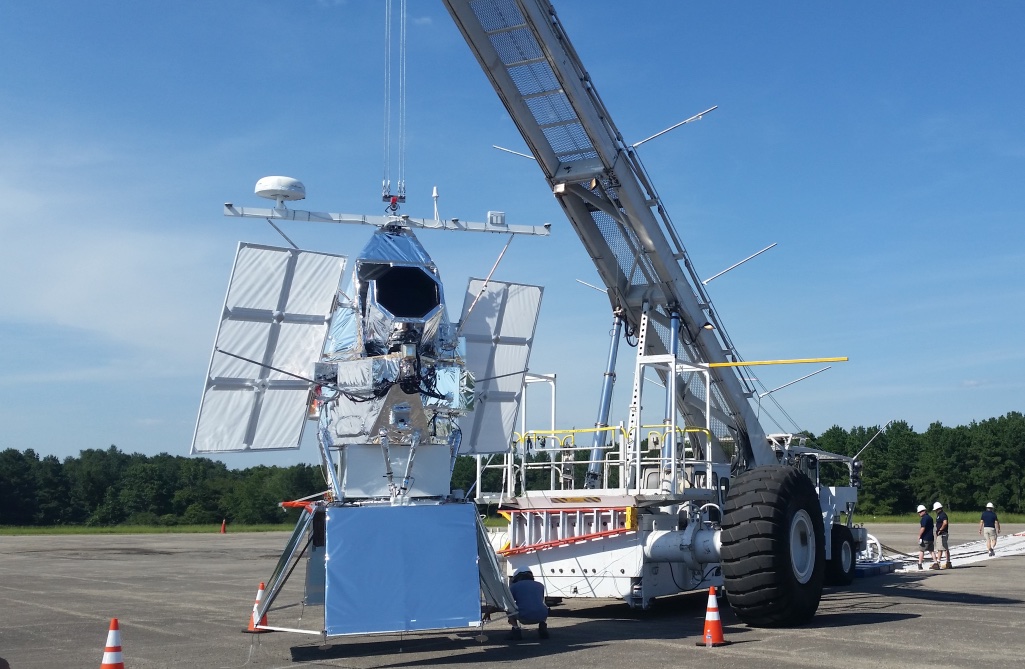}
\caption{SuperBIT is shown on the launch pad in Palestine.  During a series of test flights since 2015, the SuperBIT payload has demonstrated wide-field, diffraction limited imaging using a 0.5 meter telescope. SuperBIT is the first instrument to demonstrate sufficient 3-axis pointing stability to enable diffraction limited long exposures (up to 30 minutes) over a large field of view.}
\label{fig:superbit}
\end{figure}

\begin{wrapfigure}[14]{l}{0.55\textwidth}
\vspace{-2mm}
\centering
\includegraphics[width = 0.475\textwidth]{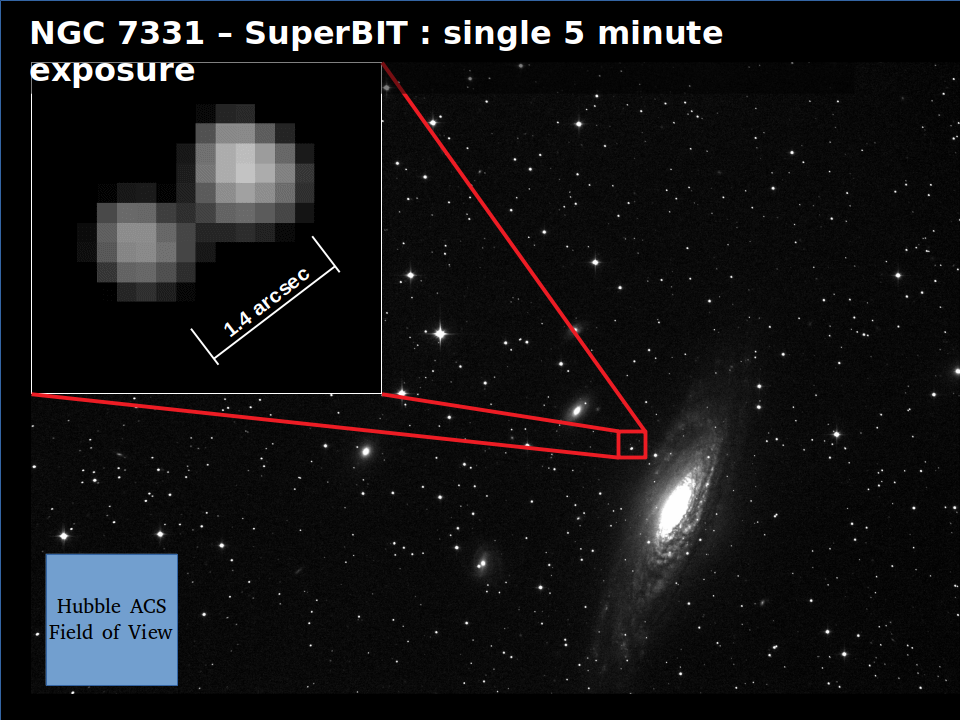}
\caption{A single 5 minute exposure in the vicinity of NGC7331, during a test flight from Palestine.  SuperBIT's FOV is 30x as large as Hubble's largest instrument, and has achieved diffraction limited imaging over the full FOV.}
\label{fig:cosmo_imaging}
\end{wrapfigure}

In these regimes, the advantages of the stratospheric platform over the best terrestrial sites are overwhelming.  The lack of atmospheric loading and interference, as well as the ability to cover a significant fraction of the full sky, allows balloon borne instrumentation access to the higher frequencies and the largest angular scales. 

These are of critical importance for characterizing the epoch of reionization, constraining the neutrino sector, and the search for primordial gravitational waves. 

\clearpage

\noindent {{\textbfit{\textcolor{goldenpoppy}{~~Cosmic Concordance}}}}

Cosmological observations at high- and low-redshift, including measurements of the Hubble parameter and of the large scale distribution of matter, provide independent tests of the standard cosmological model.  The unique capabilities of the Hubble Space Telescope (HST) have proven invaluable in the development of this field.  Long-duration mid-latitude Super Pressure Balloon flights provide a platform for wide-field imaging in the near-UV to near-IR wavelengths, offering space-quality diffraction limited imaging and spectroscopic capabilities that extend far beyond those of HST.  Balloon payloads offer an ability to field cutting edge technologies with a relatively rapid development cycle, making the program well poised to take advantage of the exponential growth in the imaging technology that is transforming the field. Importantly, these capabilities of HST in the near-UV to optical wavelengths will not be replicated in the coming generation of large missions, including JWST and NGRST, both of which will be at least as over-subscribed as HST.  Finally, intensity mapping of molecular lines in the mid- to far-infrared represents a powerful probe of structure formation - probing the large-scale distribution of matter at intermediate redshift, and tracing the formation and evolution of galaxies over cosmic time.  

\noindent {{\textbfit{\textcolor{goldenpoppy}{Beyond Standard Model Physics}}}}

Over the past twenty years the evidence that the Universe is dominated by Dark Matter and Dark Energy has become overwhelming.  At the same time, well motivated efforts to directly detect weakly interacting dark matter have come up empty handed. Therefore, there is ample motivation to explore methods of probing the dark sector via the only interaction known to exist: Gravity.  Clusters of galaxies represent the largest over-densities of matter in the Universe and as such represent a powerful laboratory for probing the nature of dark matter.   The gravitational lensing of background galaxies provides a minimally biased mechanism for tracing the distribution of dark matter within a galaxy cluster.  High resolution reconstruction of the distributions of dark matter, gas and stars within interacting clusters, and of halos within relaxed clusters, can serve as a sensitive probe of physics within the dark sector and help to shed light on a theoretical understanding of particle dark matter 
~\cite{Harvey1462,robertson17,Sokolenko_2018}.

\begin{figure}[htb]
\begin{center}
\centerline{\includegraphics[width=5in]{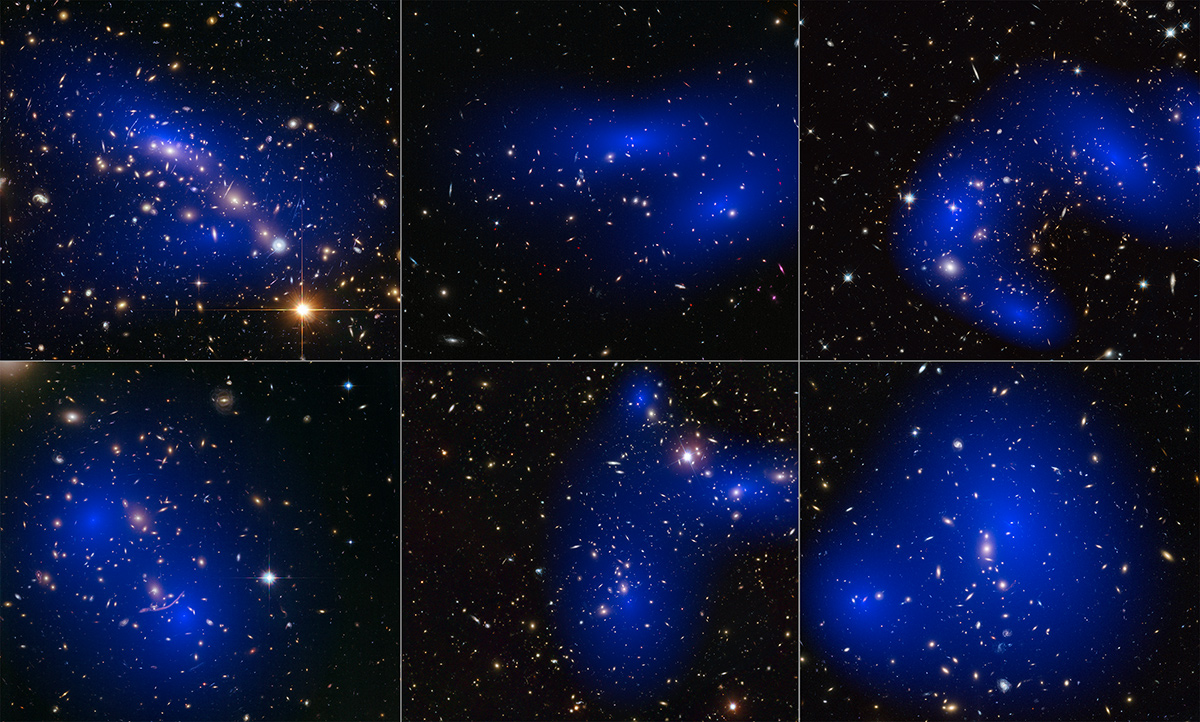}}
\caption{The distribution of luminous and dark matter resulting from galaxy cluster collisions is sensitive not only to gravity, but also any interactions within the Dark Sector. Using optical data from HST it is possible to separately map the distributions of dark and visible matter (HST NASA/STScI, D. Harvey, et al). The statistical nature of the measurement requires a large sample of clusters, making the long duration super-pressure platform an invaluable tool for the science~\cite{harvey_2019}.
\label{fig:cluster_collisions}}
\end{center}
\end{figure}

\noindent { \textcolor{royalblue}{\textbfit{Notable recent and planned payloads for Observational Cosmology}}}

The scientific potential of the stratospheric balloon platform has attracted a number of ambitious programs in the field of cosmology.  
The Spider experiment (Fig.~\ref{fig:spider}) flew on a successful 2.5 week mission, returning images of the polarized CMB with unprecedented areal coverage and sensitivity at mm-wavelengths (see Figure \ref{fig:spiderdata})~\cite{Nagy_2017}. 
SuperBIT underwent a series of engineering flights, demonstrating diffraction limited imaging in the near-UV to near-IR~\cite{Romualdez_2020}.

Similarly, a number of new payloads are under development or are preparing for science flights in the next few years.  Spider has developed a new suite of high frequency arrays to better constrain Galactic dust emission, which serves as a limiting factor in the search for cosmological B-mode polarization of the CMB~\cite{Bergman_2018}. Having completed its engineering flights, SuperBIT is ready for a mid-latitude science flight on the next available super pressure balloon.  SuperBIT will observe several clusters each night, building a uniform lensing catalog that will dramatically improve mass-observable relations that are critical for cluster cosmology.  

At the same time, the team is working on the development of a 1.5 meter class successor that will enable qualitatively new observations, capable of imaging a given area of the sky to a given depth nearly forty times faster than Hubble.  The near term scientific objective is to use strong- and weak-lensing to probe for dark matter interactions beyond gravity and the standard model (see Figure~\ref{fig:cluster_collisions})~\cite{harvey_2019,schwinn_2017}.

Experiments in development, such as EXCLAIM and TIM, are pioneering the field of intensity mapping at sub-mm and THz wavelengths, respectively. These promise to illuminate the star formation history and capture the evolution of structure over cosmic time~\cite{exclaim_2020,TIM}.  FIREBALL uses UV observations of the faint circumgalactic medium to track the baryons and study galaxy formation at intermediate redshifts~\cite{Hamden_2020}.  Each of these programs includes a significant component of student leadership, providing invaluable field experience that will ensure the scientific rigor and mission success of future NASA endeavors.

{\it It is important to note that many of these science opportunities rely upon the unique capabilities of the mid-latitude long duration balloon flights.} These require a combination of access to a large fraction of the sky, with the Sun below the horizon, over the course of long duration flight. Just as the advent of Antarctic LDB flights ushered in a new era of scientific discovery two decades ago, the mid-latitude super-pressure platform is poised to open new windows in cosmology in the coming one.

%\subsubsection{Findings and recommendations.}

%The PAG recommends continuing strong NASA support for millimeter- and submillimeter-wave balloon missions, with particular emphasis on large-scale survey efforts.

\hyperlink{TOC}{Return to table of contents.}

%  end of Astro sections

\newpage

\section{Earth Science}
\label{sec:earth}

% \begin{tcolorbox}[colback=royalblue!8!white,colframe=royalblue,fonttitle=\bfseries,title=Relevance to SMD Science Goals]

% Currently, NASA-supported Earth Science remote sensing studies primarily use aircraft or satellite observations. However, balloons provide a unique platform for observations in remote locations such as Antarctica and for in-situ studies the Earth's atmosphere. 

% ADD -  BIG questions in Earth science?
%  \textit{????} and \textit{????}

% \vspace{0.75cm}

% \textbf{Climate Change} (Sec.~\ref{sec:climate}) .... as stated in Section~..... of NASA's ????? Roadmap: 
% \begin{quote}
% The blah.
% \end{quote}
% The studies of ???, \textit{????}, through ???.

% \vspace{0.75cm}

% Studies of \textbf{Stratospheric Ozone Loss} (Sec.~\ref{sec:ozone}) directly addresses the question \textit{???} posed in ???? \textbf{Impact of Solar Radiation on the Stratosphere} (Sec.~\ref{sec:radiation_budget} answers \textit{??}

% \vspace{0.75cm}

% \textbf{High Energy Atmospheric Physics} (Sec.~\ref{sec:TGF}), a relatively young field....

% \end{tcolorbox}

\subsection{Climate Change}
\label{sec:climate}

The climate structure of the Earth is undergoing rapid and unprecedented changes.
%These changes have global impacts, demanding critically needed studies and forecasts of:
%(1) hurricane formation, intensity and trajectory, 
%(2) the rate of sea level rise, 
%(3) climate-forced changes in the large-scale dynamics of the atmosphere, 
%(4) wildfire risk, 
%(5) shortwave forcing of the climate that is the dominant uncertainty in climate forecasts, 
%(6) drought and agricultural production, 
%(7) severe storm initiation, development and geographic coverage, 
%(8) Arctic sea and glacial ice breakup, 
%(9) UV radiation increases over the central United States in summer, 
%(10) risk of flooding, 
%(11) increases in CO2 and CH4 emission globally, and 
%(12) cirrus cloud, pyrocumulus, and volcanic eruptions impact on climate. 
These changes have global impacts, demanding critically needed studies and forecasts of natural disasters, %such as hurricanes, wildfires, severe storms and drought; climate-forced changes in 
atmosphere dynamics, %Arctic glacial ice breakup and 
sea level rise, and increases in %CO$_2$ and CH$_4$ 
global carbon dioxide and methane emissions.
With direct {\it in situ}
measurements, the balloon platform provides a unique capability to answer the pressing science questions related to climate change.

%The PAG  recommends NASA focus on three key research areas that require balloon-borne observations in and from the stratosphere, addressing 
%	Response of Large Scale Atmospheric Dynamics to Carbon Dioxide and Methane. 

%The inability to accurately forecast the rate of onset of these irreversible changes to the climate system that will destabilize society in the coming decades, underscores the current large uncertainties in quantitative forecasting. 
%This in turn directly implies that we have lost control of the ability to quantitatively navigate the risks associated with rapid, irreversible changes to the Earth’s climate structure as the future unfolds. 
%Scientific balloon observations constitute an integral part of strategic innovations in climate research and associated advances in public policy.

\begin{figure}[htb!]
\begin{center}
\includegraphics[height=0.87\textheight]{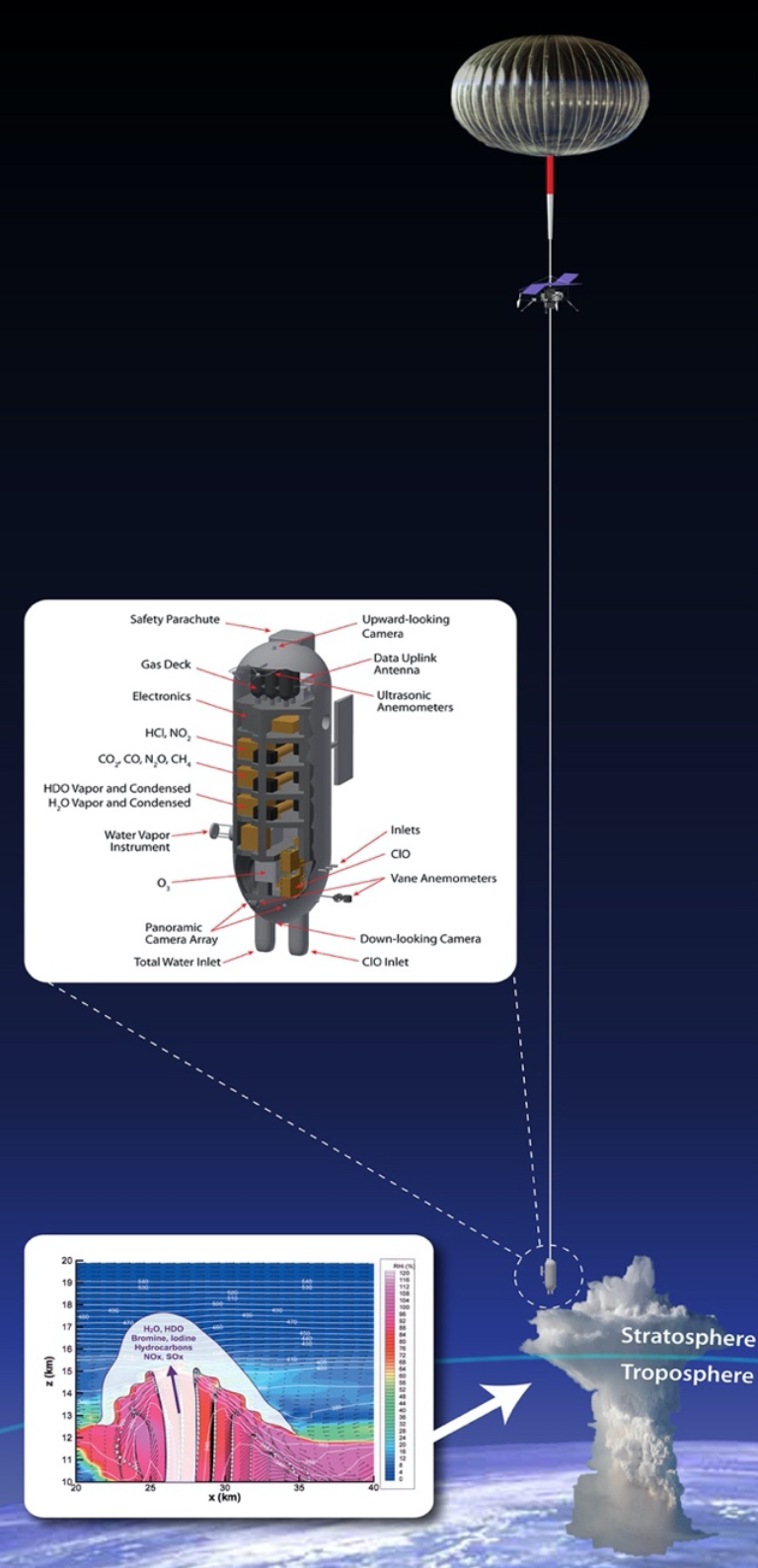}
\caption{
The StratoCruiser concept is shown in combination with the "reeldown" system, uniquely able to observe the dynamics of deep stratospheric convective injection as well as observe the photochemical evolution of the convective injection with respect to ozone loss over the U.S. in summer.}
\label{stratocruiser}
\end{center}
\end{figure}

As greenhouse gases increase in Earth’s atmosphere, the stratosphere is predicted to have two distinct responses. 
First, as the surface warms, the stratosphere cools, and both observations and models show that stratospheric cooling is a robust response to increasing radiative forcing by greenhouse gases. 
Second, chemistry-climate models universally predict that the Brewer-Dobson circulation (BDC) – the primary circulation pattern in the stratosphere – should be accelerating in response to greenhouse gas increases. 
The BDC is responsible for transporting trace gases, such as water vapor and ozone, that are critical for radiative forcing and impact surface weather and climate. 
The acceleration of the BDC predicted by models would be a positive feedback on the increased radiative forcing by greenhouse gases. 
However, rates and patterns of the BDC inferred from observations are not consistent with each other or with model predictions, raising important questions.

The large-scale dynamical structure of the global atmosphere constitutes one of the most critical factors that determine the Earth’s response to increased forcing by fossil fuel extraction, distribution and combustion. 
This is because the large scale atmosphere dynamics determines the geographic position of deserts and cryosystems in the polar regions, the trajectory and intensity of weather patterns and large scale storm systems, and the amount of water vapor entering and exiting the stratosphere. 
The amount of water vapor, in particular, critically constitutes to important climate feedback and is key to establishing the degree of stratospheric ozone loss, and thus the amount of ultraviolet radiation reaching the Earth’s surface.
Quantifying how mean ages, residence times, and transport pathways vary with latitude and altitude is critical for understanding the Brewer-Dobson circulation and its role in controlling trace gas distributions.

\noindent { \textcolor{royalblue}{\textbfit{Notable recent and planned payloads for Climate Change}}}

The StratoCruiser (Fig.~\ref{stratocruiser}) is an Earth Science payload system that has been under persistent conceptual and detailed design study for the last decade, to address a number of the climate science and related issues. 
The StratoCruiser, which is designed to fly on a superpressure balloon, engages two advances in technology: a stratospheric solar-electric propulsion system for horizontal navigation and station keeping in the stratosphere, and a “reeldown” system providing in situ observations over an altitude range of 10 to 15~km below the balloon float altitude. 
A balloon-borne stratospheric Earth-observing payload such as the StratoCruiser can systematically and comprehensively map the details of this dynamical structure globally employing an array of high 
spatial resolution {\it in situ} measurements of short, medium and long-lived atmospheric tracers in combination with highly sensitive dynamical observations of three-dimensional velocity fields, wave-breaking structures, momentum transport and heat transport as well as radiance divergence observations. 
%These are the detailed, high spatial resolution observations that are critically needed to quantitatively establish the dynamical structure of the atmosphere required to develop tested and trusted forecast models of climate response to rapidly increased forcing by carbon dioxide and methane.

\begin{wrapfigure}[16]{l}{0.35\textwidth}
\vspace{-6mm}
\centerline{~~\includegraphics[trim=0 0 0 0, clip, width=2in]{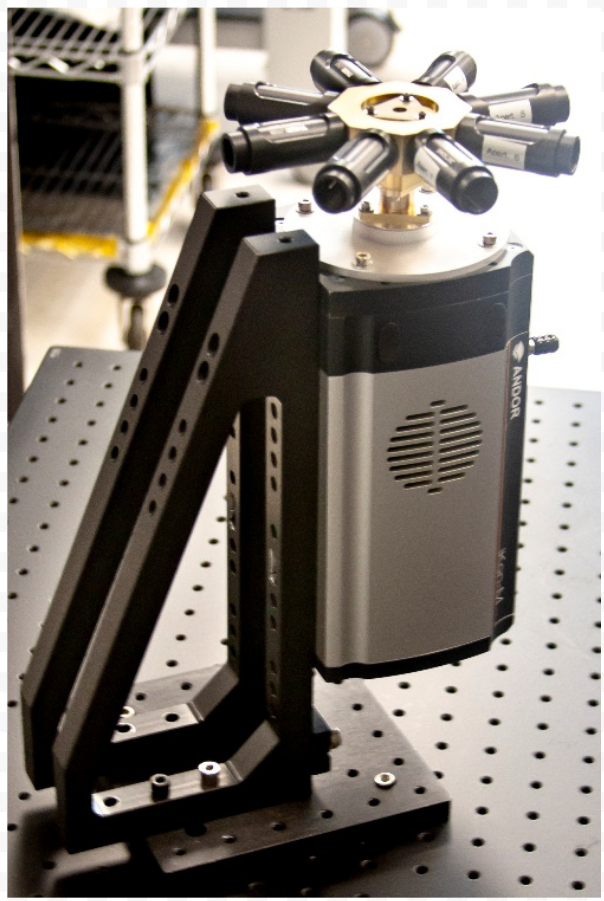}}
\caption{ The MASTAR instrument during laboratory testing in 2019.
\label{mastar}
}
\end{wrapfigure}
The Multi-Angle Stratospheric Aerosol Radiometer (MASTAR) is a balloon-borne instrument designed to measure aerosols that exist in the stratosphere, 10-40 km above the Earth’s surface.  
Stratospheric aerosols, some of which are present naturally while others are injected abruptly by volcanic eruptions and large wildfires, influence the solar heating of Earth’s atmosphere, and thus will impact the extent of climate change effects caused by greenhouse gases.  
Satellite or balloon-borne observations of these aerosols can provide the spatial sampling and vertical resolution necessary to reduce uncertainty in the modeling of radiative forcing, provide correlative measurements for other scientific missions, and monitor dynamic natural events such as volcanic plumes that have widespread and rapidly changing effects.

%MASTAR measures aerosols by looking horizontally at scattered light from the atmosphere in eight directions simultaneously.  Using multiple viewing directions gives better sampling of aerosol plumes from natural events.
The MASTAR instrument was completed for laboratory testing in 2019 (see Fig.~\ref{mastar}).  
It is designed to fit on a Cubesat-scale satellite or as a piggyback on a balloon gondola.  
A high-altitude research balloon is a valuable platform for testing the MASTAR concept because it will provide a similar viewing geometry to a satellite measurement.  
In addition, a balloon gondola offers a better opportunity to view in many directions than most aircraft.  
MASTAR has been approved for a “piggyback” flight of opportunity on a NASA research balloon.  %We are currently working to complete a thermally controlled enclosure to support this flight

The Underpass mission~\cite{westferris}, which is being developed as a citizen science project at the West Ferris Secondary School, is designed to measure CO$_2$ in the stratosphere as a way to evaluate atmospheric models of different trace gases. %From a balloon perspective, supporting and expanding the use of AirCore is critical for evaluating the models, the GHG satellites, and doing this 
These small-scale, low cost samplings of the atmosphere in an underpass of a satellite, or in coordination with aircraft, % or unusual ecological events (like fire/drought), very important 
can be use to benchmark models, and therefore can advance climate science.

\subsection{Stratospheric Ozone Losses}
\label{sec:ozone}

Emerging evidence from both observations and theoretical analyses has demonstrated that the stratosphere over the central United States in summer stands in sharp contrast both dynamically and photochemically with respect to other geographic regions globally~\cite{doi:10.1002/2013JD020931}.
%We now know, through the union of high altitude aircraft, satellite and NEXRAD weather radar observations, that the stratosphere over the US in summer is vulnerable to ozone loss as
This is a result of a series of factors unique worldwide to the central US in summer that is based on convection from the severe storms over the Great Plains.

%Several factors, as well as their coupling, are directly involved: 
%(a) convection, as a result of severe storms over the Great Plains, delivers both markedly enhanced water vapor, as well as source species from the lower troposphere, deep into the stratosphere over the US in summer, 
%(b) the depth of convective injection is sufficient to reach altitudes of rapidly increasing available inorganic chlorine that is then catalytically converted on simple, ubiquitous, binary water-sulfate aerosols, to free radical form, ClO, 
%(c) anti-cyclonic flow in the lower stratosphere over the US in summer, as a result of the North American monsoon, contains that convective injection in a gyre that provides time for photochemical catalysis to exert influence, 
%(d) it is the chlorine radical, ClO, that couples to the available BrO radical to form a catalytic cycle rate limited by
%ClO + BrO $\rightarrow$ Cl + Br + O$_2$ that constitutes the mechanism capable of removing ozone; and it is the same mechanism that contributes to ozone loss over the polar regions in winter, 
%(e) the remarkable sensitivity of increases in the skin cancer incidence in the U.S. to fractional decreases in ozone column concentration places strict demands on the quantitative understanding required to accurately forecast ozone losses in a changing climate, 
%(f) 

The frequency and intensity of severe storms is increasingly tied in the scientific literature to increased forcing of the climate by increasing levels of CO$_2$, CH$_4$ and N$_2$O in the Earth’s atmosphere. 
As a result, climate forcing is mechanistically linked to forecasts of ozone reduction. 
\begin{wrapfigure}[27]{r}{0.47\textwidth}
\includegraphics[width=0.47\textwidth]{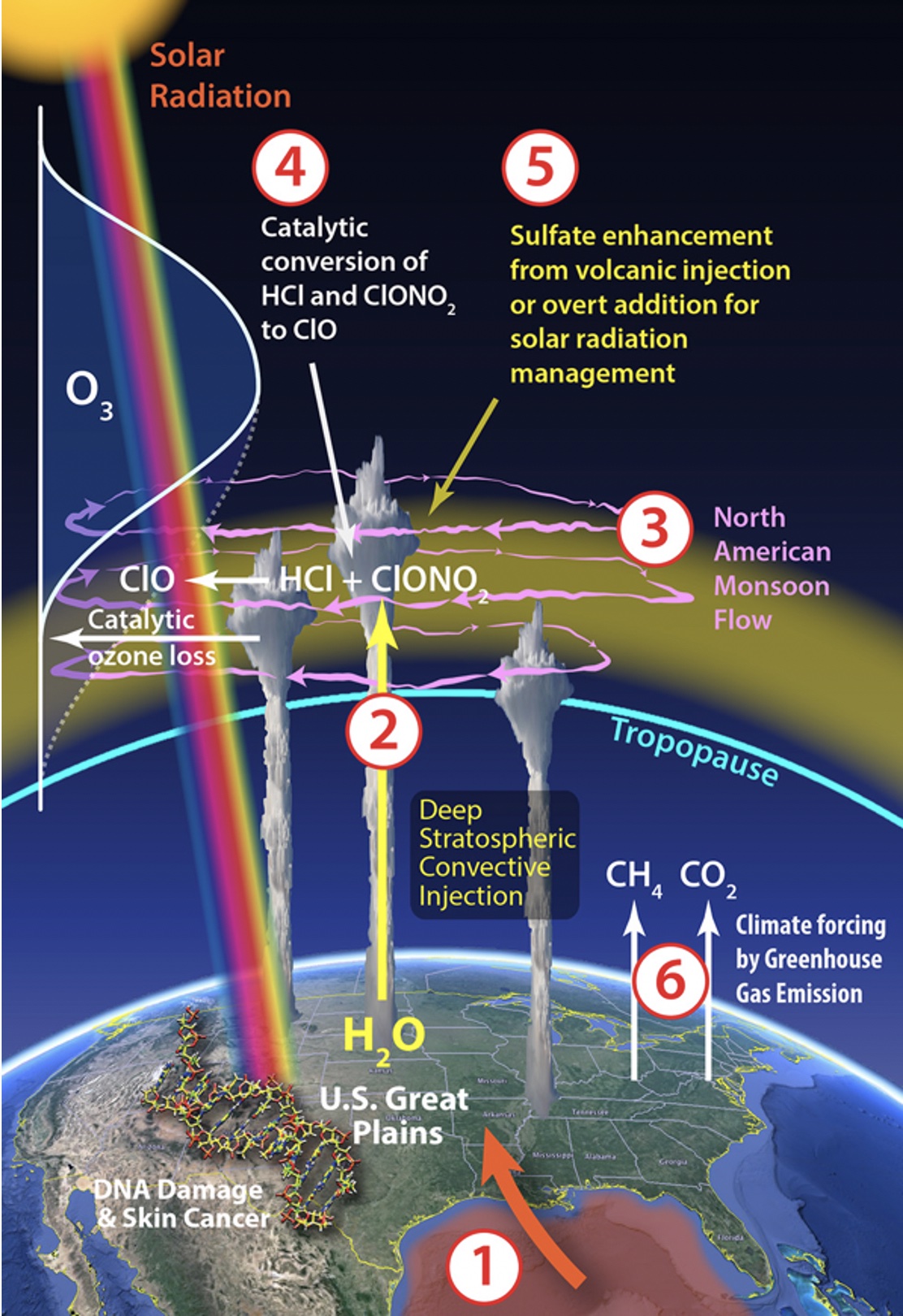}
\caption{Graphical display of the coupling of increasing surface temperatures of the Gulf of Mexico with the increased frequency and intensity of convective injection into the stratosphere that in turn triggers the catalytic loss of stratospheric ozone over the Great plains of the United States in summer.}
\label{fig:ozoneloss}
\end{wrapfigure}
With increased forcing of the climate in the coming years, it is therefore expected that there will be an increased frequency and intensity of convective storms over the Great Plains. 
Unfortunately, these links show that the convective injection into the stratosphere is inherently irreversible and so too is the reduction in the ozone column concentration.

This represents a particularly acute public health issue because of the extreme sensitivity of humans to increases in UV radiation. 
Epidemiology studies in the United States has demonstrated that for every 1\% reduction in stratospheric column ozone concentration, a 3\% increase in the incidence of skin cancer results.
With the current 3.5 million new cases of skin cancer annually in the US alone, this corresponds to an additional 100,000 cases annual for a 1\% decrease in column ozone.
%Therefore, this deep stratospheric convective injection over the U.S. in summer links human health, NEXRAD radar observations, solar radiation management and ozone loss from volcanic injection.

This coupled problem requires a new balloon-borne observing strategy to advance renewed efforts to understand the stratosphere-troposphere response to increased climate forcing by CO$_2$, CH$_4$, N$_2$O, and other greenhouse gases. 
%Critically important quantitative information is needed on the structure of the 3-D velocity fields within the storms, the structure of the wave-breaking events associated with the convective intrusion, the concentration of species co-injected with water by the convective event, both the vapor phase and condensed phase H$_2$O and HDO concentrations injected by the storm, the concentration of (a) the reactive chemical precursors that are normally removed by photolysis and oxidation in the troposphere, (b) chemical tracers injected by the convection, subsequent chemical kinetics of the injected chemical cocktail that tracks the time evolution of the source species, reactive intermediates, isotopes and catalytic free radicals that attack ozone.
Balloon-borne measurements could markedly improve our understanding of the effects of deep convection on the dynamics and chemistry of the lower stratosphere over North America. 
Ultimately, these measurements can elucidate the convective processes that inject lower tropospheric photochemical, aerosol and water vapor into the stratosphere, identify the source regions, and quantify the effects of convection on the composition of the lower stratosphere above the US.

\subsection{Solar Radiation Management}
\label{sec:radiation_budget}

Efforts to engineer the climate by Solar Radiation Management are under serious consideration because reduction in incoming shortwave radiation is one of the only methods that can reverse the rapid irreversible loss of the Earth’s cryosystems.
Cryosystems that exert an inordinate control over the global climate structure of the entire planet.
However, the addition of sulfate to the lower stratosphere directly engages the same heterogeneous and homogeneous catalytic removal of ozone in the lower stratosphere that is mentioned in the previous section. 
This results because an increase in the reactive surface area of sulfate aerosols in the stratosphere shifts the critical temperature for conversion of inorganic chlorine to free radical form to higher temperatures. 

In order to directly test the impact of sulfate addition to the lower stratosphere, it is necessary to first seed a cylindrical volume of the lower stratosphere with a variety of sulfate-water vapor mixtures. 
The seeding platform must then withdraw from the test domain, tracking that volume with a Lidar system that can continuously track the position of the seeded region. 
Multiple vertical soundings of the seeded region are then used to follow the time evolution of the photochemistry that results from the sulfate-water perturbation. 
The key rate limiting steps involved in the use of stratospheric sulfate aerosols to reduce solar forcing of the climate involve both the UV radiation field as well as the presence of heterogeneous surfaces unique to the lower stratosphere. 
Thus observations within the lower stratosphere itself are required for the controversial public policy questions associated with Solar Radiation Management.

\subsection{Geophysical \& Planetary Acoustics}

\begin{wrapfigure}{r}{0.415\textwidth}
\vspace{-3mm}
\centerline{~~\includegraphics[trim=0 0 0 0, clip, width=2.75in]{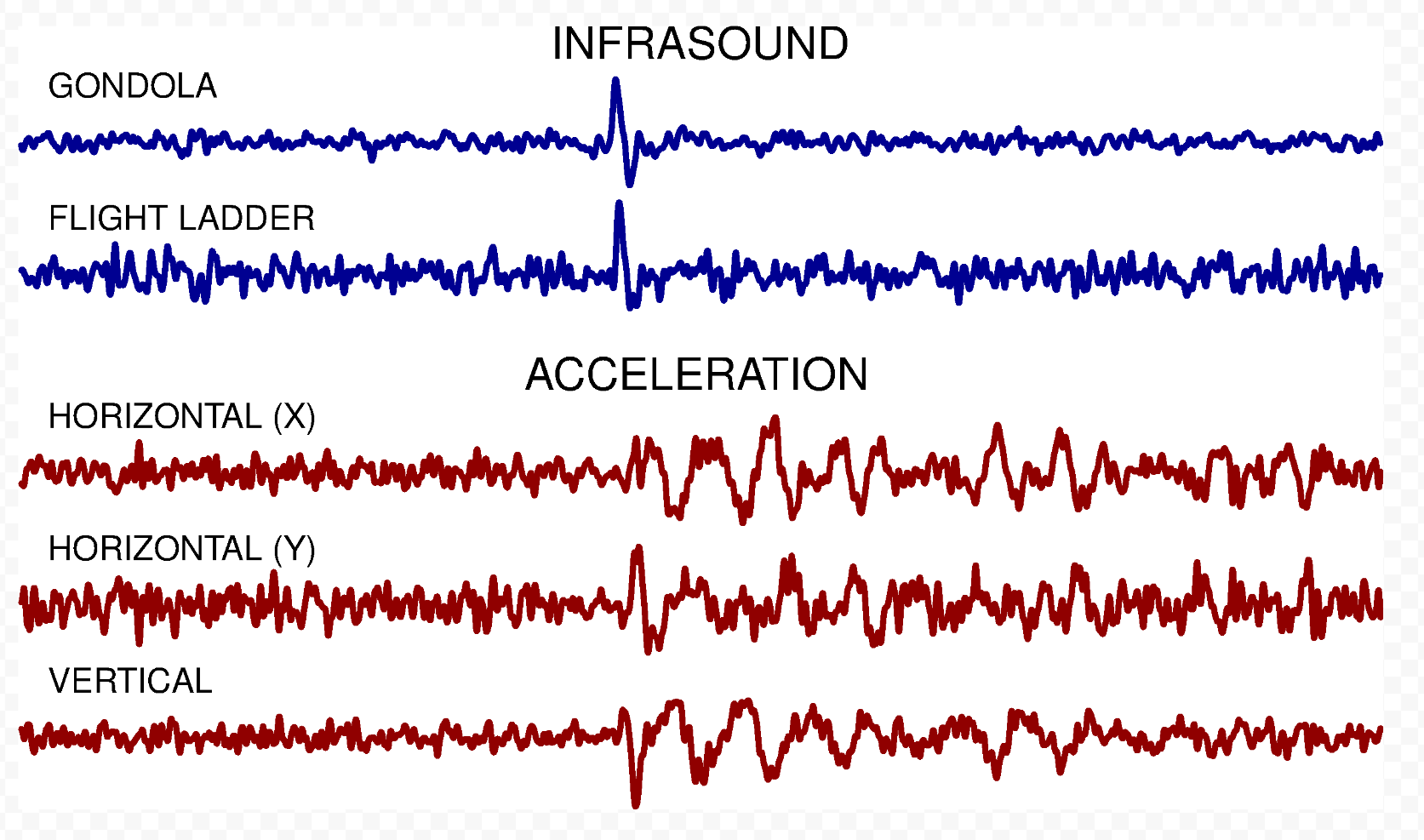}}
\caption{A faint acoustic pulse causing an 11.8 million cubic ft zero pressure balloon to shift perceptibly.  This signal, recorded on the Carolina Infrasound payload on the 2016 High Altitude Student Platform, is what led to the SUPERSEIS concept.
\label{infrasound}
}
\end{wrapfigure}
Natural events and human activity often generate acoustic waves capable of traveling tens to tens of thousands of kilometers across the globe.  
Ground-based acoustic sensors are limited to dry land and often suffer from wind noise.  
In contrast, balloon borne acoustic sensors can cross oceans, polar ice caps, and other inhospitable areas, greatly expanding sensor coverage.  Since they move with the mean wind speed, their background noise levels are exceptionally low.  
In the last six years, such sensors have recorded sounds from colliding ocean waves, surface and buried chemical explosions, thunder, wind/mountain interactions, wind turbines, aircraft, and possibly meteors and the aurora.  
These results have led to new insights on acoustic heating of the upper atmosphere, the detectability of underground explosions, and directional sound fields generated by ocean waves.
Traditional pressure sensors give a scalar result, and thus cannot determine the direction of arrival of a signal.  
However, since balloons are suspended in fluid, acoustic waves cause the entire flight system to shift, providing a vector back to the signal source (Fig. ~\ref{infrasound}).

\noindent { \textcolor{royalblue}{\textbfit{Notable recent and planned payloads for Geophysical \& Planetary Acoustics}}}

The Superpressure Balloon Seismometer (SUPERSEIS) instrument is a directional acoustic sensor that leverages the motion of acoustic waves in the atmosphere to geolocate distant sound sources~\ref{infrasound}.   
Since it only weighs a few kilograms, it is an ideal piggyback for many NASA missions, and was previously launched with the COSI mission from Wanaka, New Zealand in 2016. 
Continued flight opportunities will assist in characterizing of background noise levels and refining signal source location methods.  
Eventually, the SUPERSEIS instrument will be placed on constellations of commercial balloons for global geophysical monitoring as well as included on payloads destined for extraterrestrial environments such as the clouds of Venus.

\subsection{High Energy Atmospheric Physics}
\label{sec:TGF}

\begin{wrapfigure}{r}{0.3\textwidth}
\vspace{-3mm}
\centerline{~~\includegraphics[trim=0 0 0 0, clip, width=2in]{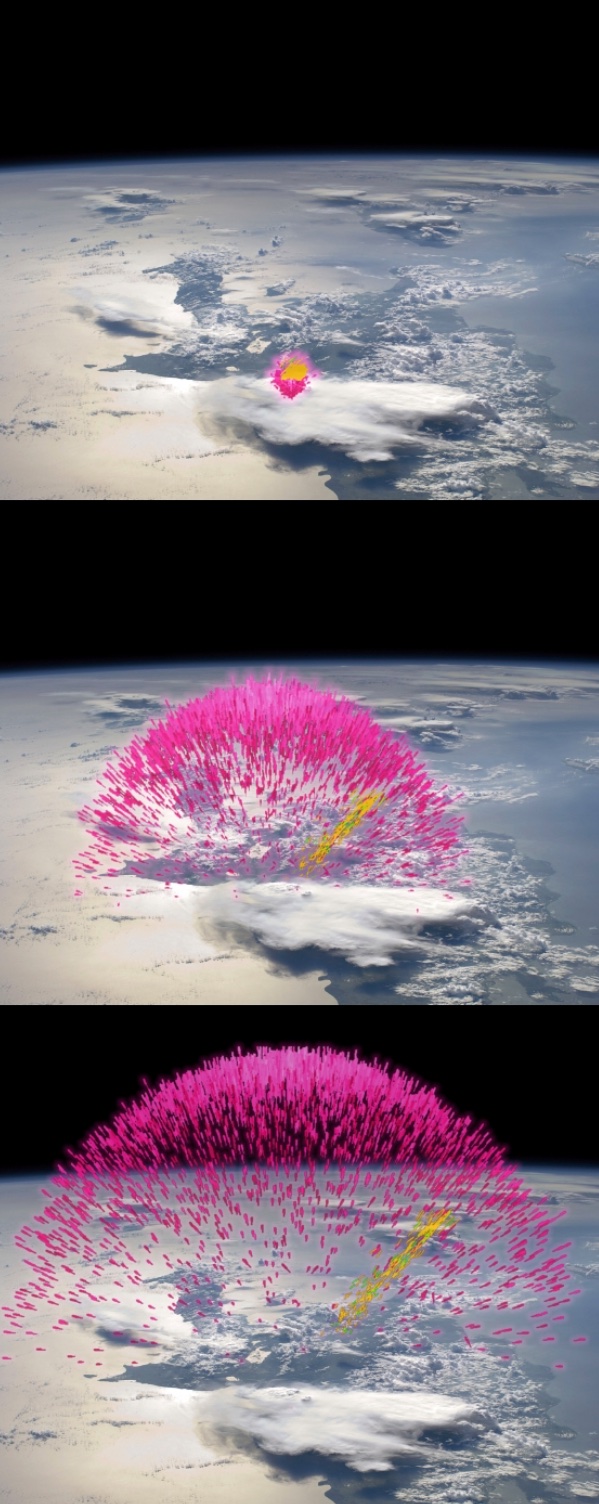}}
\caption{ Visualization of the evolution, from top to bottom, of a TGF from the top of a thunderstorm.
High energy electrons are in yellow, and gamma-rays are in magenta.(Credit: NASA Goddard Media
Studios.)
\label{fig:TGF}
}
\end{wrapfigure}

Intense sub-millisecond bursts of gamma-rays in excess of tens of MeV have been observed from terrestrial sources by satellites for 25 years~\cite{Fishman1313,2009JGRA..114.2314G,Smith2005,Briggs2010,2012SSRv..173..133D}. These Terrestrial Gamma-ray Flashes (TGF, Fig.~\ref{fig:TGF}) are associated with individual lightning flashes, but only about 0.1\% of lightning produces TGF. TGF produce a high enough radiation dose to endanger passengers and crew on aircraft \cite{Dwyer2010}, and evidence suggests that there are no weak or subluminous TGF, i.e., there is some threshold for their production, and they are an all-or-nothing event.  The evidence for this tentative conclusion comes from spacecraft, aircraft, and ground data \cite{smith16, Smith2011, smith18}, but balloon observations provide the best combination of sensitivity (much closer to the lightning than a spacecraft) and exposure (spending more time over a given storm than an aircraft, which has to pass through quickly and turn around).

In addition to searching for subluminous TGF, a balloon payload would be extremely sensitive to x-ray emission from sprites, blue jets, and gigantic jets, all of which involve electric fields which make electron acceleration and x-ray production plausible, although none has yet been observed. These phenomena occur directly at balloon altitudes, so the observations would have the full sensitivity of in-situ exploration. Positron beams, predicted by the leading TGF production model and observed by the Fermi spacecraft \cite{Briggs2011}, are a result of stratospheric pair production. Unlike aircraft, balloon payloads can travel over thunderstorms, intercepting upward traveling beams of gamma rays and particles, providing a link between local and satellite observations and placing new constraints on TGF generation models. To date, only a handful of balloon experiments have been launched to study these phenomena. The most recent, a NASA-funded USIP payload LAFTR (Light And Fast TGF Recorder), was funded primarily as an educational project. However, there is an opportunity for growing the use of balloon-based science payloads in this emerging field of high energy atmospheric physics. 

\hyperlink{TOC}{Return to table of contents.}

\newpage

\section{Planetary Science}
\label{sec:planetary}

At the time of this writing NASA has no proposal opportunities that could fund balloon missions in planetary science. This obstacle is specific to the Planetary Science Division among the four divisions within the Science Mission Directorate.

\begin{tcolorbox}[colback=royalblue!8!white,colframe=royalblue,fonttitle=\bfseries,title=Relevance to SMD Science Goals]

The 2011 Planetary Decadal Survey \textit{Vision and Voyages} states:
\begin{quotation}
Balloon- and rocket-borne telescopes offer a cost-effective means of studying planetary bodies at wavelengths inaccessible from the ground. 
Because of their modest costs and development times, they also provide training opportunities for would-be developers of future spacecraft instruments. 
Although NASA’s Science Mission Directorate regularly flies balloon missions into the stratosphere, there are few funding opportunities to take advantage of this resource for planetary science, because typical planetary grants are too small to support these missions. 
A funding line to promote further use of these suborbital observing platforms for planetary observations would complement and reduce the load on the already over-subscribed planetary astronomy program.
\end{quotation}

Dankanich et al. (2016)~\cite{Dankanich2016PlanetaryBS} surveyed the current Planetary Decadal Survey to see which proposed Decadal investigations were well-suited to terrestrial balloon missions. 
Of the roughly 200 important questions listed in the Decadal Survey, the authors identified 44 that are at least partially addressable from balloon platforms, % in their Tables 3.1 through 3.10, showing links to Decadal 
particularly addressing questions for Primitive Bodies, Inner Planets, Mars, Giant Planets and Satellites. 

\end{tcolorbox}

\begin{figure}[htb!]
\begin{center}

\begin{tabular}{cc}
\includegraphics[width=3.25in]{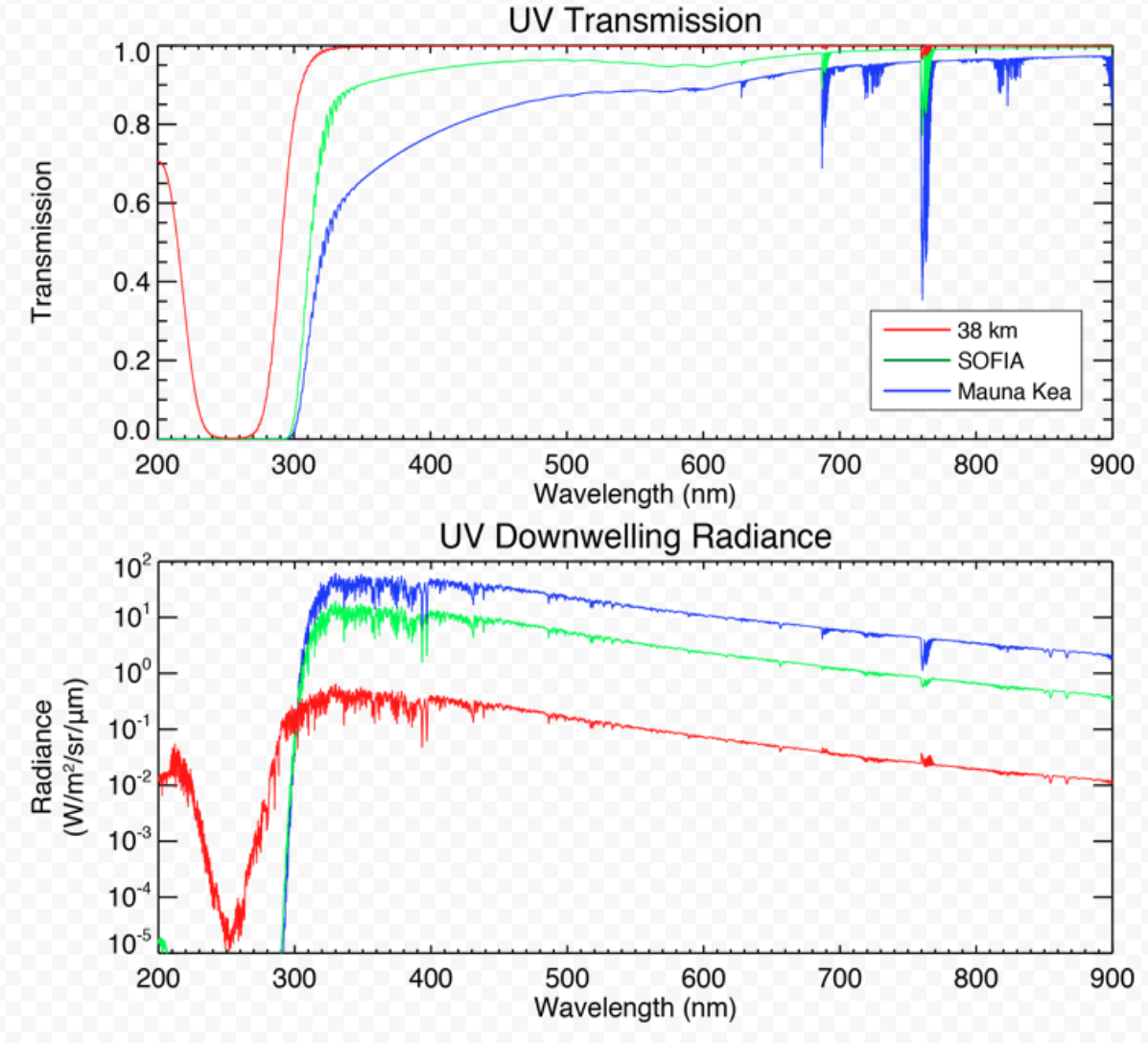}&\includegraphics[width=3.25in]{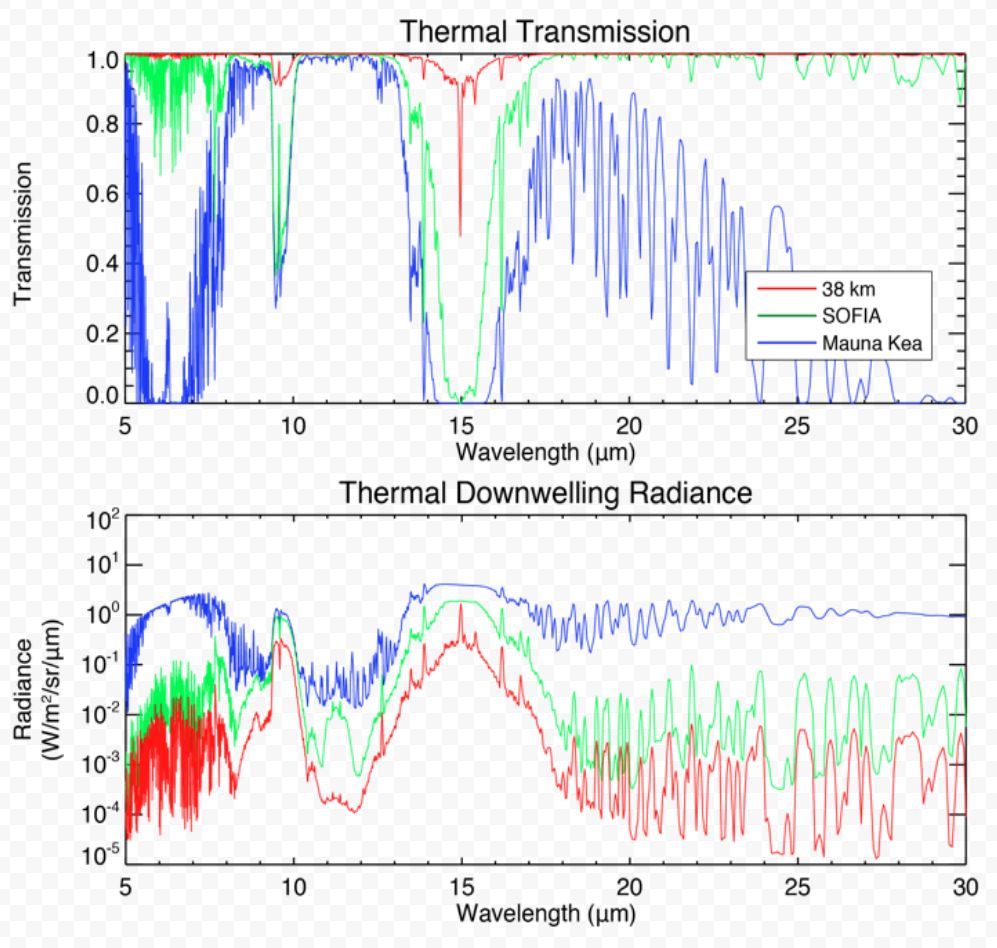} \\
\end{tabular}
\begin{tabular}{c}
\includegraphics[width=3.25in]{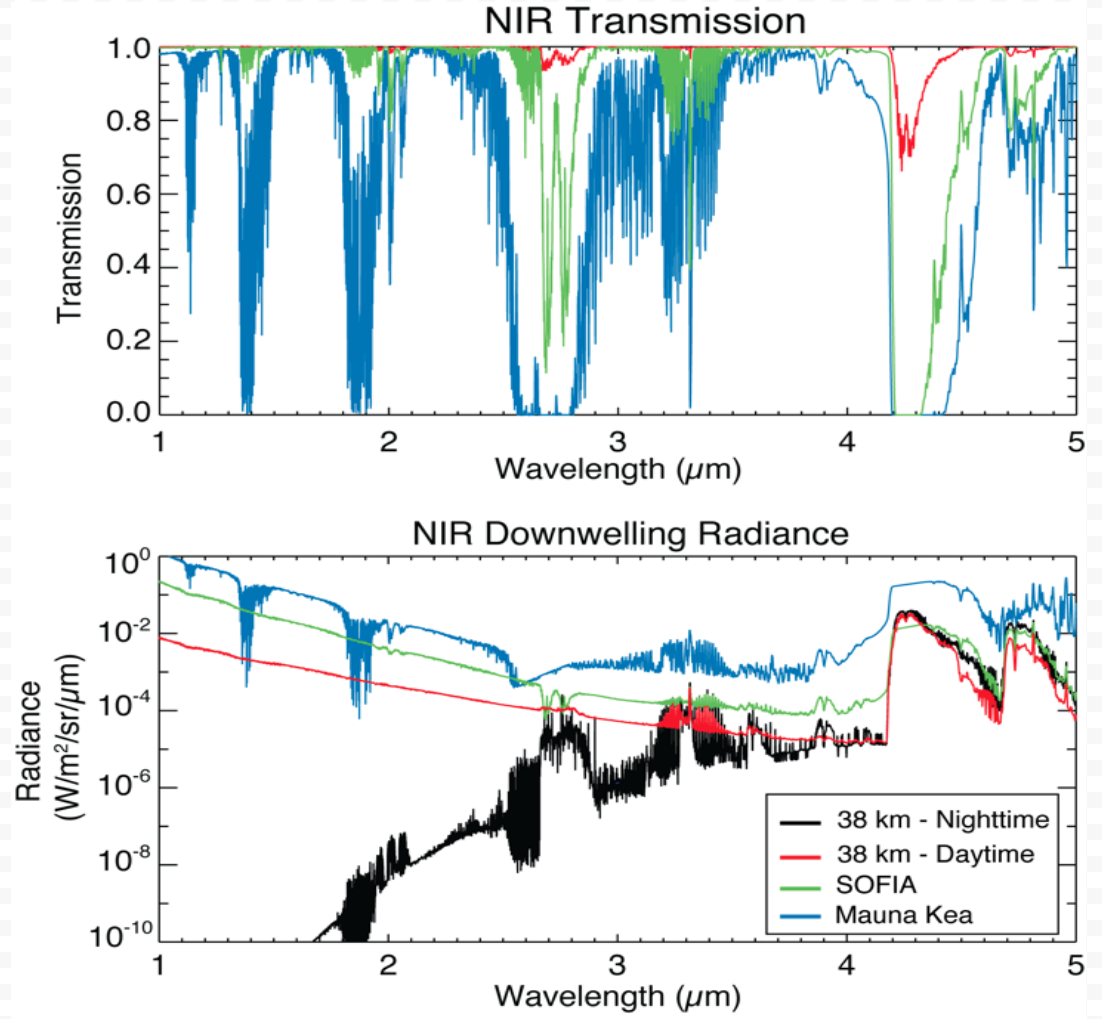}\\
\end{tabular}

\caption{Models of transmission through the telluric atmosphere above 4.2 km (Mauna Kea), 13 km (SOFIA) and 38 km (a stratospheric balloon) show the advantage of the balloon platform for key wavelengths. The downwelling radiance represents summer daytime conditions. 
\label{radiance}}
\end{center}
\end{figure}

The advantage of the balloon platform for Planetary Science derives from three main capabilities: (a) access to key wavelengths (Fig.~\ref{radiance}), (b) time-domain advantages afforded by super-pressure flights, and (c) diffraction-limited imaging at UV and visible wavelengths, where ground-based adaptive optics (AO) systems have poor Strehl ratios.

The recent GHAPS/SIDT report (Gondola for High Altitude Planetary Science/Science Instrument Definition Team) outlines many planetary investigations in which a balloon mission would yield low-hanging fruit~\cite{ghaps}.
The report looked at several classes of questions in detail, assuming the Design Reference Mission \#4 (DRM4) outlined in the Dankanich report -- a 1-m aperture balloon-borne telescope capable of 0.02'' stability in the focal plane.  
For twelve classes of questions (e.g., asteroid and KBO binaries, atmospheric chemistry and dynamics on Ice Giant planets), the GHAPS/SIDT report described the motivating science and quantified the expected versus required SNRs (signal-to-noise ratios) from the DRM4 balloon-borne telescope.

\subsection{Atmospheric Dynamics}

The gas giants (Jupiter and Saturn), the ice giants (Uranus and Neptune), Mars, Titan and Venus all have observable clouds that can be tracked to study atmospheric dynamics. 
There is a rich history of dynamical studies on these objects~\cite{CHOI200735, BARBARA2021114095, Peralta_2018}, either from spacecraft or from high-acuity assets like Keck/AO or HST (Hubble Space Telescope).
Because different wavelengths sound different altitudes, it has been useful to combine HST imaging (at $\lambda$ = 0.2 to 1 $\mu$m) with ground-based adaptive optics (typically $\lambda$ = 1 to 2.5 $\mu$m). 
One problem has been the availability of observing time: the total number of days available on Keck or HST may be less than the number of days needed for an investigation (e.g., weeks of cloud tracking to characterize planetary scale waves or other long-lived phenomena).

\noindent {{\textbfit{\textcolor{goldenpoppy}{Tracking Venus’s cloud tops at UV wavelengths}}}}

The two most recent missions to Venus, VEx (Venus Express) and Akatsuki, were both equipped with UV cameras to image Venus at 283 (sensing SO$_2$) and 365 nm (the unknown UV absorbers). Both missions could image Venus’s full disk at these wavelengths with spatial resolutions of about ~20 km, although VEx’s eccentric polar orbit only obtained full-disk images of Venus’s southern hemisphere and Akatsuki’s eccentric equatorial orbit provides irregular sampling with coarser resolutions over most of its orbit. Compare these to a terrestrial 1-m aperture balloon-borne telescope: its diffraction-limited PSF width is 1.22 $\lambda$/D = 0.075” (at 365 nm), commensurate with a spatial resolution of 30 km on Venus from a distance of 0.55 AU. Larger apertures could actually exceed the spacecrafts’ resolutions. A Venus balloon mission could carry additional UV filters (e.g., with sensitivity to SO), and, unlike the spacecraft missions, it could obtain images at 10-min intervals over a 100-day baseline to help characterize Hadley cell-like circulation, convection, gravity waves (some launched by Venus’s mountains) and planetary scale waves.

\begin{figure}[htb!]
\begin{center}
\vspace{2mm}
\centerline{\includegraphics[width=5.5in]{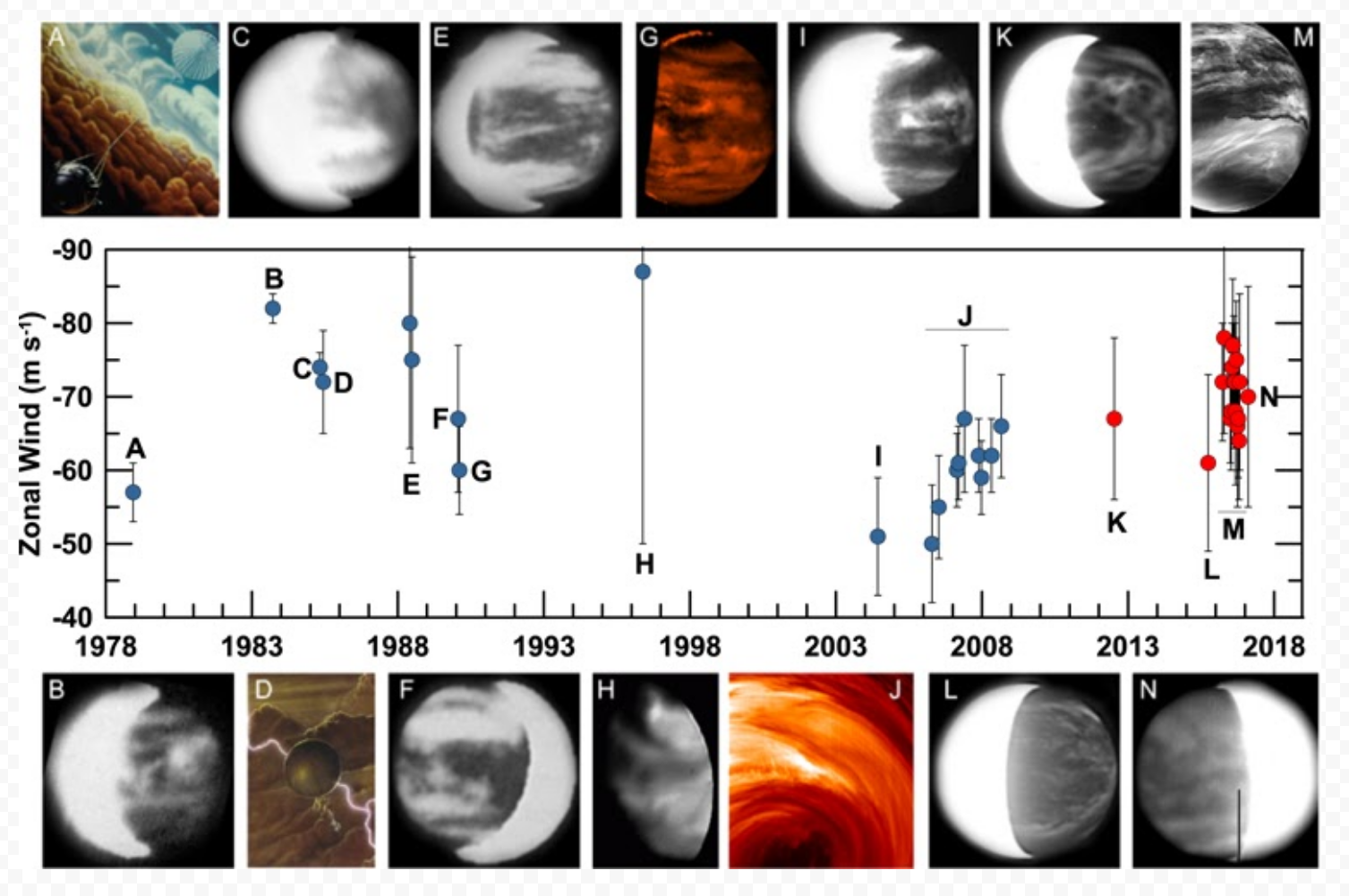}}
\caption{Decadal variation of the zonal winds at the nightside lower clouds of Venus. Data correspond to time averages of the zonal winds obtained between 30$^{\circ}$S and 30$^{\circ}$N with cloud tracking, except for panels (A) and (D). Blue dots represent time averages using wind speeds provided in previous publications. New data presented in this work and based on cloud tracking measurements are shown by the red dots and were obtained from the TNG/NICS (K), IRTF/SpeX (L and N), and Akatsuki/IR2 images (M). More information at the paper: http://iopscience.iop.org/article/10.3847/1538-4365/aae844/meta.
\label{zonalWinds}}
\end{center}
\end{figure}

\noindent {{\textbfit{\textcolor{goldenpoppy}{Tracking Venus’s lower and middle cloud decks}}}}

Venus’s CO2 atmosphere has transmission windows at 1.74 $\mu$m and 2.2 – 2.6 $\mu$m, where thermal emission from the hot surface and lowest scale heights of the atmosphere show reveal clouds from 48 – 55 km as silhouettes (Fig.~\ref{zonalWinds}). Cloud tracking in these wavelength windows determines wind fields that constrain Venus’s circulation, and the distribution of cloud opacities, particle sizes and chemical composition constrain the coupled radiative-convective-chemical processes that control Venus’s atmosphere.
At 1.74 and 2.3 $\mu$m, the diffraction limit of a 1-m aperture is 0.44 and 0.58”, respectively – hardly better than good ground-based sites. Image sequences from the IRTF can resolve cloud motions with errors at the 3 m/s level in good seeing conditions, but the goal should be 0.5 m/s errors, in order to resolve meridional winds that are expected to be ~1 m/s. A large aperture – about 3-m in diameter – and a 6-hr observing interval are necessary to achieve the 0.5 m/s goal.

\noindent {{\textbfit{\textcolor{goldenpoppy}{New technologies needed for cloud-tracking UV and Infrared telescopes}}}}

Using Venus as a proxy for cloud-tracking observations in general, it is clear that balloon-borne telescopes need to advance in several areas. At UV and visible wavelengths, no balloon-borne telescope has yet achieved diffraction-limited performance. We recommend studies in three key areas to support this goal: thermal stabilization of OTAs, wavefront sensors to generate real-time in-flight wavefront errors, and a wavefront correction  scheme (e.g.,deformable mirrors) to compensate for OTA aberrations.
At IR wavelengths, there is a need for large aperture telescopes, although not necessarily ones with diffraction-limited image quality. We recommend studies of lightweight and affordable mirror materials, to see if D > 3-m apertures can be flown when combined with wavefront control measures.

\subsection{Comets}

\noindent {{\textbfit{\textcolor{goldenpoppy}{Comets in UV and IR Wavelengths}}}}

The composition of cometary jets can help determine where a comet formed in the early solar nebula, based on the idea that different volatiles condensed at different distances from the Sun. Two wavelengths are particularly useful: OH lines at 308 – 311 nm, which serve as a proxy for water production (McKay et al. 2018), and the CO2 band at 4.3 $\mu$m. The former can be observed from the ground, but with telluric extinctions around 50\% (Fig.~\ref{radiance}), the latter cannot.

\noindent {{\textbfit{\textcolor{goldenpoppy}{Cooled OTAs for IR Wavelengths}}}}

Heat transfer in the stratosphere is dominated by radiative and conductive processes. There is an advantage to reducing the thermal emission from the telescope optics and the OTA enclosure: on the BOPPS mission, thermal emission from the primary and secondary mirrors accounted for approximately two-thirds of the background counts at 4.3 $\mu$m, despite the fact that the primary mirror was coated with low-emissivity gold. We recommend studies on ways to cool the OTA, beginning with sun- and earth-shields. Preliminary thermal models (Young et al. 2015) suggest that passive shielding can reduce the daytime temperature of a telescope by ~100 K, which translates to nearly a factor of 6x reduction in thermal background from a telescope’s own mirrors.

\subsection{Asteroids and Trans-Neptunian Object Satellites}

Satellite detections and their orbit characterizations can potentially (a) determine the mass of the central body, (b) constrain formation scenarios and (c) rule out classes of events (e.g., impacts, close encounters) in an object’s history. These detections can be challenging, since they often entail identifications of faint objects adjacent to bright objects. This means that the PSF halo is of paramount importance. Cloud tracking (or other imaging of low-contrast fields) is relatively tolerant of power in a PSF’s halo, as long as the PSF’s core is narrow. Satellite detection requires low power in the PSF halo to ensure that the satellite is not swamped by light from the central body. The best wavelength region (in terms of the best SNR) is often in visible wavelengths because of the peak in solar flux, but can be in specific absorption bands if the central object is covered in certain constituents (e.g., water ice).

\hyperlink{TOC}{Return to table of contents.}

%\subsection{Findings and recommendations for planetary science.}

%[THERE ARE NO DOUBT MORE RECOMMENDATIONS HERE BUT THEY NEED TO BE CALLED OUT SPECIFICALLY]

%Recommendations for detections of faint companions. The PAG recommends studies to improve image performance on balloon-borne telescopes, including temperature control, pointing stabilization, wavefront sensing and wavefront correction.

\newpage

\section{Solar \& Space Physics (Heliophysics)}
\label{sec:helio}

Balloon experiments have a rich history in both solar and magnetospheric physics. In addition to producing important stand-alone science in a range of areas, these experiments have worked in tandem with larger NASA missions to augment their science return and have contributed significantly to the development of Explorer-class satellite missions.

\begin{tcolorbox}[colback=royalblue!8!white,colframe=royalblue,fonttitle=\bfseries,title=Relevance to SMD Science Goals]

Balloon-borne solar and space physics directly addresses three of the key science goals laid out in the 2013 Heliophysics Decadal Survey, \textit{Solar and Space Physics: A Science for a Technological Society}.  

\vspace{0.75cm}

\textbf{Solar Physics} (Sec.~\ref{sec:solar}) studies contribute to Key Science Goal 1, \textit{Determine the origins of the Sun’s activity and predict the variations in the space environment}, by improving our understanding of solar flares and coronal mass ejections.

\vspace{0.75cm}

Balloons provide an ideal platform to address Key Science Goal 2, \textit{Determine the dynamics and coupling of Earth’s magnetosphere, ionosphere, and atmosphere and their response to solar and terrestrial inputs}. Studies of \textbf{Particle Precipitation} (Sec.~\ref{sec:precipitation}) into the atmosphere characterize the loss of relativistic electrons from Earth's radiation belts and quantify the impact on our atmosphere. Balloon-based measurements of the \textbf{Large-scale Magnetospheric Electric Field} (Sec.~\ref{sec:efield}) have a long history and provide a global view of solar-wind-magnetosphere-ionosphere coupling. \textbf{Thermospheric and Mesospheric Studies} \ref{sec:ITM} probe processes in the upper atmosphere that are driven both by terrestrial weather from below and space weather from above.
\vspace{0.75cm}

 Balloon-based studies in all of these areas address Key Science Goal 4 to \textit{Discover and characterize fundamental processes that occur both within the heliosphere and throughout the universe}. Balloon observations have made important contributions ranging from particle acceleration and magnetic reconnection in solar flares to wave-particle interactions in space plasmas to plasma-neutral coupling in atmosphere-ionosphere systems.

\end{tcolorbox}

\subsection{Solar Physics}
\label{sec:solar}

\begin{figure}[htb]
\begin{center}
\centerline{\includegraphics[width=5.5in]{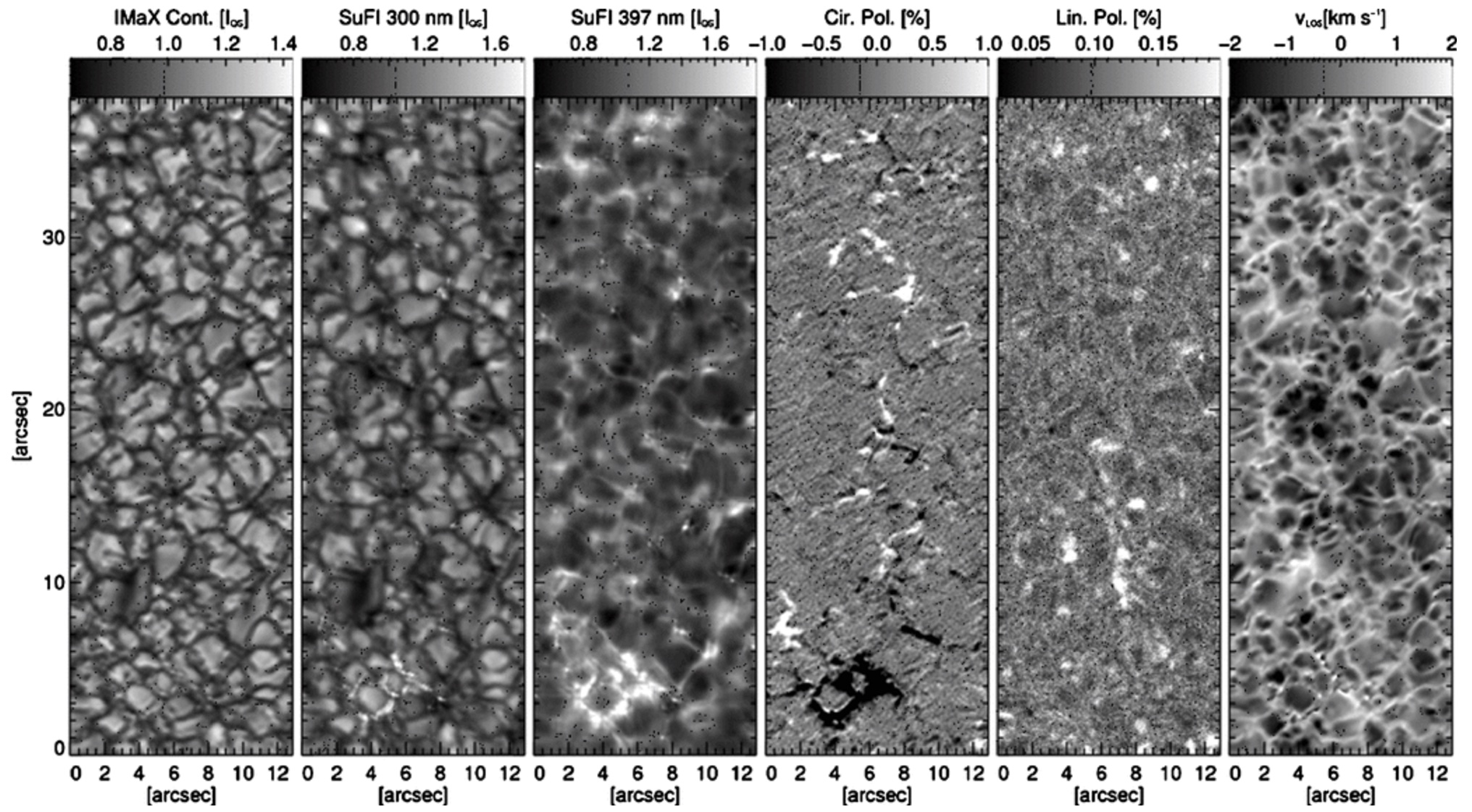}}
\caption{SUNRISE Images of solar granulation. From left to right: continuum intensity at 525.04 nm, intensity in the 300 nm band, intensity in the Ca II H line core, net circular polarization deduced from Fe I 525.02 nm, total net linear polarization and LOS velocity. Figure reproduced from \cite{Solanki_2010}.
\label{fig:granules}}
\end{center}
\end{figure}

Balloons can carry large and powerful optical telescopes high above the atmosphere where they make ultra high-resolution observations of the Sun at $\sim$50 km spatial scales (Figure~\ref{fig:granules}), three times better than the state-of-the-art Hinode space mission [NRC Committee on NASA’s Suborbital Research, 2010]. Balloons also enable measurements at UV and IR wavelengths unreachable from the ground due to atmospheric absorption. 

Gamma-ray observations of the Sun must also be done above the atmosphere and require heavy instruments, making balloons an ideal platform. Gamma-ray imaging indicates that solar flares accelerate and transport ions differently from electrons. Long-duration balloon flights with solar gamma-ray instruments are critical for catching infrequent transient events such as solar flares. 	

\noindent\textcolor{blue}{\textbfit{Notable recent and planned payloads in Solar Physics}}

Utilizing the economical large lifting capability of balloon payloads, the Gamma-Ray Imager/Polarimeter for Solar flares (GRIPS) payload was flown in Antarctica in 2016~\cite{GRIPS}, achieving angular resolution significantly better than any observatory to date, including the Reuven Ramaty High Energy Solar Spectroscopic Imager (RHESSI) spacecraft, in the hard X-ray to gamma-ray range of 20 KeV - 10 MeV. GRIPS is designed to answer key questions in solar energetic particle acceleration, investigating correlations between energetic electrons and ions, and anisotropies in the distributions of relativistic electrons in the corona \cite{Shih2012}.

The SUNRISE solar telescope was built by an international team led by Max Planck Institute but with U.S. science participation and balloon operations support from NASA. SUNRISE investigates the physics of the solar magnetic field and convective plasma flows in the solar atmosphere, both crucial for understanding solar activity. The instrumentation includes both UV and IR spectropolarimeters and an imager-tunable magnetograph. The 1-meter solar telescope includes image stabilization to provide near diffraction-limited images and a spatial resolution of less than 100 km on the solar surface (Fig.~\ref{fig:granules}). The 3rd flight of Sunrise - Sunrise-3 - is scheduled for launch in June 2022 from ESrange in Sweden. 

\begin{wrapfigure}{r}{0.4\textwidth}
\centerline{~~\includegraphics[trim=0 0 0 0, clip, width=2.6in]{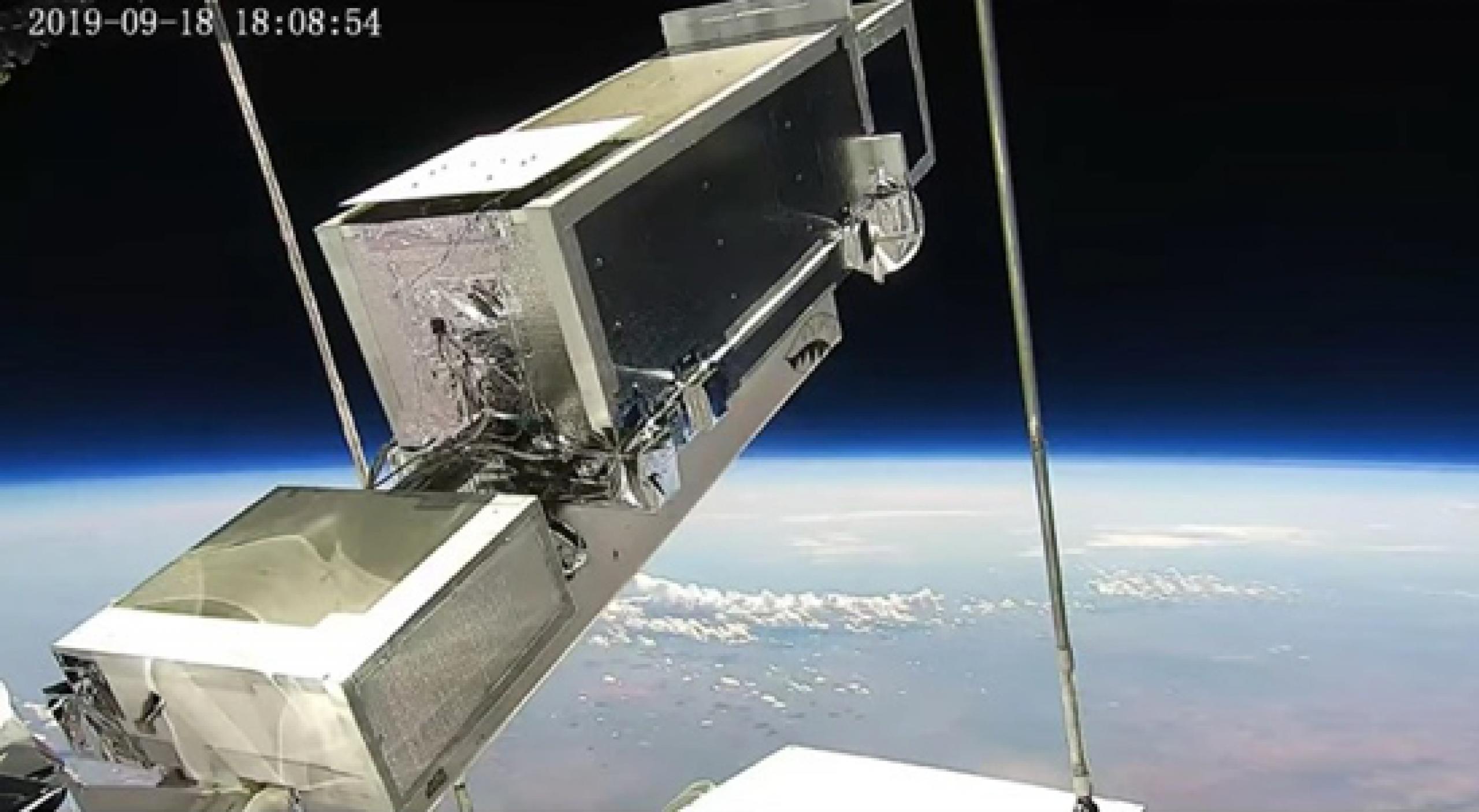}}
\caption{The BITSE coronagraph is pointed toward the Sun during its September 2019 flight.
\label{bitse}
}
\end{wrapfigure}

The Balloon-borne Investigation of Temperature and Speed of Electrons (BITSE, Fig.~\ref{bitse}) flew a test flight in September 2019 and demonstrated the feasibility of passband ratio imaging with a coronagraph to obtain 2-D maps of electron temperature and flow speed, in addition to density, from 3-9 solar radii. BITSE successfully utilized the Wallops Arcsecond Pointer (WASP), and has now proven technologies needed for a future mission on the International Space Station, called Coronal Diagnostic Experiment (CODEX), which is planned to fly in 2023.

\begin{figure}
    \centering
    \includegraphics[width=5in]{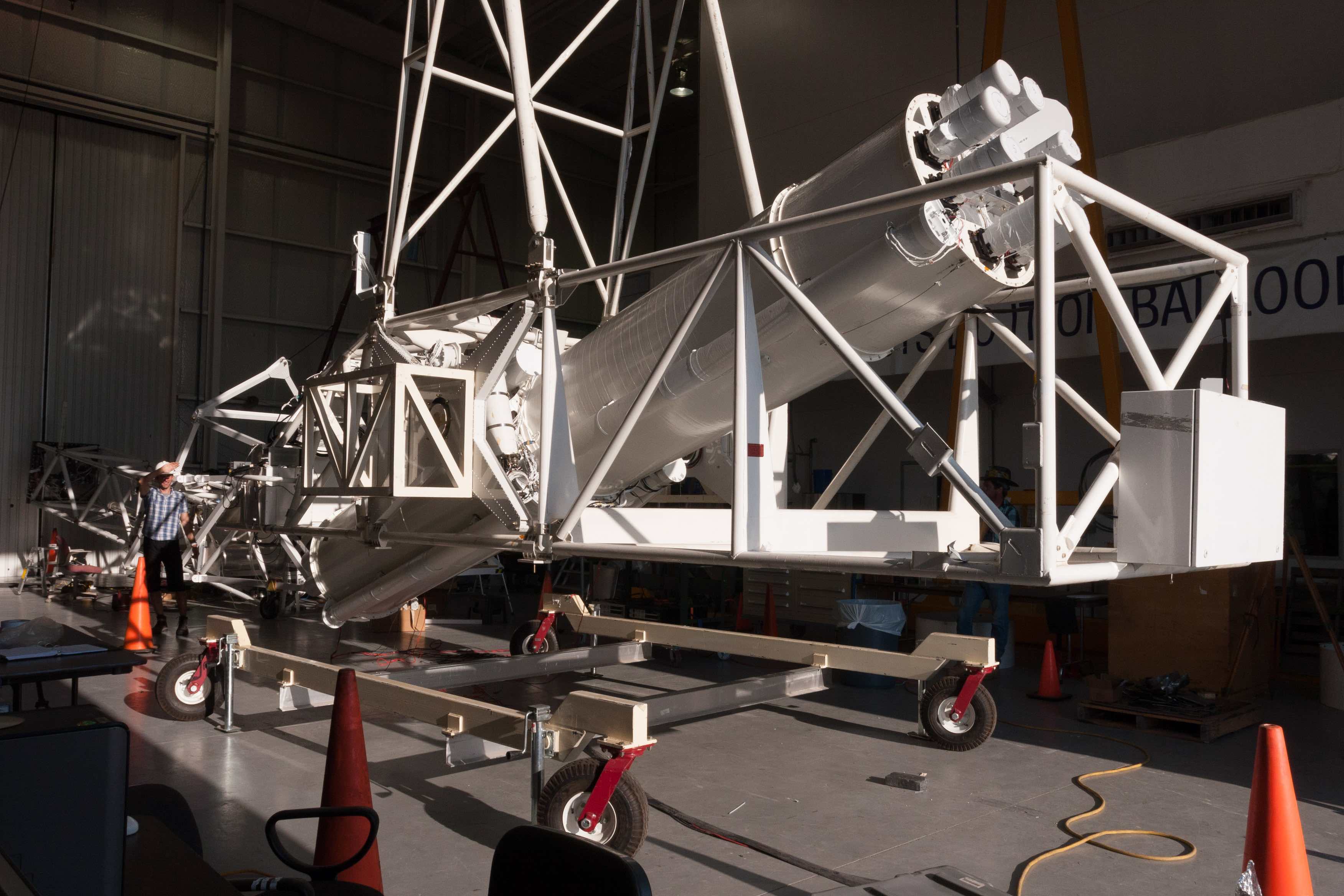}
    \caption{The HEROES hard X-ray telescope fully integrated prior to flight from Fort Sumner in 2013. HEROES was funded with both solar and astrophysics science goals and completed observations of a quiescent active region of the sun, the black hole binary GRS~1915+105, and the Crab Nebula during its 27 hour flight.}
    \label{fig:heroes}
\end{figure}

The High Energy Replicated Optics to Explore the Sun (HEROES) hard X-ray telescope was quite unique in that it was funded with both solar and astrophysics objectives~\cite{HEROES}. HEROES was built off of the success of the MSFC-developed HERO payload and consists of multiple grazing-incidence hard x-ray telescopes with a 6~m focal length, coupled to gas scintillation proportional counter detectors. During a flight from Fort Sumner, NM, in September 2013~(Fig.~\ref{fig:heroes}), HEROES observed the quiescent sun during the day and astrophysical sources at night, with fine pointing and excellent angular resolution.

\subsection{Particle Precipitation into Earth’s atmosphere}
\label{sec:precipitation}

Particles with energies spanning more than four decades - from auroral electrons (100 eV  - 10 keV) to relativistic electrons from the Van Allen belts (100 keV - MeV) - rain down onto Earth’s upper atmosphere from space. Auroral precipitation controls the nighttime conductivity of the ionosphere while relativistic precipitation plays an important role in the dynamics of the Van Allen radiation belts \cite{Millan2007}, and may also be important for the creation of HOx and NOx that destroy ozone \cite{Randall2005}.  

Electron microburst precipitation was discovered with balloons \cite{Anderson1964} and is known to be an important loss mechanism for the radiation belts. These rapid ($\sim$100 ms) bursts of precipitation provide an ideal target for studying the physics of non-linear wave-particle interactions in space plasmas. Balloon-borne payloads, which are essentially stationary beneath the precipitating region, can easily distinguish the spatial and temporal variations of these bursts which is difficult from a LEO satellite. Multi-point experiments using multiple small balloons have probed the large-scale spatial structure of energetic electron precipitation, and experiments are under development to perform high-resolution imaging, only possible from a balloon platform. 
 
\noindent\textcolor{blue}{\textbfit{Notable recent and planned payloads in Particle precipitation into Earth's atmosphere}}

\begin{figure}[htb]
\begin{center}
\centerline{\includegraphics[width=6in]{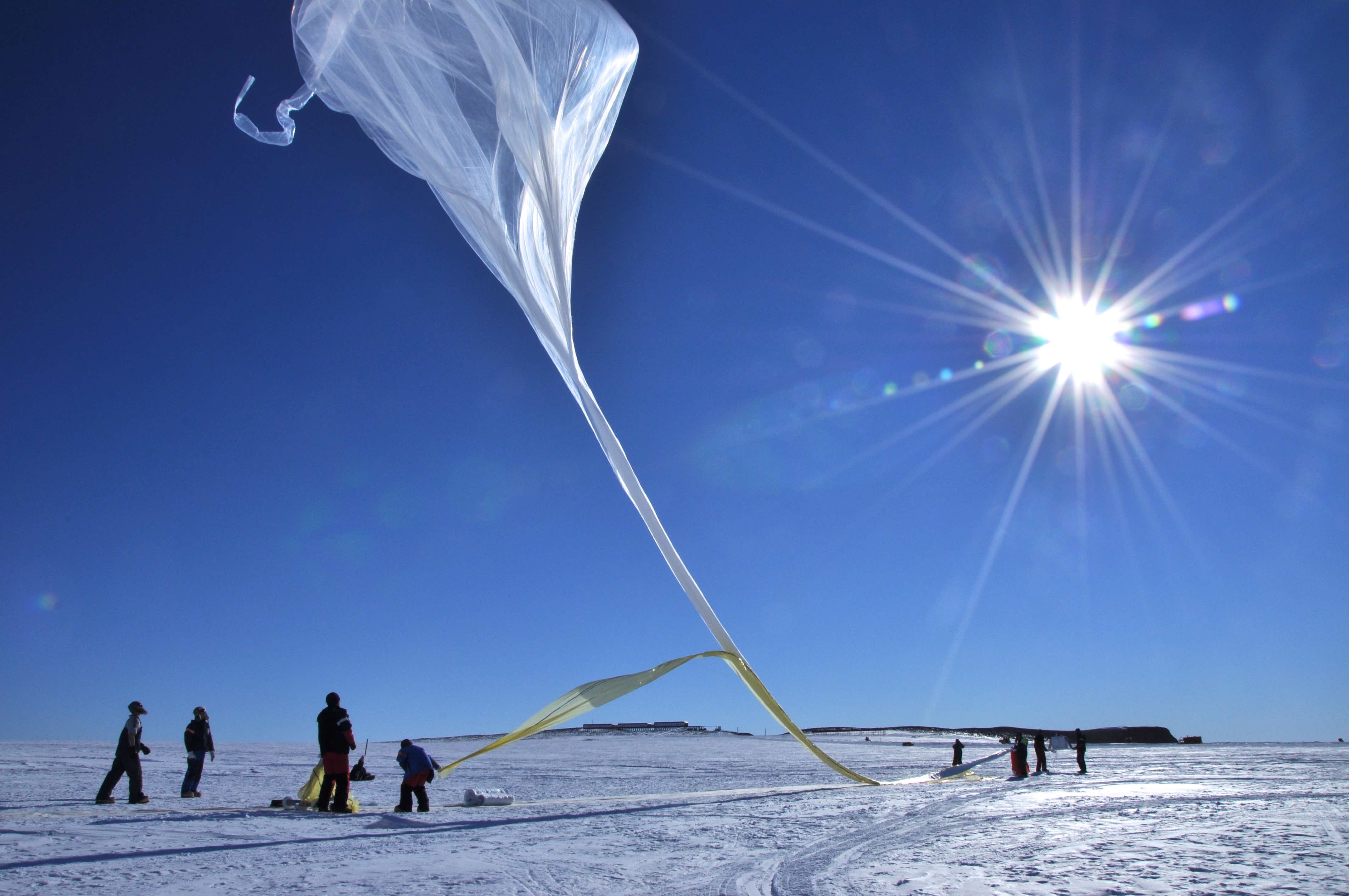}}
\caption{BARREL hand-launch at the South African Antarctic Station SANAE IV in 2013. 
\label{fig:barrel}}
\end{center}
\end{figure}

The Balloon Array for Radiation belt Relativistic Electron Losses (BARREL) is a multiple-balloon experiment that worked in tandem with NASA’s twin Van Allen Probes to study atmospheric loss of radiation belt electrons \citep{Millan2013}. A total of 58 BARREL science flights were conducted during 7 balloon campaigns between 2013-2020, from launch locations in Antarctica and Sweden. Each $\sim$25 kg payload carries a 3"$\times$3" sodium iodine scintillator that measures bremsstrahlung x-rays produced by energetic ($\sim$20 keV-10 MeV) electrons as they precipitate into Earth's atmosphere. The majority of BARREL payloads were hand-launched on 300,000 cubic foot zero-pressure balloons (Figure~\ref{fig:barrel}), with the exception of one payload flown as a piggyback on the HIWIND payload, and the 2018-2019 and 2019-2020 ultra-long duration flights, which flew on small superpressure balloons from McMurdo (Section~\ref{ch:tech}). BARREL showed that electromagnetic ion cyclotron plasma waves cause duskside relativistic precipitation \cite{Li2014, Blum2015} and made new discoveries about the spatial scale of energetic precipitation \cite{Breneman2015, Anderson2017}. BARREL primarily probed the large-scale structure of precipitation but also provided some tantalizing evidence of spatial structure at smaller scales \cite{Sample2020}.

\vspace{3mm}

  The COSI Compton telescope (Sec.~\ref{sec:astroparticle}), designed for high-energy astrophysics, made serendipitous observations of electron precipitation during its 2016 flight from Wanaka, New Zealand (See Figure 12 in \cite{kierans2016}). These observations could provide the first detailed images of duskside precipitation, providing insight into wave-particle interaction processes occurring in the radiation belts. However, it has so far proven difficult to obtain data analysis funding to address Heliophysics science objectives using data from an Astrophysics experiment. Like HEROES, the COSI observations provide another example of the connection between high-energy astrophysics and heliophysics experiments where observing techniques often overlap. These examples highlight the opportunities for collaboration between the divisions within SMD, such as support for dual-purpose missions and cross-disciplinary guest investigator proposals.

Balloon Observations Of Microburst Scales (BOOMS) will make high-resolution images of fast ($\sim$100~ms) microburst electron precipitation thought to be caused by nonlinear wave-particle interactions. These measurements will, for the first time, quantify the distribution of electron microburst scale-size, revealing important information about the connection with plasma wave spatial scales near the equator. BOOMS is planned for launch from Sweden in June 2022.

Also in development is a Near Infrared (NIR) InGaAs camera that will be capable of imaging aurora on the dayside and sunlit regions \cite{Zhou2017}. Imaging aurora in daylight is notoriously difficult, but modeling indicates that it is possible to image the aurora at NIR wavelengths from sufficient altitudes, making balloons the ideal platform. The work is currently funded by the NASA ROSES Heliophysics Technology and Instrument Development for Science program, and could open up a new window for studying auroral particle precipitation processes on the dayside of Earth's magnetosphere, including the cusp region. In addition, this work will enable lower energy auroral precipitation to be studied using long-duration balloon flights in Antarctica, which occur during the summer months.

\subsection{Thermosphere and Mesosphere studies}
\label{sec:ITM}

Joule heating in the high latitude regions of Earth has an important impact on the global circulation of the thermosphere, a part of Earth’s upper atmosphere. The magnetosphere, driven by the solar wind, is now understood to be an important source of energy input to the thermosphere. 
However, the details of the energy transfer process are not well-quantified, and the effects of geomagnetic activity, season, and geomagnetic latitude are not well-studied. 
Balloon-borne experiments are extremely effective for monitoring these thermospheric neutral winds, which are observed in closer proximity than is possible from low earth orbit. Such experiments will help quantify the coupling of the magnetosphere, ionosphere and thermosphere and will improve thermospheric density models.

\begin{wrapfigure}{r}{0.45\textwidth}
\centerline{~~\includegraphics[trim=0 0 0 0, clip, width=3in]{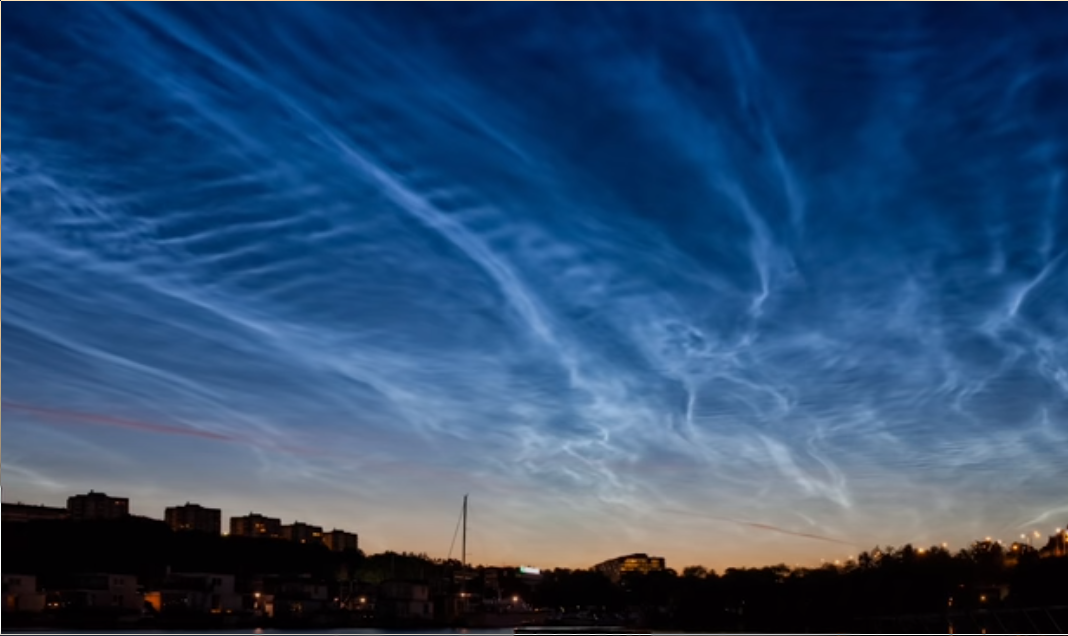}}
\caption{ Noctilucent clouds as observed from ground level.
\label{fig:pmc}
}
\end{wrapfigure}

Balloon payloads also afford excellent platforms for the study of mesospheric phenomena that are still poorly understood:
the formation and development of polar mesospheric clouds (PMC), also sometimes called noctilucent clouds as shown in Fig.~\ref{fig:pmc}. 
These tenuous and rare clouds show turbulent microstructure that requires high-resolution imaging to fully determine the structure and rapid changes as lower-altitude energy is transported into this region.

\noindent\textcolor{blue}{\textbfit{Notable recent and planned payloads for Thermosphere and Mesosphere investigations}}

The High altitude Interferometer Wind observation (HIWIND) payload from the High Altitude Observatory, National Center for Atmospheric Research in Boulder, Colorado, is designed to measure thermospheric winds by monitoring the Doppler shift in the airglow emission O 630 nm induced by the neutral winds. 
The payload first flew most recently in 2018 from Kiruna, Sweden, and continues in active development. 

Also in 2018, The Polar Mesospheric Cloud Turbo (PMC Turbo) payload flew successfully from ESrange Sweden, to Canada, capturing
$\sim 120$~Tbyte of images of PMC. 
PMC Turbo data will be used to investigate the dissipation of lower altitude instabilities and turbulence through their coupling to the mesosphere. 
These processes may be important in influencing weather and climate over the entire atmosphere, and imaging of PMC from balloon altitudes offers a unique window on these effects not otherwise available.

\subsection{Large-scale Magnetospheric Electric Field. }
\label{sec:efield}

Unlike the magnetospheric magnetic field, which penetrates to the ground, the dynamics of the large scale electric field in the ionosphere and magnetosphere cannot be deduced from the ground. A BARREL-like flotilla capable of measuring vector electric fields would allow for an instantaneous determination of the large scale magnetospheric electric field. There is no other cheap way to get direct, real- time, mapped-in-situ, large scale magnetospheric electric fields with 1 to 10 second resolution, and global, instantaneous coverage. 

Magnetospheric science is a system science often requiring multi-point measurements and coordinated measurements from different platforms. This holds true for studies of the global electric field, but also for other areas of research including particle precipitation, and ionospheric studies.

%\subsection{Findings and recommendations for Heliophysics}

%\begin{itemize}
%\item Recommendations for solar physics. The PAG recommends continued development of ultra-long duration flight capabilities which will improve the chances of detecting infrequent transient events such as solar flares.

%\item The PAG recommends continued operations from locations that support research in auroral and radiation belt physics, and high latitude magnetosphere-, ionosphere-, and thermosphere-coupling, requiring magnetic latitudes ranging from 55-70 degrees (e.g., Kiruna, Sweden).

%\item The PAG recommends continued support and development of small hand-launch balloon capability, in particular to support simultaneous operation of multiple (dozens) small, long-duration payloads. 
	
%\item The PAG recommends that NASA develop and support hand-launch (science payload 50 – 75 pounds) balloon infrastructure including tracking, balloon fill, telemetry, and ground support that is lightweight and able to the launched in the vicinity of thunderstorms.
%\end{itemize}

\hyperlink{TOC}{Return to table of contents.}

\chapter{Capabilities of the Balloon Platform}
\label{ch:tech}

Since the founding of NASA scientific balloon program, and particularly within the last two decades, the sophistication and complexity of balloon payloads has steadily increased. 
Flight durations for launch sites in Antarctica and more recently in mid-latitudes from Wanaka, New Zealand, have steadily increased on average to the point where the long-standing goal of 100 days at float now appears to be possible within the next decade of the program. The combination of more sophisticated payloads combined with these long flight durations now has the potential to yield scientific returns that can rival or exceed what can be accomplished in a much more expensive orbital mission. 

In this chapter, we will describe the state-of-the-art balloon platforms, launch facilities, relevant technologies, and commercial capabilities that enable the groundbreaking science discussed in the previous chapter.

\section{Conventional, Long Duration, and Superpressure\\Balloons}

What can ultimately limit the type and quality of the science performed from a balloon platform is the altitude of the carrier balloon. 
In particular, high energy astrophysics experiments often require both high altitude and high suspended weight, as well as long flight times to increase the likelihood of detections, as discussed in Section~\ref{sec:astroparticle}. 
Similar statements can also be made for overcoming the absorptive and scattering/seeing effects of the atmosphere for payloads operating in the UV and optical. 
Furthermore, in the infrared and far-infrared ranges, atmospheric water vapor limits sensitivity. 
In all cases, the deleterious effects of the atmosphere decrease exponentially with height, making even modest increases in the
balloon maximum altitude capability yield significant increases in science return.

\begin{figure}
    \centering
    \begin{subfigure}{0.22\textwidth}
        \centering
        \includegraphics[height=2.2in]{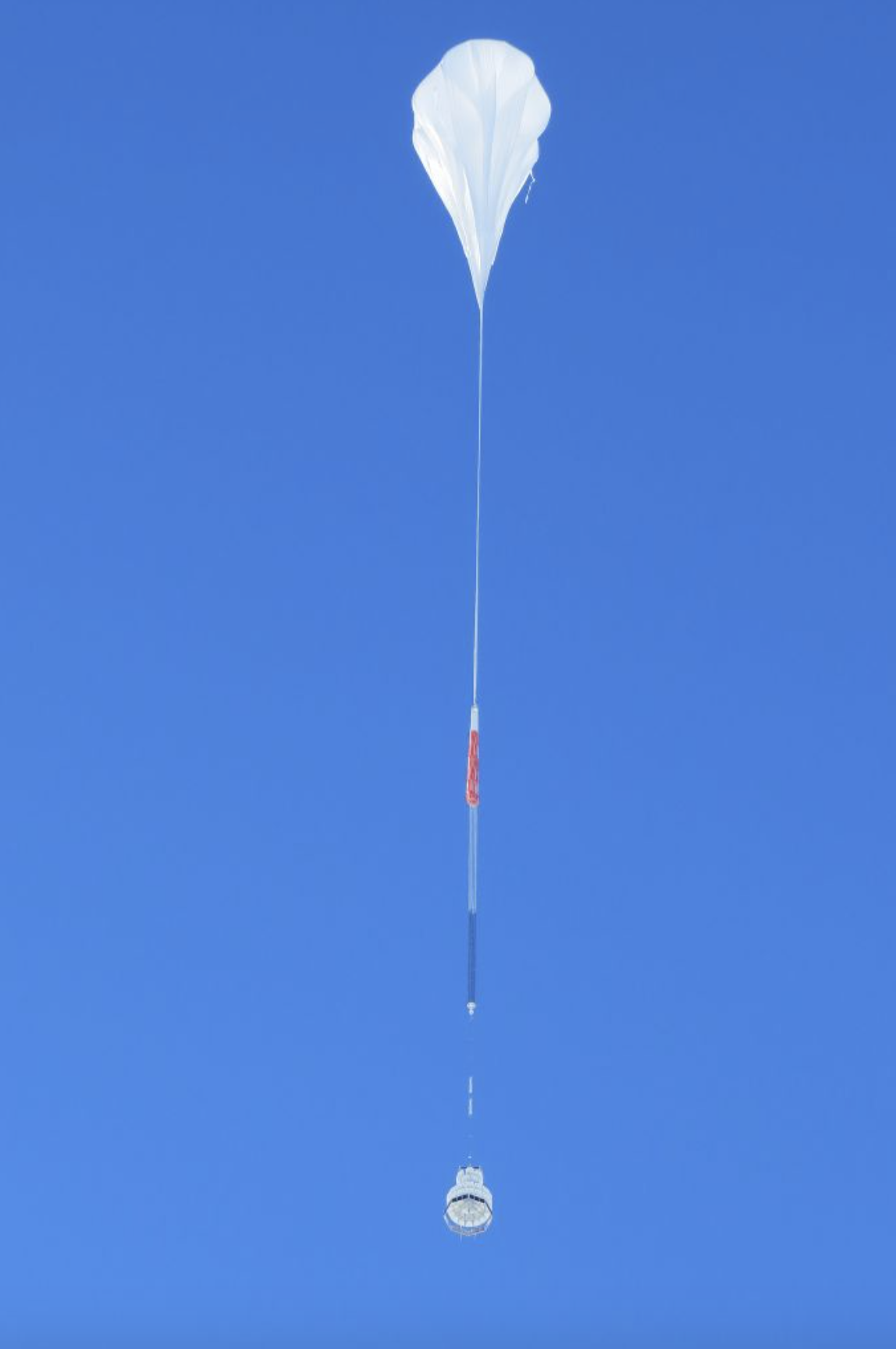}
        \caption{}
    \end{subfigure}
    \hfill
    \begin{subfigure}{0.37\textwidth}
        \centering
        \includegraphics[height=2.2in]{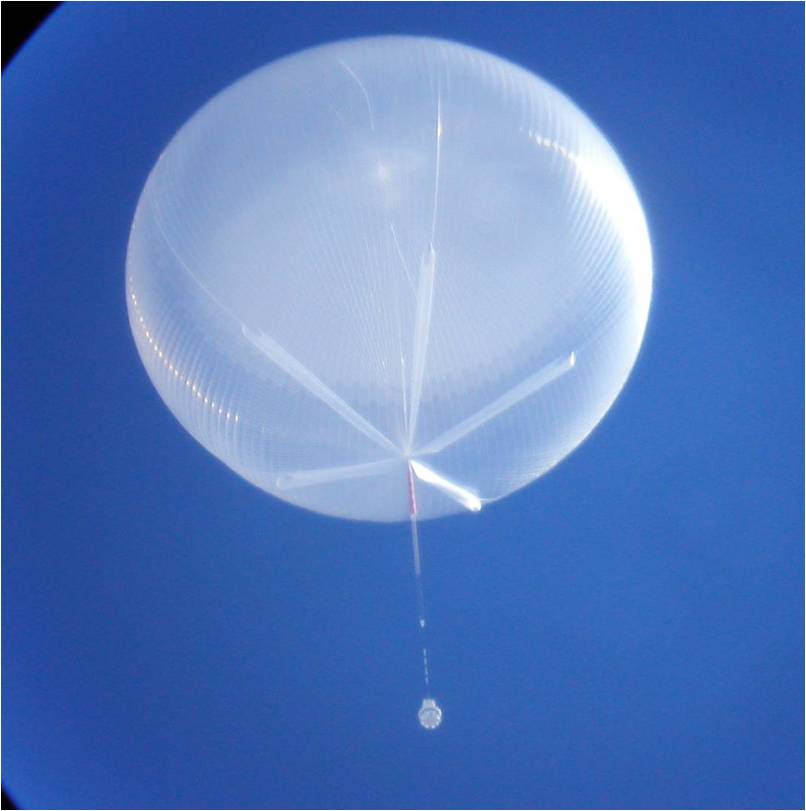}
        \caption{}
    \end{subfigure}
    \hfill
    \begin{subfigure}{0.37\textwidth}
        \centering
        \includegraphics[height = 2.2in]{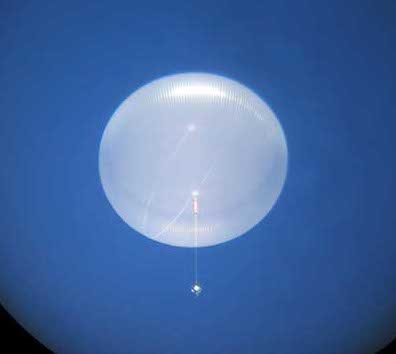}
        \caption{}
    \end{subfigure}
    \caption{(a) Zero pressure balloon from ANITA flight in 2008 shortly after launch. As the balloon rises, the atmospheric pressure drops, and the helium inside the balloon expands to fill the volume. (b) A 29~MCF zero pressure balloon at float has a fully inflated diameter of 130~m. (c) The superpressure balloon at float is shaped more like a pumpkin, and for the 7~MCF balloon shown here, it has a fully inflated diameter of 83~m~\cite{csbfwebsite}.}
    \label{fig:zp_vs_spb}
\end{figure}

NASA's scientific balloons come in two general types: the standard zero-pressure balloon (ZPB) and the superpressure balloon (SPB).
During the launch procedure, these balloons are only partially filled with helium so that once the platform reaches float altitude, the helium expands to fill the whole volume, as shown in Fig.~\ref{fig:zp_vs_spb}.
In the case of a ZPB, once the target altitude is reached, any excess helium is vented such that the balloon is brought into pressure equilibrium with its surroundings. 
If the temperature of the atmosphere drops, the pressure within the balloon will also drop, causing it to lose volume. 
Following Archimedes Principle, the balloon will then sink and ballast must be released to maintain altitude. 
Likewise, if the atmospheric temperature increases, the balloon’s internal temperature will increase. 
The balloon’s volume will then increase and the balloon will rise; to maintain a specific altitude, it may then be necessary to vent helium. 
This use of expendables (ballast and helium) fundamentally limits the time at float for a ZPB. % to approximately 1 to 2 months; however, more drastic day-night temperature cycles, like those found at mid-latitudes, will shorten this duration. 
The ZPBs are used for short (1-2 days) conventional domestic flights, 
as well as long duration balloon (LDB) flights that occur in the austral summer from McMurdo Station, Antarctica (up to 8 weeks).

\begin{figure}[thb]
\begin{center}
\centerline{\includegraphics[width=5in]{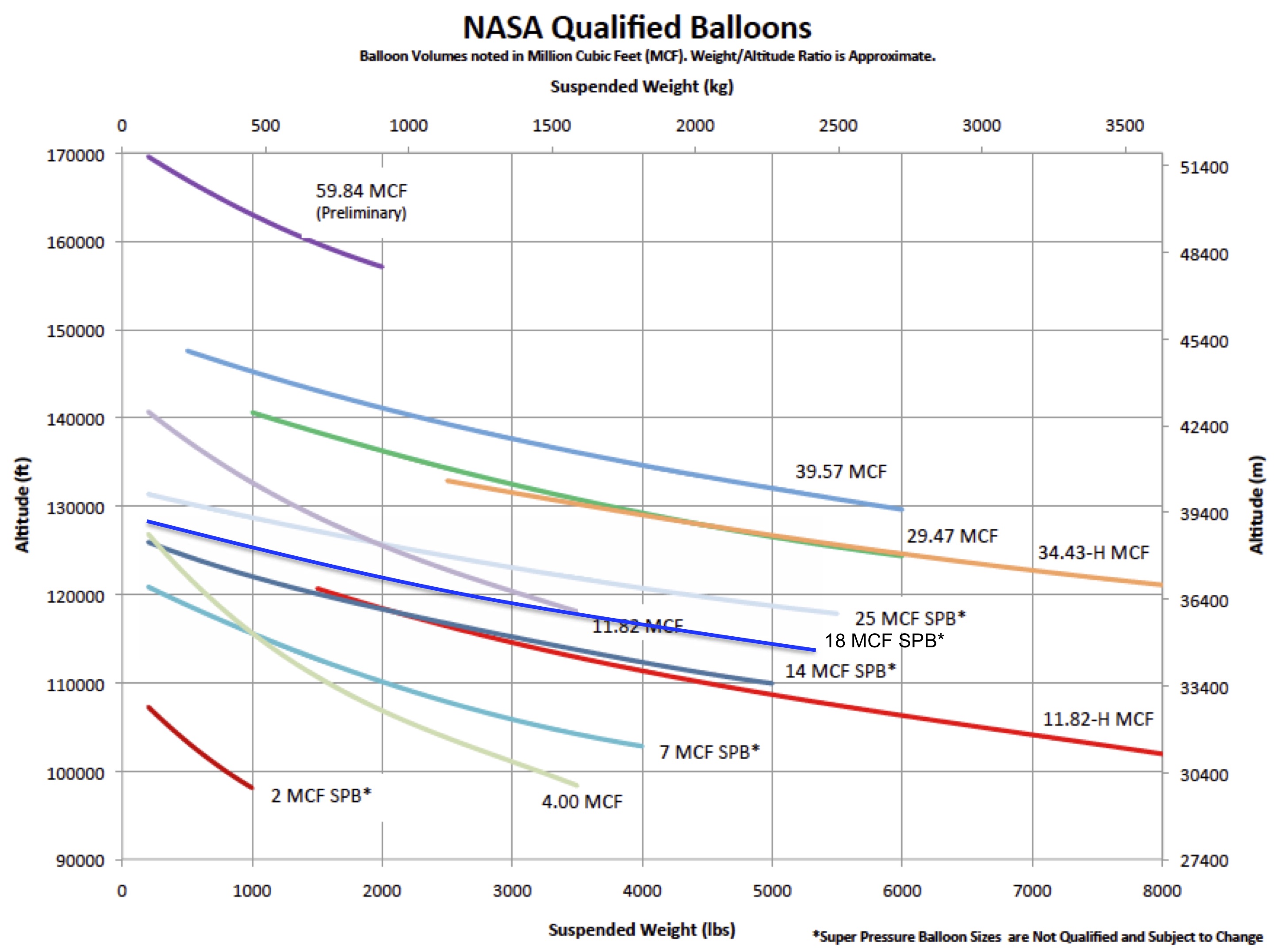}}
\caption{Summary of the range of stratospheric balloons that have flown in the NASA balloon program.
\label{fig:balloons}}
\end{center}
\end{figure}

The SPB is designed to maintain the same altitude through day-night cycles, ultimately allowing for ultra-long duration balloon (ULDB) flights of up to 100 days. 
The balloon is made strong enough such that it can maintain a specific volume even when the helium within it is at a slightly higher pressure than the surrounding atmosphere, hence the term ``superpressure.'' 
This ability to provide a volumetric margin against changes in the temperature of the ambient atmosphere holds the altitude of the balloon constant without the use of expendables, thereby dramatically increasing a SPB’s time at float.

Figure~\ref{fig:balloons} shows the achievable altitude versus suspended weight for the current generation of NASA Qualified Balloons.
ZPBs can take as much as 4~tons into the low stratosphere (~100,000 ft), but less than a ton into the high stratosphere (> 160,000 ft).  
The figure also shows that SPBs have only achieved modest sizes and suspended payload weight with the 18~MCF balloon being the only size used for science flights thus far (see Sec.~\ref{sec:astroparticle}), restricting their practical use to modest altitudes (~110,000 ft). In order to realize the full scientific and cost saving potential of scientific ballooning, the development of larger zero pressure and super pressure balloons should be considered a high priority.

\section{Hand launch and Piggyback Payloads}

\begin{wrapfigure}[12]{r}{0.45\textwidth}
\centerline{~~\includegraphics[trim=0 0 0 2in, clip, width=3in]{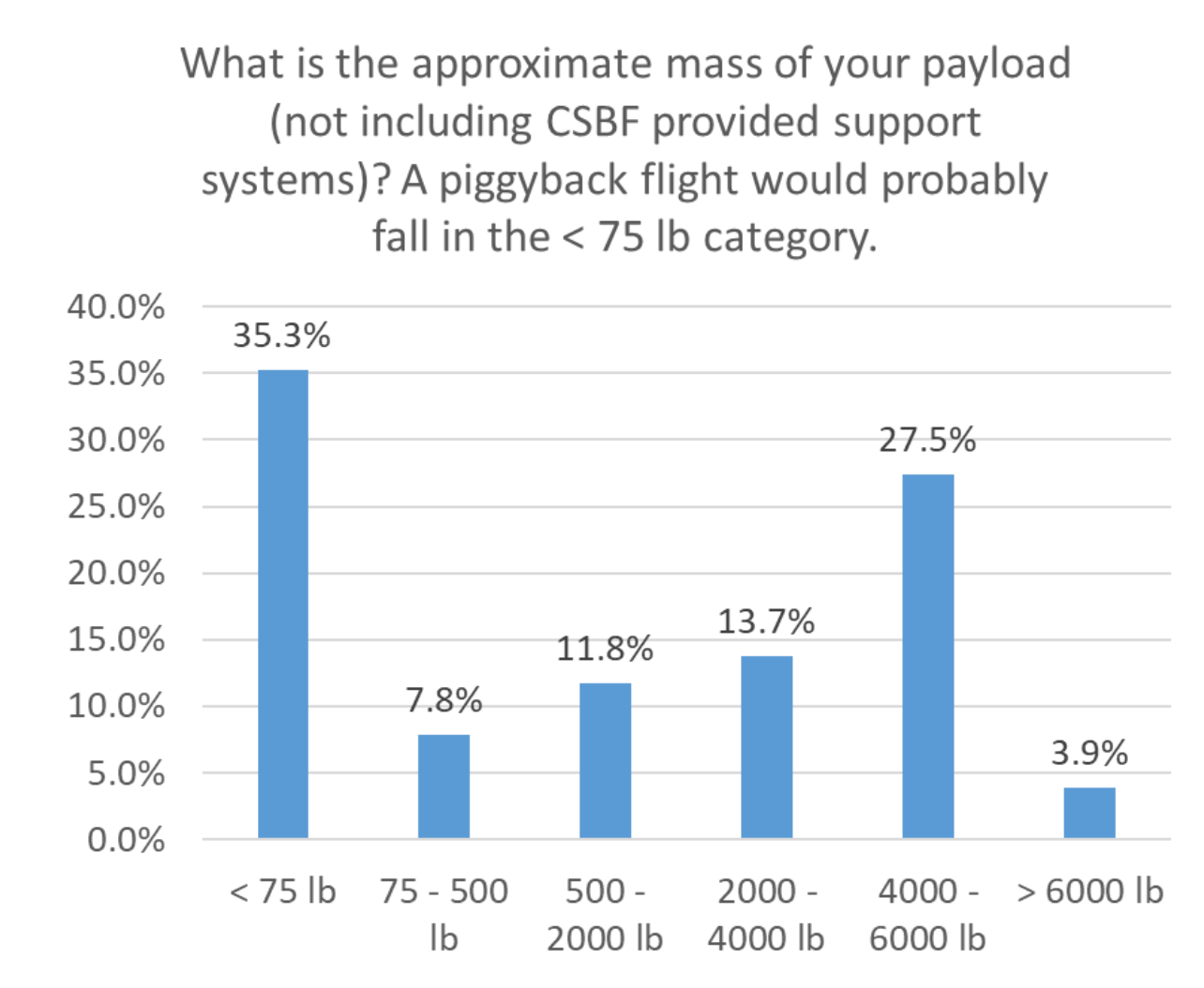}}
\caption{NASA BPO survey results showing percentage of respondents versus payload weight. Figure from 2020 survey, courtesy of NASA BPO.
\label{fig:survey_mass}
}
\end{wrapfigure}

In a recent survey conducted by the BPO, roughly 35\% of the 51 respondents indicated that their payload weighs less than 75 lbs (Figure~\ref{fig:survey_mass}). These smaller payloads can be flown on 
either "hand-launch" balloons or hosted as piggyback experiments on other science or technology development payloads. 

Piggyback experiments have been flown for educational purposes (Sec.~\ref{sec:pipeline}), flight testing of new instrument technologies, and for stand-alone science. The BPO and CSBF continue to make great efforts to find suitable rides for piggyback experiments, often accommodating them on qualification flights of new balloons such as the 60M ZPB and SPBs. Improvements to the balloon program websites that are currently underway and continued NASA support of workshops for the balloon science community are important for helping new PIs navigate the process for finding a flight for their piggyback experiment.

\begin{wrapfigure}[17]{l}{0.4\textwidth}
\centerline{~~\includegraphics[trim=0 0 0 0, clip, width=2.7in]{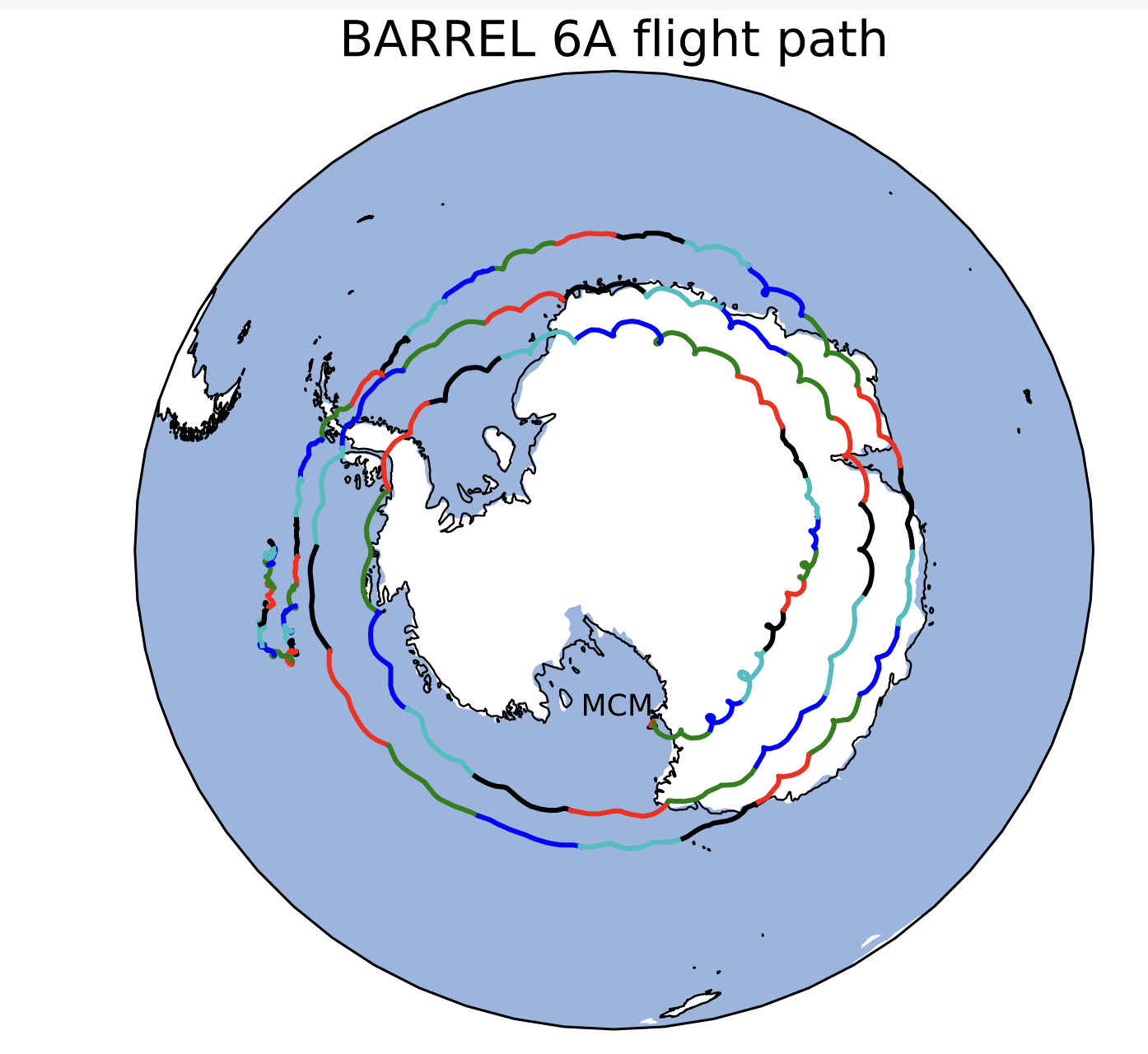}}
\caption{BARREL ground track for 2018-2019 75 day SPB flight in Antarctica. Each color represents one day.
\label{fig:small_SPB}
}
\end{wrapfigure}
Small balloons, with volume less than about 500,000 cubic feet, can be launched by hand, requiring only simple launch equipment and less restrictive launch criteria (e.g., low level winds). For suitable payloads, small balloons can provide different observational capabilities than a single large balloon, such as balloon arrays for multi-point measurements like BARREL (Sec.~\ref{sec:precipitation} or the Polar Patrol experiment carried out by Japan from Syowa station in the 90's \cite{Ejiri1993}. Since small balloons take less time to launch, it is possible to carry out multiple launches in a single day. Recently, the NASA balloon program has carried out test flights using small (600 kcu ft) superpressure balloons. During both the 2018-2019 and 2019-2020 Antarctic campaigns, a spare BARREL payload was launched on a small superpressure balloons enabling science data collection for 73 and 101 days respectively (Figure~\ref{fig:small_SPB}).

\section{Launch Sites and Facilities}
\begin{figure}[tb]
    \centering
    \includegraphics[width = 0.9\textwidth]{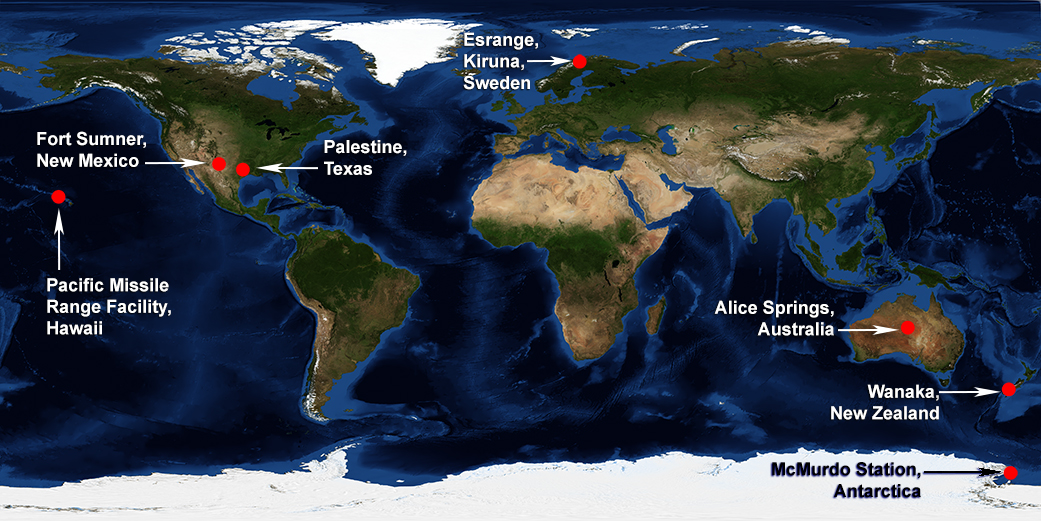}
    \caption{NASA/BPO has seven facilities world wide for balloon launches (Credits: NASA/Zell).}
    \label{fig:launch_locations}
\end{figure}

Figure~\ref{fig:launch_locations} shows a world map of currently available launch locations for
NASA scientific balloon missions. 

In the continental US, \textbf{Palestine, Texas}, the site of the Columbia
Scientific balloon Facility where many payloads are initially integrated for other sites, provides 
limited opportunities for short flights, subject to significant population-flyover constraints due to the increasing population density of the area. Palestine's facilities, while still serviceable, suffer a high volume
of use during the year, and are badly in need of maintenance and upgrades to ensure that payload
integration continues as productively as it has in the past.

\textbf{Ft. Sumner, New Mexico}, provides a more remote continental US location, with 
the best conditions occurring in the early Fall season. The launch site is at Fort Sumner Municipal Airport, 
with a large World War II
historical aircraft hangar used in part for payload staging and integration, and a NASA-fabricated high-bay building 
supplying primary payload staging, telemetry, and flight controls. 
Ft. Sumner is again primarily used for test flights,
and flight duration is typically at most around 24 hours, though longer in some cases.
One of the issues with operations in Ft. Sumner is that the WWII hangar facility is aging 
and requires significant maintenance and upgrades to avoid negative impact on the productivity 
of scientific groups that must use it for payload preparation.

\textbf{The Pacific Missile Range Facility (PMRF) at Barking Sands Kauai}, in the Hawaiian Islands, was the site
for launch within the last decade of the JPL Mars Science Laboratory’s Low Density Supersonic Decelerator (LDSD).
This mission was developed to test Mars reentry methodologies for a payload dropped from stratospheric
altitudes. LDSD utilized a large tower to enable a static launch, an unusual balloon launch configuration
where the payload was released from high up in the tower so that during its pendulum arc it would not impact
the ground, a function that is normally accomplished using a launch vehicle to minimize the pendulum motion.
PMRF can support long northern over-ocean flights at latitudes of around $25^{\circ}$, but the payload
must survive an ocean landing and recovery must be done via ship.

\textbf{ESRange (European Space Range), in Kiruna, Sweden}, provides both sounding rocket and balloon launch capabilities.  ESRange balloon launches usually produce northern polar westward trajectories, often crossing Greenland into Northern 
Canada, although in winter, flights to the east are also possible. From mid-May to mid-July, the summer polar stratospheric vortex is established, providing stable east winds, which can allow for 
full around-the-world trajectories are possible, but require agreement for overflight of Russia, which
has limited such opportunities.

\textbf{The Australian balloon launch Station (ABLS) located in Alice Springs} is 
operated by the University of New South Wales, and had complete payload support hangars and flight operations
facilities. The ABLS was used extensively for decades for balloon flights in the 
NASA balloon program up until a mishap during the abortion of a launch in 2010 led to a suspension of 
NASA balloon operations there for a year. Since then the ABLS has only had limited use in balloon flights,
but remains available as an excellent facility.

\textbf{Wanaka, New Zealand} is most recent addition to 
NASA-supported launch sites. The base is located at a local airport and has been 
in use since 2015 for the launch of Superpressure balloons in support of midlatitude missions that
provide day-night profiles. Flights traverse primarily over ocean, and are terminated in the southern
regions of South America, such as Patagonia, where population density is very low. The facilities at
Wanaka remain in active development, as this promising new location in support of the SPB program grows in scientific reach and importance.

\textbf{The Long Duration Balloon (LDB) facility near Williams Field, Antarctica} supports daytime-only flights
which can circle the Antarctic continent for several orbits over the typically two-month duration 
of the stable South Polar Vortex. Flight durations have reached 8 weeks in some cases. The facilities
at LDB are supported out of the US Antarctica Program's (USAP) McMurdo Station, Ross Island, which is where science crews are housed.
Transportation to the payload integration and launch site, which is about 7 miles out on the Ross
Ice Shelf, is done with vehicles provided by USAP. There are two large payload buildings, along with
several other supporting buildings on skis, and a third temporary payload building that can be
deployed for support of three-payload launches, which are possible in good seasons. In the past
a third payload building has been approved by NASA, but the necessary USAP support for this
building has not yet been negotiated. The establishment of this third payload building would
enable much better third-payload support during those seasons where it is practical, and implementation
of this third building should continue to be pursued.

Even now the demand for balloon flights outstrips the ability of NASA to launch them, either from mid-latitude sites (e.g. Ft. Sumner, Palestine, or Wanaka) or polar regions (e.g. Antarctica and Sweden).  At any given time, there is often a backlog of missions waiting for flight, which both delays the delivery of science to the community and increases its associated cost. The BPO addresses
these issues by searching for new viable launch sites, both within the continental US (CONUS),
and around the world. Recently, the BPO has identified the Pacific Northwest as a potential
CONUS region of interest, and evaluation of sites there may yield new CONUS options for
the program, an important and commendable effort.

%\section{CSBF and BPO Technology}
% A very brief summary of standard power and telemetry capabilities for conventional and ULDB. CIP/SIP/MIP shown in \ref{fig:bpo_capabilities}. WASP. 

\section{CSBF and BPO Engineering Support}
The Balloon Program and CSBF provide substantial support for science groups. 
The main engineering support is in the form of the Support Instrument Package (SIP), which acts as the command and data acquisition system for long duration flights. 
The SIP is a self-contained electronics package that can be easily configured to be compatible and configured for different science payloads.
For conventional flights which only require Line of Sight (LOS) telemetry, the Consolidated Instrument Package (CIP) is used. For small hand launch payloads, the Micro Instrument Package (MIP) has been developed. A summary of the balloon mission profiles and support packages is given in Table~\ref{fig:bpo_capabilities}.

\begin{table}[hbt!]
    \centering
     \caption{Capabilities and support for balloon missions. Credit: NASA/BPO~\cite{bpo_capabilities}}
    \includegraphics[width = 0.95\textwidth]{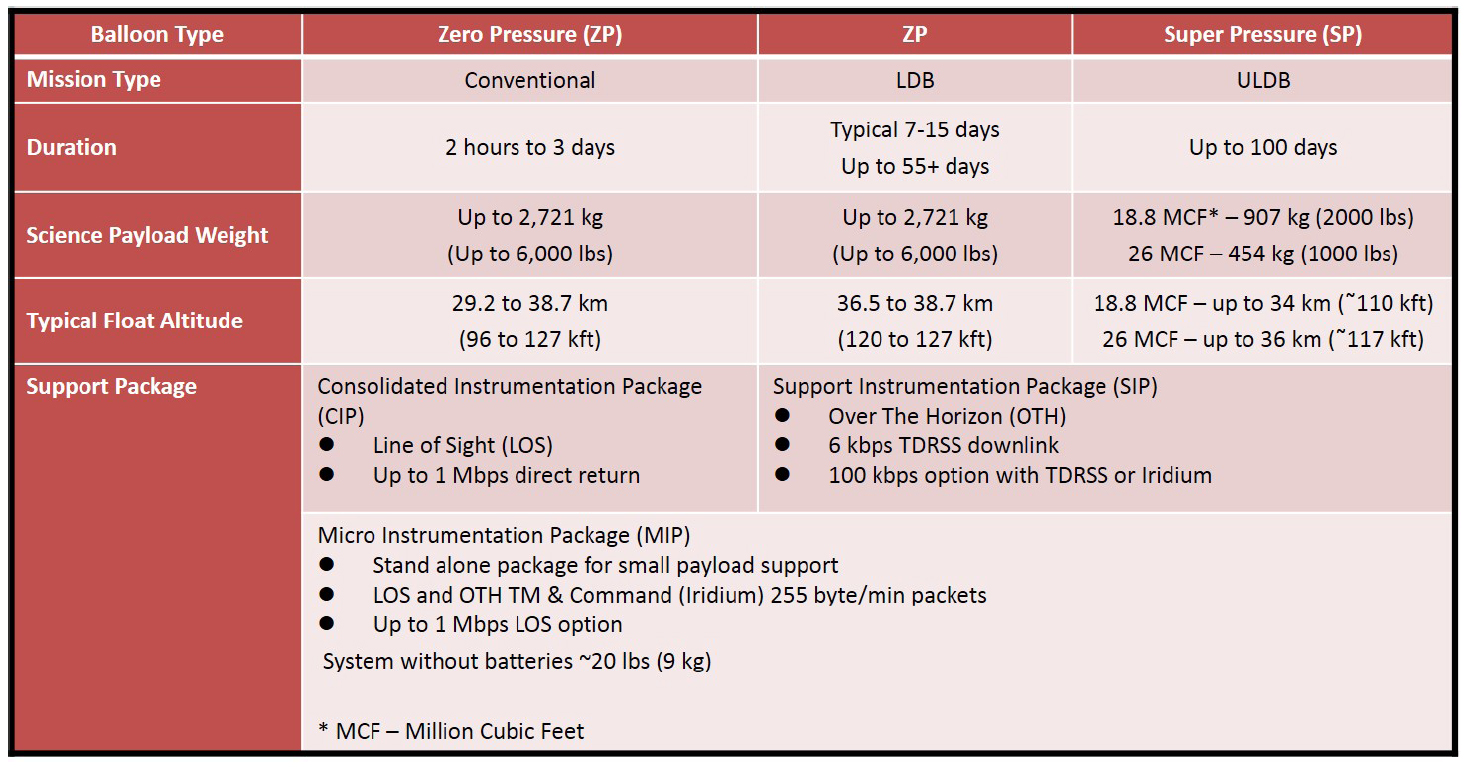}
    \label{fig:bpo_capabilities}
\end{table}

Each science group is responsible for providing their own power system; however, there is some support from CSBF. 
For conventional flights, the power system consists of relays to turn on/off the instrument, and batteries, where these batteries are commonly provided by CSBF. 
For LDB flights, the power system also includes photo-voltaic (PV) solar panels and a charge controller. 
The Balloon Program Office and CSBF provide standard designs for the power system and can consult for new science groups as they develop their own.
Additionally, CSBF provides the NASA Solar Pointing System (SPS), or rotator, which achieves rough pointing to maintain sun exposure on PV arrays in LDB flights. 

\subsection{Telemetry}
CSBF provides for beyond-the-horizon telemetry to the payload from all of the launch sites for the duration of each long duration balloon flight.  
Since the last report, BPO and CSBF have invested in increasing the reliability, flexibility and bandwidth of the telemetry.  The increasing complexity of the payloads continues to push the limit of this network, particularly with regard to the downlink of the science data.
Currently, the available telemetry includes the Iridium short burst data, which consist of 255 byte packets at a rate as high as once per minute.  The omni-directional TDRSS system can provide bandwidths below 10 kilobits per second (kbps), while the pointed high-gain TDRSS can provide 92 kbps through a standard transceiver.  The high gain system requires course pointing of the antenna, which imposes limitations on science payloads that require stable pointing with even modest azimuthal slew rates.

For the past several years, the Iridium Pilot system provided a unique and valuable capability for science groups. Pilot provides for an ethernet based connection to the payload with rates as high as 60 kbps.  While not always available, the ability to directly log into payload computers significantly mitigates a great deal of risk over the course of a long duration flight.  Unfortunately Iridium is scheduled to phase out the service in the coming few years, making replacement of that capability a high priority.

A new TDRSS transceiver supports rates from 150 kbps to 1 Mbps through an Ethernet interface using UDP packets.
Looking to the future, BPO and CSBF are developing a Ka-band transmitter that can achieve 20 Mbps through TDRSS for approximately 2 hours each day. It is important to note that payloads have already flown
which have produced more than 100~Tbyte of data, even in a short flight 
(see PMC Turbo description in section~\ref{sec:ITM}) . The inability of
any current telemetry offering for balloons to even begin to address the risk-reducing need
of a complete data downlink during flight is a current weakness within the program.

Unless the telemetry shortfall is addressed with 
order-of-magnitude improvements within the next decade, when
mission data volumes could approach the Exabyte ($10^{15}$~bytes) level, the problem
will only become more acute. Direct recovery of data drives is an advantage of the balloon program
compared to spacecraft missions, but the risk of mission failure through loss of data drives
is still significant and must be reduced.

\subsection{Wallops Arc Second Pointer}

\begin{figure}[thb]
    \centering
    \includegraphics[width=5.75in]{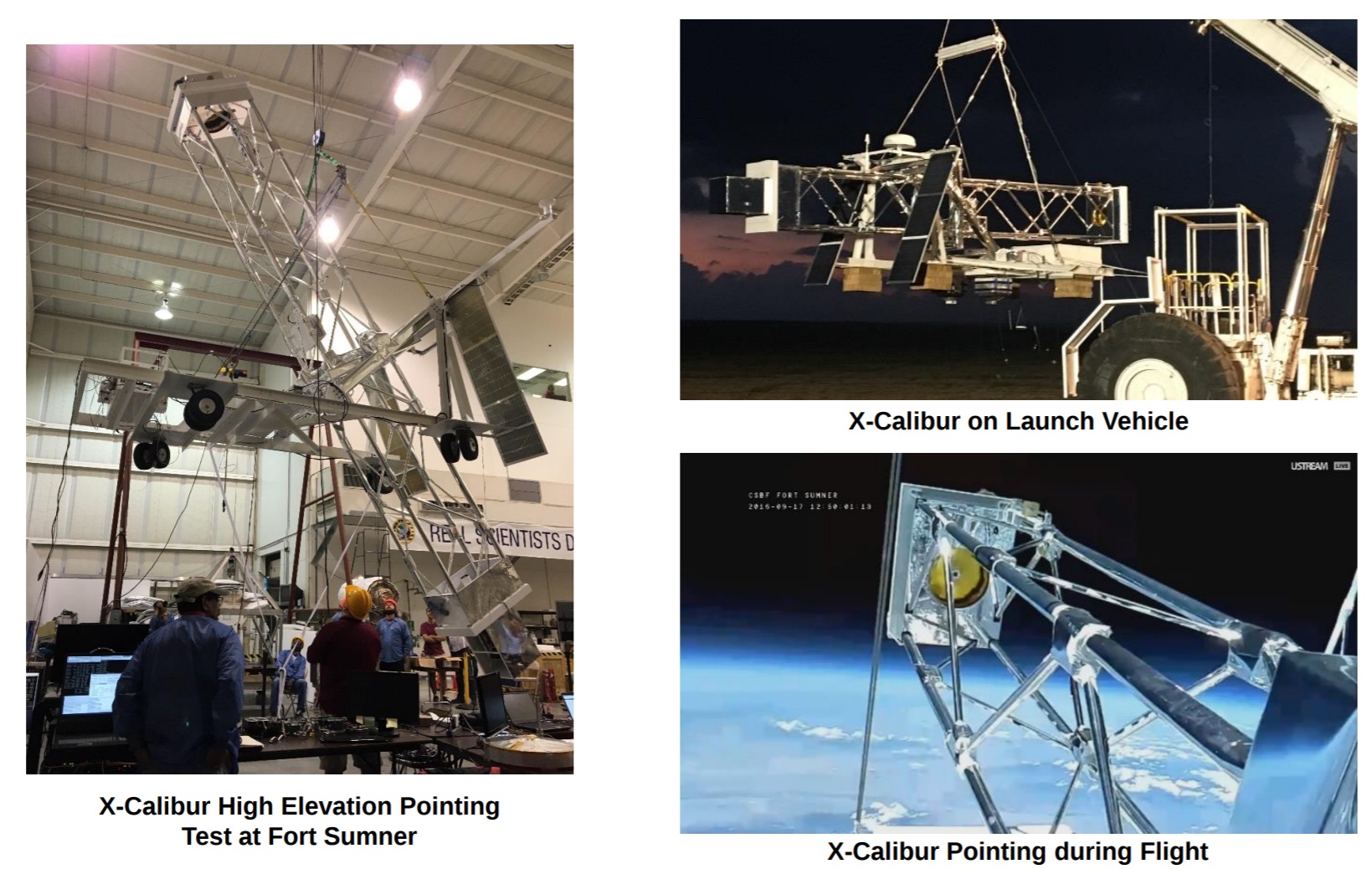}
    \caption{The Wallops Arc Second Pointer during integration, launch preparation and flight during
    the 2016 X-Calibur test flight at Ft. Sumner.}
    \label{fig:wasp}
\end{figure}

The Wallops Arc Second Pointer (WASP, Fig.~\ref{fig:wasp}) is a NASA-provided support system that can point science instruments on balloon gondolas at targets with arcsecond accuracy and stability. 
The system was developed at Wallops Flight Facility and it underwent test flights in 2011 and 2012 before supporting its first science flight in 2013. It has demonstrated RMS pointing stability to about 
0.2 arcsec.

WASP is a flexible system that can be used to support a variety of science instruments to meet accurate pointing requirements requirements and it has since flown on a number of payloads within the past few years, namely X-Calibur and Picture-C.
With major components of the WASP system being reusable, the overall costs to BPO and to the science users is reduced.

WASP represents an important development for optical and infrared astronomy within the balloon 
program. This first-generation pointing system does require a substantial footprint
on the payload, and a substantial portion of the mass budget, but it is an excellent
start for what we hope will be a multi-generational program with improved pointing stability,
lower mass and a smaller physical scale as the program continues.

 \section{Funding Opportunities}
% APRA, Pioneers, MOO.
Currently, NASA offers four different opportunities for competitive science proposals
utilizing stratospheric balloons: 
\begin{enumerate}
    \item The Astrophysics Research and Analysis (APRA) program, one of many offerings
    within the omnibus Research Opportunities in Space Science (ROSES) program,
    provides the bulk of opportunities for new balloon payloads. Funds for new suborbital
    investigations (these may include sounding rockets) in recent years range from \$6-8M
    in the first year, and the number of selections has ranged from 9-13. New APRA
    suborbital investigations can be planned for up to 5 years, but are capped
    at a total of \$10M as of the most recent offering.
    
    \item Also within ROSES is a new offering this year, the Astrophysics Pioneers.
    This program allows investigations involving both suborbital and orbital scope,
    and has a \$20M cost cap. Pioneers investigations have several additional requirements
    beyond APRA investigations: they must begin with a 6-month formulation phase,
    which is cost-capped at \$600K, and involves a gate review of a Concept Study
    Report to get to the
    implementation phase, which begins 9 months after the start. The entire project, including a test-flight,
    must be complete within 5 years.  This timeline is very constrained for a balloon payload; it remains to be seen if the schedule and funding profile are consistent with the integration and test schedule, including a test flight, for a complex long duration balloon mission. The report of the Astrophysics Advisory Committee from March 2020 states, {\it \lq\lq The short timeline may restrict the class of experiments that can be accommodated, as well as the complexity of the payloads... The committee believes that the astrophysics Pioneers program will foster investigations that are qualitatively different than the current ecosystem of large APRA investigations\rq\rq}~\cite{apac_march_2020}.
    
    \item The Heliophysics Low-Cost Access to Space (H-LCAS) seeks to both advance technologies
    and directly address key heliophysics science questions. 
    H-LCAS typically funds 6-7 new payloads, with funds of \$5-6M shared among these for the first year, and a maximum of four years duration for an investigation. As the announcement states:
    \textit{LCAS is expected to lead the way in the development of much of the instrument
concepts for future solar, heliospheric, magnetospheric, and ionosphere-thermosphere mesosphere (ITM) missions. LCAS-investigations provide unique opportunities not only
for executing intrinsically meritorious science investigations, but also for advancing the
technology readiness levels of future space flight sensors and supporting technologies
and for preparing future leaders of NASA space flight missions, such as junior
researchers and graduate students.} 
    
    \item Missions of Opportunity (MoO) provide the final option for balloon
    payload funding. These offerings are provided in parallel from
    Astrophysics, Earth Science, and Heliophysics {\it Explorers class} space mission announcements, 
    or on occasion within other disciplines, such as NASA's
    Living With A Star Heliophysics MoO, which funded the BARREL
    mission discussed above in the Heliophysics section. 
    Cost caps are typically \$35M, sometimes larger. MoO payloads can either be proposed
    in support of a space mission, or as standalone science missions.
    These missions must follow NASA flight-project safety and mission assurance
    protocols, although these can be tailored to the equivalent Class-D level 
    for balloon payloads.
    The GUSTO mission is currently in the implementation phase of an Explorers
    Mission of Opportunity. 
\end{enumerate}

It is evident from this list that while Astrophysics and Heliophysics
have consistent opportunities for new scientific payloads, and even more so with
the new Pioneers offering, the opportunities in Earth and Planetary Science are
far more sparse, since they depend on the several-year cadence of the NASA Explorers
class offerings, rather than the yearly cadence of the ROSES program.
In fact, currently we know of no opportunities to propose
balloon payloads within the NASA Planetary Science discipline,
despite the fact that we find the science potential for such payloads, as
discussed in chapter 2, to be compelling. 
We find this lack of consistent opportunities for new investigations in
these central NASA scientific disciplines to be concerning, and one of our highest priority
recommendations is that NASA seek ways to provide such opportunities within the next 
decade.

\section{Commercial launch capabilities and opportunities}
\label{sec:commercial}

As part of our charge, the PAG has considered commercial offerings for balloons and launch services that may be relevant to the science goals considered in our report. We provide here an overview of our findings; specific details of the various vendors are easily found via standard search. The PAG did not attempt to assess the cost-benefits of these offerings, since vendors require fairly detailed specifications to provide explicit cost estimates.

The PAG finds few if any commercial providers who offer services and capabilities that are comparable to those available 
within the NASA Balloon Program, with the exception of lightweight payloads, for which there are substantial offerings, including
those from the current provider of NASA balloons, Raven Aerostar. These latter offerings appear to be limited to payloads
in the 50 kg and below range however, at least for altitudes required for the large majority of the science
investigations considered here. Heavy-lift balloon offerings appear to be limited to a few vendors whose primary goal
is to provide launch services to sounding or orbital rockets, similar to the ``rockoons'' employed in the early days
of scientific ballooning. At least one vendor did offer large payload capabilities with significant dwell time at lower altitudes,
of order 24~km, and it may be possible that this vendor could provide custom services to higher altitudes, but the vehicle
in this case does appear to be a zero-pressure balloon.
 
 The PAG notes that none of these vendors specifically addressed whether their systems could be tailored to a launch in 
 Antarctic regions, so it is unclear if there are any offerings that could provide LDB-comparable services, assuming
 that the US Antarctic Program agreed to allow a commercial vendor access to launch locations.
 
 \subsection{Tethered Aerostats}
 An interesting exception in lighter-than-air technology for which there are substantial commercial offerings is that of
 tethered aerostats. To our knowledge such vehicles have not been used for NASA science investigations, but we note that 
 there was a significant study of a aerostat-based large telescope, the POlar Stratospheric Telescope (POST)
 to be deployed above the Antarctic plateau, made in the late 1990's, although the proposal never went forward. The later
 development of SOFIA also supplanted much of the capability offered by POST. Requirements for the telescope stabilization
 drove the required technology at the time. The viability of a tethered aerostat-based telescope might be of interest to
 revisit after two decades of advancement in technology.
 
There are a variety of commercial offerings for tethered aerostats with wide payload mass limits, ranging from tens of kg to several thousand kg,
certainly encompassing a range of scientific interest. Altitudes are however, limited to typical several km at most, though
there are untethered, station-keeping hybrid vehicles that may reach 10km altitudes. The science applications for these
lower altitude platforms are thus only a subset of those current in this report, but at least one published study~\cite{Besson12} and a more recent update~\cite{Cosmin20} has indicated that for ultra-high energy neutrino detection, a tethered aerostat could have order-of-magnitude higher sensitivity in a neutrino energy range that is at the heart of the cosmogenic neutrino spectrum. Clearly, significant development would be needed for such vehicles to become viable lighter-than-air science offerings, but the evidence of their potential seems rather compelling.

\hyperlink{TOC}{Return to table of contents.}

%\section{Findings and recommendations for balloon technologies.}

%\begin{itemize}

%\item
%The PAG recommends that NASA increase the capacity of launch facilities, and more importantly, the number of ground crews that can support them.  Ground crews and their associated facilities can only support a limited number of launches within a given launch window (as determined by prevailing weather conditions). This situation can result in missed launch opportunities and avoidable fatigue to all concerned. Therefore, an increase in ground crews and launch facilities should be given the highest priority. 

%\item
%The PAG finds that commercial lighter-than-air offerings, while having some overlap with NASA scientific balloons for very light payloads, still do not show any likelihood of satisfying the needs of the larger portion of the NASA scientific ballooning community.

%\item The PAG does note that commercial offerings of tethered aerostats, and possibly untethered station-keeping airships may provide scientific opportunities that could be compelling, and may also engender new ideas for investigations. Since these vehicles appear to be developing an economy of scale for both military and non-military applications, the PAG recommends that NASA study these vehicles, specifically how they might be integrated with the existing program, and the requirements for implementation within the larger program.
%\end{itemize}

\chapter{The Workforce Pipeline and Education through Scientific Ballooning}
\label{ch:education}

\begin{wrapfigure}{r}{0.35\textwidth}
\centerline{~~\includegraphics[trim=0 0 0 0, clip, width=2.15in]{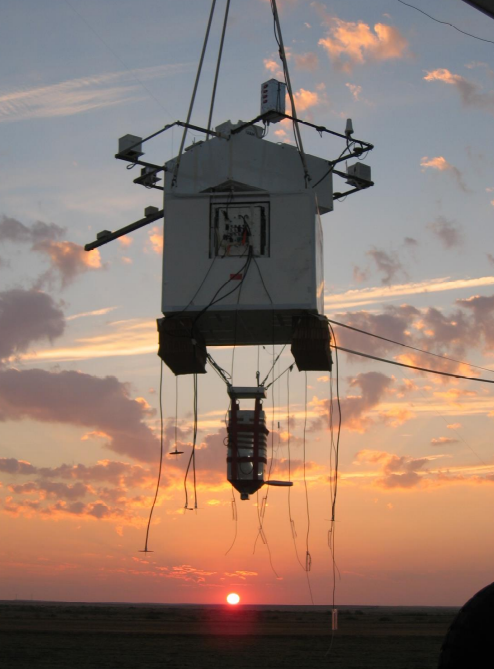}}
\caption{The High Altitude Student Platform (HASP) which carries multiple student development payloads to the edge of space was highlighted in the 2018 COSTEM report as an 
exemplary example of transdisciplinary learning.
\label{fig:epo1}
}
\end{wrapfigure}

During the first decade of this century numerous studies (e.g. \cite{EPO1,EPO2,EPO3}) raised great concern that the declining number of students entering the workforce as engineers and scientists was a major U.S. national security and economic issue. 
More recently, hearings by the U.S. Senate Commerce, Science, and Transportation Committee in 2019 continued to emphasize STEM literacy and experiential activities to engage and prepare ``a diverse workforce needed for the United States to lead and prosper in an increasingly competitive world driven by advanced technology~\cite{EPO4}.'' 
Additionally, the latest report by the Committee on STEM Education (COSTEM) of the National Science \& Technology Council published in December 2018, ``Charting A Course for Success: America’s Strategy for STEM Education,'' identified three major objectives for NASA: 1) foster STEM ecosystems that unite communities, 2) increase work-based learning and training through educator-employer partnerships, and 3) encourage transdisciplinary learning~\cite{EPO5}.

While ballooning contributes to all three objectives it is the last one where ballooning has the most impact on STEM education. 
In fact, the COSTEM report highlighted the NASA Balloon Program Office (BPO) / Louisiana Space Grant Consortium (LaSPACE) High Altitude Student Platform (HASP) balloon program as an exemplary example of transdisciplinary learning (Fig.~\ref{fig:epo1}). 
Further, both the NASA Strategic Plan 2018~\cite{EPO6} and the NASA Science Mission Directorate (SMD) Explore Science 2020-2024 Vision (Science Plan)~\cite{EPO7} have multiple objectives where ballooning can contribute.

\begin{tcolorbox}[colback=royalblue!8!white,colframe=royalblue,fonttitle=\bfseries,title=Relevance to SMD Science Goals]
Scientific ballooning contributes to multiple objectives in the NASA Science Mission Directorate by improving scientific literacy, attracting and retaining a diverse group of students to STEM fields of study, and developing a skilled, technical workforce that is essential to NASA mission success.

\vspace{0.5cm}

The \textbf{Workforce Development Pipeline} (Sec.~\ref{sec:pipeline} and \ref{sec:training}) that is intrinsic to scientific ballooning directly addresses NASA SMD goals for education and training.
As stated in the current Science Plan: 
\begin{quote}
SMD is also investing in students and early career faculty to help them grow into leaders of the future. 
SMD has been particularly focused on developing a new cadre of mission Principal Investigators through workshops and hands-on training as part of existing mission teams.
\end{quote}
Scientific ballooning uniquely enables and supports students and early-career scientists and engineers specifically through its short project timescales and hands-on training.

\vspace{0.5cm}

Improving \textbf{Diversity and Inclusion} (Sec.~\ref{sec:diversity}) in its workforce and programs is a major goal at NASA. 
The NASA SMD Science Plan states that ``… diversity is a key driver of innovation and more diverse organizations are more innovative.'' 
Further, the plan specifies Strategy 4.1 to ``Increase the diversity of thought and backgrounds represented across the entire SMD portfolio through a more inclusive environment.''

\end{tcolorbox}

\begin{wrapfigure}{r}{0.43\textwidth}
\centerline{~~\includegraphics[trim=0 0 0 0, clip, width=2.55in]{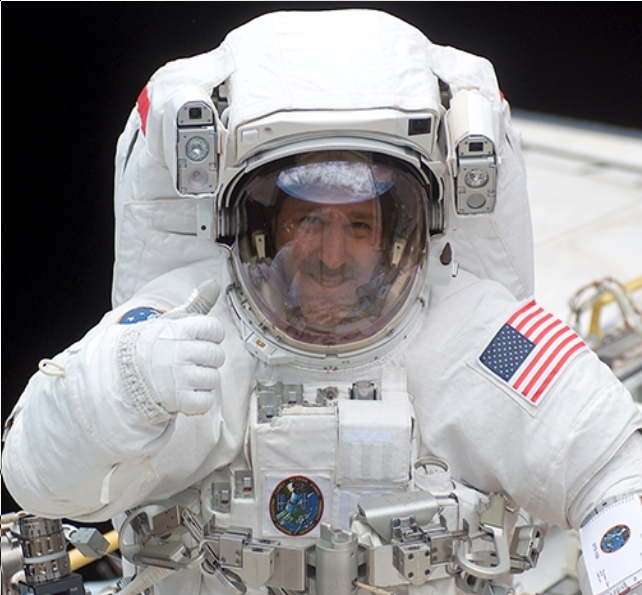}}
\caption{Dr.~John M. Grunsfeld in space shuttle Columbia's cargo bay during a 2002 spacewalk.}
\vspace{-1cm}
\label{fig:epo2}
\end{wrapfigure}
Ballooning is an ideal platform for achieving the NASA STEM engagement and technical workforce development objectives. 
Over the years many leading scientists and engineers began their careers in the balloon program. 
An outstanding example is Dr. John M. Grunsfeld (Figure~\ref{fig:epo2}) whose career was comprised of five space missions as an astronaut, including three missions to repair and upgrade the Hubble Space Telescope, as well as such distinguished appointments as NASA Chief Scientist, Deputy Director for the Space Telescope Science Institute, and Associate Administrator of the Science Mission Directorate at NASA Headquarters. 
Dr. Grunsfeld attests to the value of the balloon program to his own professional development: 
\begin{quotation}
As an undergraduate and as a graduate student I had the great fortune to perform experiments in high-energy astrophysics using high-altitude balloons as a platform for access to space. The NASA scientific ballooning program provided me with the complete and quintessential scientific experience, going from concept to hardware, observations, and scientific analysis of the results—all in the time frame of a few years. The rich environment that NASA’s sub-orbital program supports not only enables top quality science, but is also crucial as a training program.
\end{quotation}

Thus, in addition to supporting multiple significant scientific objectives, ballooning contributes to the aerospace workforce development pipeline, enhancing public awareness of NASA programs, and training the next generation of scientists and engineers.

\section{The Workforce Development Pipeline}
\label{sec:pipeline}

Ballooning is a unique NASA capability with characteristics that enable and support workforce development from middle school through early career faculty, training scientists and engineers to serve the NASA mission at the Agency and in the space science industry. 
One key characteristic of this flight platform is that many balloon payloads are recovered, enabling payload development to be iterated over multiple flights. 
This is important for student programs where the failure of a balloon payload is used as a ``teaching moment'' and the recovered payload is then improved prior to the next flight. 
Another important characteristic is the short development cycle for balloon payload development.
Typically, a professional level scientific payload can go from concept to first flight in as little as 5 to 6 years, while smaller, lighter, limited student payloads are typically developed in less than one year. 
Such time scales are consistent with the times that academic students, from middle to graduate school, can participate and gain from the full development cycle experience. 
Finally, balloon systems provide low cost access to the near space environment. 
Balloon systems that cost less than a few thousand dollars can carry light payloads to an altitude in excess of 100,000 feet, above 99\% of the Earth’s atmosphere, and where the blackness of space and curvature of the Earth can be observed. 
Such systems are affordable by high schools, colleges, and universities enabling teams of students across the country access to a space flight experience. 
The relatively low cost and limited project life cycle timeframe coupled with the high likelihood of recovering the physical payloads makes ballooning a remarkable testbed for project payloads in development to eventually fly on much more expensive platforms. 
This combination of characteristics is unique among NASA flight vehicle capabilities and, as a result, multiple balloon flight programs have been development to encourage and excite student involvement in the NASA mission at all grade levels, ultimately feeding the workforce development pipeline.

One of the enabling organizations that has played a key role in the development, and continued support, of authentic student balloon flight experiences across the United States is the \textit{National Space Grant College and Fellowship Program} (Space Grant). 
Every state, plus Puerto Rico, Washington DC, Guam, and the Virgin Islands, has a Space Grant program that, among other associated activities, provide special opportunities to support STEM workforce development in their jurisdictions. 
Dr.~Angela Des Jardins, Director of the Montana Space Grant Consortium and former Chair of the National Council of Space Grant Directors stated in her letter to the Balloon Roadmap PAG~\cite{EPO8} that the vast majority of aerospace opportunities available to rural, small-population states such as Montana are the ones created by the jurisdiction’s Space Grant program. 
Dr.~Des Jardins goes on to state that Montana Space Grant has
\begin{quotation}
…put the biggest emphasis on high altitude ballooning. The decision to emphasize ballooning reflects the value it brings: interdisciplinarity, beginning-to-end project perspective, multi-faceted teamwork, access to space-like environments, consideration for safety responsibilities, and career opportunities. Indeed, many of these values reflect the skills that NASA and other STEM employers hope their employees bring with them, but often do not.
\end{quotation}

Dr.~Des Jardins’ perspective and commitment is not unique among Space Grant jurisdictions and most support a variety of authentic flight experiences for students in their state across all grade levels. 
\textbf{Thus, taking advantage of the Space Grant network would be an effective way for the NASA Balloon Program Office to augment available resources for development of the workforce pipeline from middle and high school students through university undergraduate and graduate students.}

\begin{wrapfigure}{r}{0.45\textwidth}
\centerline{~~\includegraphics[trim=0 0 0 0, clip, width=2.75in]{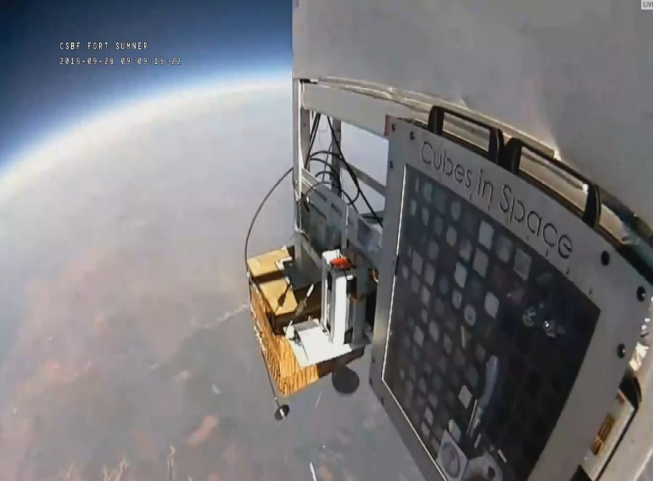}}
\caption{The Cubes in Space\texttrademark carrier on a NASA scientific balloon at float including about 100 individual experiments from grade 6 to 12 student teams.
\label{fig:epo3}
}
\end{wrapfigure}

A particular program designed for middle and high school students just entering the workforce development pipeline is the ``Cubes in Space\texttrademark'' program managed by \textbfit{idoodledu inc.}~\cite{EPO9}.
Cubes in Space\texttrademark enables students 11-18 years of age to design and propose experiments to launch into space on a sounding rocket or to the near space environment as a piggyback payload on a scientific balloon flight. 
The program takes students in this age group through a process where they design an experiment that would fit within a 4~cm cube and submit a proposal for flight. 
The proposals are reviewed and a selection is made for either a sounding rocket flight or a flight as a piggy-back payload on a NASA scientific balloon. 
A cube is shipped to the successful team and the students build, test, and install the experiment in the provided cube. 
The cube is shipped back to the vendor who then installs it in a carrier designed to easily integrate with the NASA rocket or balloon vehicle. 
Figure~\ref{fig:epo3} shows the Cubes in Space carrier loaded with about 100 student cubes at float altitude aboard a NASA scientific balloon flight in 2016. 
Following flight the cube is returned to the student team who then analyze the flight results and submit a final report. 
As stated by Amber Agee-DeHart, the Cubes in Space Program Director, the program mission is to ``Help students embrace their curiosity, develop logical and methodical thought, engage in creative problem solving, and experience the joy of learning something new that ignites their imaginations and fuels their determination for success~\cite{EPO9}.'' 
As of summer 2020, the Cubes in Space carrier has flown four times as a piggy-back payload on NASA scientific balloon vehicles.

\begin{wrapfigure}[24]{l}{0.35\textwidth}
\centerline{~~\includegraphics[trim=0 0 0 0, clip, width=2.in]{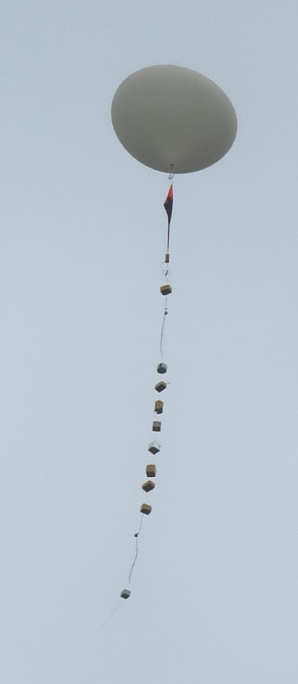}}
\caption{Latex sounding balloon launch with multiple student payloads.}
\label{fig:epo4}
\end{wrapfigure}
A common method used by many high school and university groups to reach high altitudes is the lightweight 2,000-to-3,000 gram latex sounding or ``weather'' balloon.
These types of vehicles are easy to configure to conform to FAA regulations so that no NOTAM or flight waiver is required. 
In such a configuration, the vehicle can carry up to 12 pounds total suspended weight to an altitude in excess of 100,000 feet before the flight is terminated and the suspended payloads are parachuted back to the surface of Earth for recovery. 
Figure~\ref{fig:epo4} shows an example of such a vehicle shortly after launch. 
These hand-launched vehicles use compact GPS and low wattage radio transmitters, many times using the APRS protocol~\cite{EPO10}, to provide real-time latitude, longitude, and altitude information directly to chase vehicles on the ground or to APRS network grounds stations that repeat location information to mapping websites.

\begin{figure}[t]
\centering
\includegraphics[width = 5in]{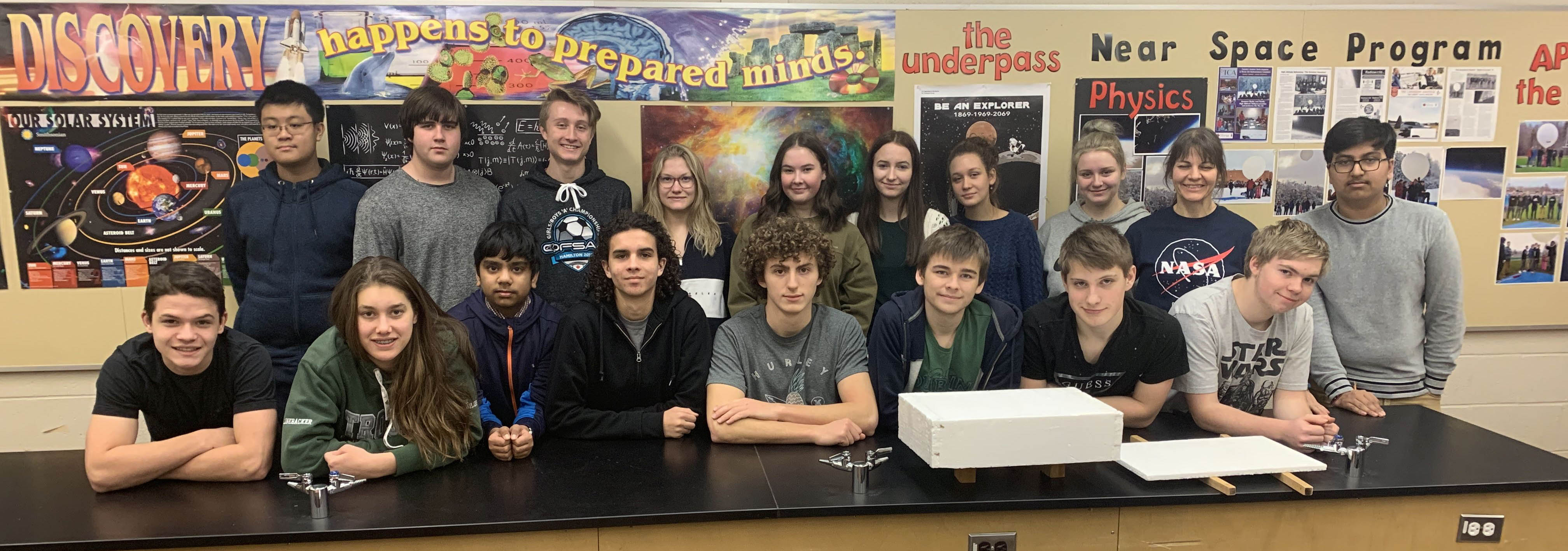}
\caption{Grade 11 students at the West Ferris Secondary School in Ontario, Canada have built a hand-launchable payload for a citizen science project that will perform atmospheric measurements of CO$_{2}$ with simultaneous measurements from NASA's OCO-2 and and OCO-3 instruments.}
\label{fig:westferris}
\end{figure}

Another reason for the popularity of these kinds of high altitude ballooning vehicles is their relative low cost. 
Typically only a few thousand dollars of investment is needed to support all the equipment and materials needed for ground and flight operations. 
This low cost is within reach of many high schools, colleges, and universities enabling authentic flight payload development experiences for students in this age group. 
One exemplar model a high altitude balloon program in a high school is the West Ferris Near Space Program, see Fig~\ref{fig:westferris}, where students have designed and built a payload to measure CO$_{2}$ in the atmosphere to test models used by NASA satellites~\cite{westferris}.
In a response to the Balloon PAG, Grade 12 teacher Kelly Shulman clearly stated the benefits of high altitude ballooning for student education:
\begin{quotation}
The balloon platform is well-suited to atmospheric experiments at the high school level. 
It is relatively inexpensive when set against its incredible impact on students involved in the program. 
As a learning tool it is truly authentic problem-solving that promotes academic risk-taking. 
This is an awesome project that not only advances the technical skills of the students, but brings context and meaning to the work they are doing that extends their view far beyond the classroom to the existential issues that will be the challenge for their generation. 
\end{quotation}

Many such programs, some with the support of the jurisdiction Space Grant, have been developed across the country and implemented as part of a STEM curriculum or extracurricular activity. 
In several cases, these programs are associated with course materials such as lectures and activities building basic skills in electronics, programming, and project management. 
Some of these programs have even been adapted to support K-12 teacher in-service workshops. 
This is another area where the \textbf{NASA Balloon Program Office could partner with the Space Grant program to support continued development of lightweight hand-launched balloon vehicles, and expand entry-level student and K-12 teacher training programs.} 

For more advanced students, the ballooning program offers a number of options including the NASA Undergraduate Student Instrument Project (USIP), piggy-back payloads, and the High Altitude Student Platform (HASP). 
The USIP program, which has its mission managed conducted by Wallops Flight Facility, is limited to undergraduate students and is intended to ``provide a hands-on flight project experience to enhance the science, technical, leadership, and project skills'' and ``to fly a science and/or technology investigation relevant to NASA strategic goals and objectives on a suborbital-class platform~\cite{EPO11}.'' 
To date there have been two cycles of the USIP program: one in 2013 that involved 10 teams and one in 2015 that supported 47 teams. 
The significant increase in the number of teams for the 2015 cycle was due to an augmentation of available SMD funding by the Space Grant program. 
Of these 57 teams, 8 were flown as piggy-back payloads on a scientific balloon, and 10 incorporated small, hand-launched balloon vehicles such as previously discussed (Fig.~\ref{fig:epo4}). 
The remainder projects were CubeSats (23), sounding rockets (6), Parabolic / small reusable launch vehicles (7), and Aerial vehicles (3)~\cite{EPO17}. 
%During the 2015 USIP cycle over 1,200 undergraduate students, 94 graduate students, and 137 faculty were involved. 
%As of April 2020, almost all of the USIP balloon teams have flown their payload and completed their final reports.

Cubes in Space and some of the USIP payloads were ``piggy-backs'' on regular and test scientific balloon flights. 
Rides for such payloads are negotiated with the primary payload PI and might or might not be supplied with payload power, telemetry bandwidth, or other resources. 
These kinds of payloads are also referred to as “Mission of Opportunity” payloads and a Balloon Flight Support Application needs to be submitted to the Columbia Scientific Balloon Facility (CSBF) at least six months prior to integration with the primary payload. 
Since 2015, the NASA 
\begin{wrapfigure}[13]{l}{0.45\textwidth}
\centerline{~~\includegraphics[trim=0 0 0 0, clip, width=2.65in]{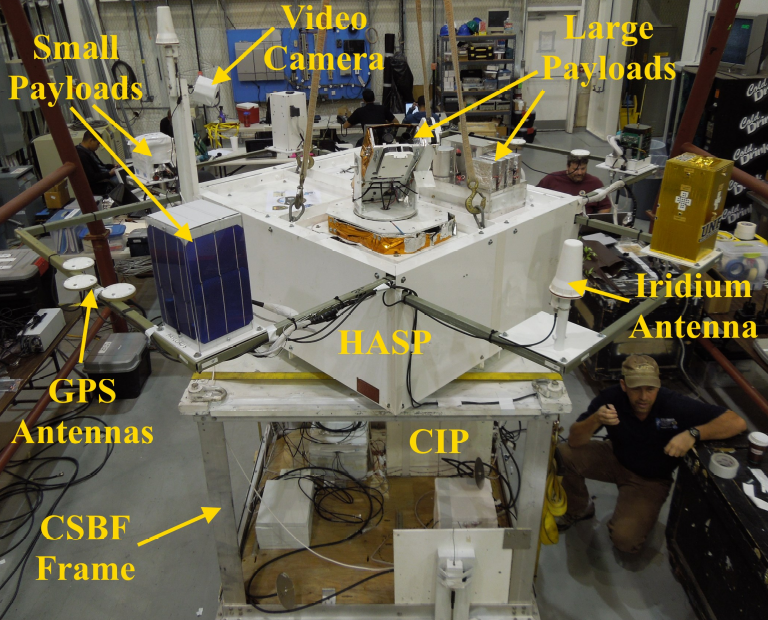}}
\caption{The High Altitude Student Platform can carry up to 12 advanced student payloads.
\label{fig:epo5}
}
\end{wrapfigure}
Balloon Program Office has flown 57 ``piggy-back'' payloads. 
Of these, 18 were designated as ``student'' payloads and most of those were from the USIP program plus four flights of the Cubes in Space carrier. 

In 2005-2006 the NASA Balloon Program Office and the Louisiana Space Grant Consortium (LaSPACE) developed the High Altitude Student Platform (HASP) to provide a dedicated platform for engaging advanced high school and university level students with NASA related science and technology, as well as to provide an authentic experience, with associated skill development, that models a spaceflight mission from concept to flight result analysis and reporting~\cite{EPO12}.
HASP, shown in Fig.~\ref{fig:epo5}, supports up to 12 student payloads simultaneous with power, downlinked data, and uplinked commanding. 
The HASP program typically opens in October of a given year. 
Student teams are required to submit a flight application similar to a preliminary design review document in December. 
All applications are reviewed by BPO, LaSPACE, and CSBF to determine selections for flight slots, which are announced in January. 
During payload development student teams submit monthly reports and a series of design documents to track their progress. 
The completed student payloads are integrated with HASP and thermal vacuum tested during July in preparation for flight in the beginning of September. 
A final project report is due within three months following flight.  
Since 2006, HASP has had 14 flights for an accumulated time at float of close to 190 hours, accepted more than 160 student teams into the program, involved more than 1,400 student participants, and successfully flown more than 73\% of the accepted payloads.

\begin{wrapfigure}[16]{l}{0.45\textwidth}
\centering
\includegraphics[width=2.8in]{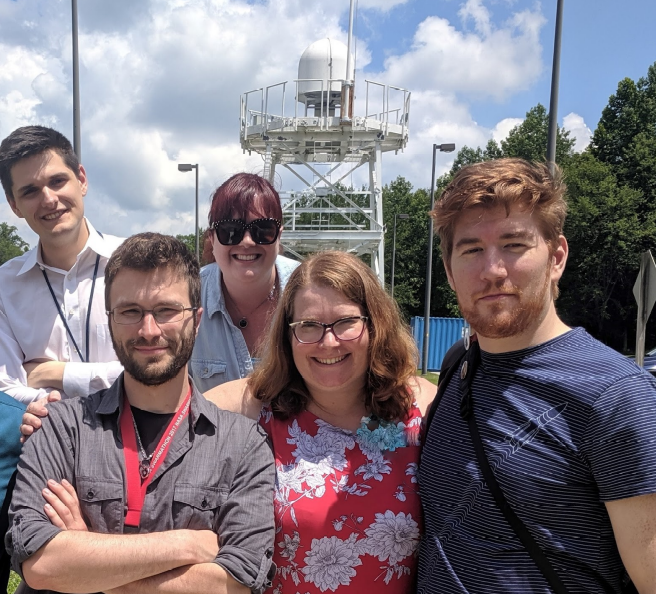}
\caption{Former HASP participants now engaged in careers supporting space exploration and applications. Back row: Ryan Theurer, Meredith Murray; Front row: Daniel Koris, Julie Hoover, Jimmy Acevedo.
\label{fig:epo6}
}
\end{wrapfigure}

\textbf{The workforce development balloon programs described here have been shown to be successful in engaging students across many grade levels in authentic flight projects that provide experience and skills directly applicable to NASA workforce needs. }
An example of this success is shown in Fig.~\ref{fig:epo6} which depicts a group of former HASP participants (2017, 2018, and 2019 flights) who are actively advancing their space related careers.
Ryan Theurer is currently working as a Satellite Data Visualization Programmer within NOAA (National Oceanic and Atmospheric Administration) as part of NOAA's STAR (Satellite Applications and Research) division under the NESDIS (National Environmental Satellite Data and Information Service) branch in partnership with NASA. 
Meredith Murray is currently a summer intern at NASA JSC working on the Micro-g NExT project in the Office of STEM Engagement. 
She graduated from the University of North Carolina at Charlotte in May 2020 with a degree in Communications Studies. 
Daniel Koris continues to work year-round as a sustained intern with NASA Goddard’s Satellite Servicing Projects Division (SSPD). 
He is currently studying computer science at the University of Colorado at Boulder. 
Julie Hoover is the Program Coordinator for the NASA Goddard Office of the Chief Knowledge Officer (OCKO), the center point of
contact for the Academy of Program/Project \& Engineering Leadership (APPEL) knowledge services and the GSFC representative for Apollo resources on the LaRC Tranquility Basement project. 
Jimmy Acevedo is the Senior Educational Outreach and STEM Engagement Specialist for NASA Goddard’s Exploration and Space Communications Projects (ESC). 
He manages the intern program for the Space Communications and Navigation (SCaN) Internship Project, a program which hosts approximately 65 students at Goddard each year.

\textbf{The NASA Balloon Program Office should continue to support the programs described here, or similar programs, to expand the opportunities available to students across the workforce development pipeline.}

\section{Enhance the Public Awareness of Ballooning}

\begin{wrapfigure}[20]{r}{0.36\textwidth}
\centerline{~~\includegraphics[trim=0 0 0 0, clip, width=2.25in]{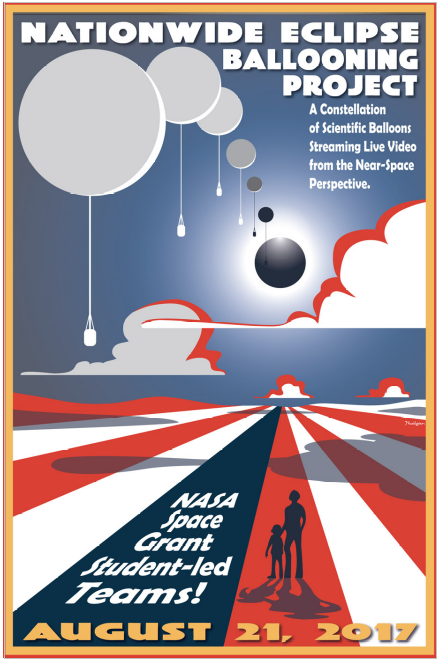}}
\caption{The Space Grant 2017 solar eclipse balloon program involved close to 50 student teams launching from Oregon to South Carolina.
\label{fig:epo7}
}
\end{wrapfigure}
During particular astronomical events ballooning offers an opportunity to engage the public with the NASA mission, space exploration, and science in general. 
One such event was the August 21, 2017 North American total solar eclipse. 
During this event the Moon’s shadow passed across the continental United States from Oregon to South Carolina with the eclipse maximum in western Kentucky. 
To take advantage of the eclipse as a student engagement and public outreach event, the Montana Space Grant Consortium organized a Nationwide Eclipse Ballooning Project (Fig.~\ref{fig:epo7}) to capture and stream real-time video of the Moon’s shadow during an eclipse from the edge of space, and to establish a nationwide structure of organizations including the BPO, NASA centers, businesses, and Space Grant jurisdictions to support the project~\cite{EPO13}.

During the years leading up to the eclipse a standard set of sounding balloon vehicle components were developed. 
These components included a lightweight flight HD video transmission system, an over-the-horizon Iridium tracking beacon, and an associated ground receiving station. 
Prior to the eclipse in 2017 more than 50 student teams were trained during a series of workshops in the use of this equipment and several practice events were conducted.

\begin{wrapfigure}[13]{l}{0.5\textwidth}
\centerline{~~\includegraphics[trim=0 0 0 0, clip, width=2.85in]{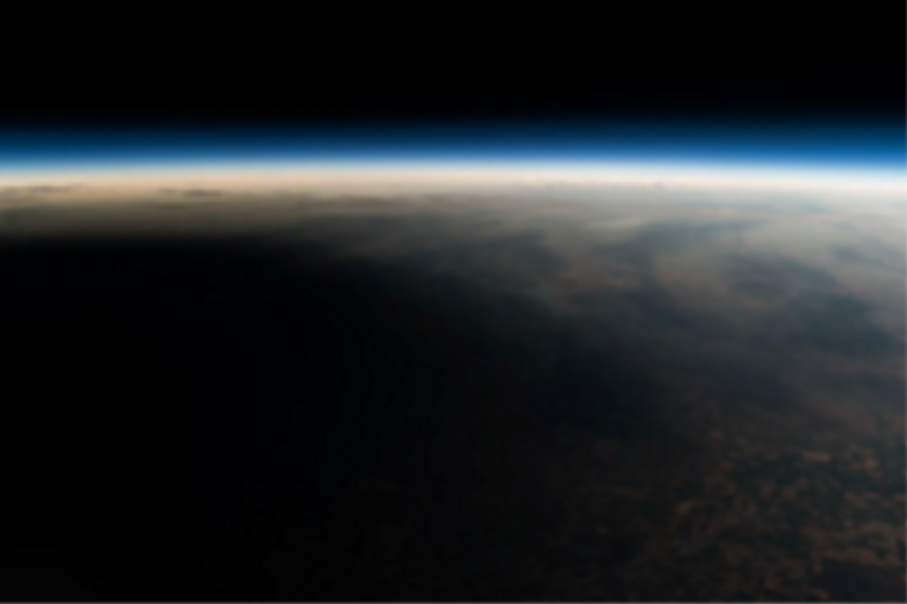}}
\caption{Image of the moon’s shadow on the Earth during the solar eclipse of August 21, 2017 by the Montana State Univ. Student Team balloon payload.
\label{fig:epo8}
}
\end{wrapfigure}
On the day of the eclipse 54 teams including 919 students from 32 states representing 75 institutions participated in the balloon launch~\cite{EPO13}. 
Launch locations and times were coordinated along the full length of the path of totality from Oregon to South Carolina so that the balloon payloads were carried to an altitude up to ~100,000 feet at the correct time to downlink HD video of the moon’s shadow on Earth (Fig.~\ref{fig:epo8}) that was live streamed over the internet. 
Some intrepid teams in the coastal states even enlisted the services of the U.S. Coast Guard to launch their balloon at sea to position the payloads at the extreme ends of the sun’s path of totality. 
The effort was an amazing success and hundreds of national and local balloon project specific stories were distributed prior to the eclipse and reached 27 million people. 
The balloon event was featured as part of NASA’s eclipse coverage, which reached 600 million people on TV platforms. 
These kinds of projects are uniquely positioned to engage the public, while simultaneously teaching students, exciting their imaginations, and encouraging them to learn more about NASA ballooning in particular and the NASA mission overall. 

\begin{figure}[htb]
\centering
\includegraphics[width=5in]{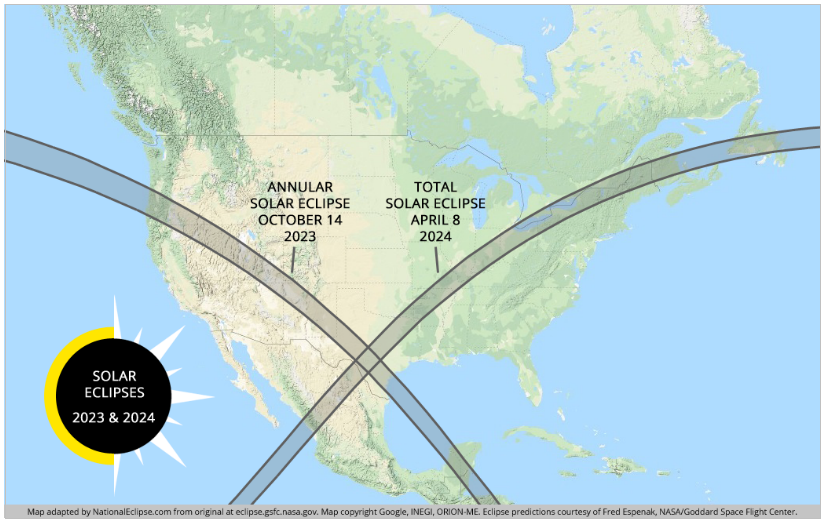}
\caption{The next Nationwide Solar Eclipse balloon project is targeting the annular eclipse of 2023 (left) as well as the total eclipse of 2024 (right).}
\label{fig:epo9}
\end{figure}

The resounding success of the 2017 national eclipse balloon project is being leveraged for a future ballooning project of national scope. 
The next eclipse cycle that will be accessible in the United States is during 2023-2024 (Fig.~\ref{fig:epo9}). 
Planning is already underway for an even larger Nationwide Eclipse Balloon project that will build upon the first mission’s lessons learned and improve the previous balloon platform~\cite{EPO14}. The anticipated project will engage both the science of eclipse generated atmospheric gravity waves, and enable an ``engineering'' payload that will include a next generation of the popular real-time HD video streaming capability. 
It is anticipated that the project will include 30 atmospheric science teams and 40 engineering teams with each team comprising of 6 to 30 students with up to 6 mentors per team. 
The time leading up to the eclipse will be devoted to developing the payload components, and recruiting/training the student teams. 
A trial launch will be scheduled for the annual solar eclipse on October 14, 2023 in preparation for the total solar eclipse of April 8, 2024.

Note that in Fig.~\ref{fig:epo9} both eclipse paths intersect in southern central Texas and the 2024 total eclipse path runs just west of the NASA Columbia Scientific Balloon Facility as noted by Dr.~Des Jardins in her letter to the Balloon Roadmap PAG~\cite{EPO8}:
\begin{quotation}
The path of totality of this eclipse passes very near the Columbia Scientific Balloon Facility in Palestine, TX. It is very exciting to think about a large NASA balloon flight from the Facility that could possibly intersect the path of totality. If such a flight were possible, many scientists would compete to have a payload on the balloon. Saving spaces for student-led academic payloads, however, would be a fantastic opportunity.
\end{quotation}

\textbf{The NASA Balloon Program Office should take advantage of the upcoming 2023-2024 solar eclipse cycle to publicly illustrate the capability and advantages of balloon platforms.}
In particular, NASA BPO should work with the Nationwide 2023-2024 Eclipse Balloon Project and provide a platform to carry multiple student payloads across both the 2023 and 2024 eclipse paths in southern central Texas.

\section{Training Next Generation Scientists \& Engineers}
\label{sec:training}

Balloon platforms are a critical vehicle for supporting PI-class academic research due to the A) short time scale between conception and flight results, B) relatively low cost, and C) recoverability. 
This combination of characteristics enables universities to support research programs that develop near-space flight hardware while engaging future scientists and engineers. 
As most balloon experiments are recovered and reflown the acceptable risk on individual balloon payloads should be high relative to spacecraft. 
By accepting risk on balloon payloads, quality assurance requirements are reduced which, in turn, will decrease development time and overall cost. 
As spacecraft instrumentation are unrecoverable, have high launch costs, and are generally managed for low risk, student access to the spacecraft systems is generally highly restricted resulting in limited training opportunities. 
However, the high acceptable risk aspect of balloon payloads allows post-docs, graduate students and undergraduate students to have direct interaction with the payload hardware and software and develop skills that will be applicable throughout their careers.

\begin{figure}[tb]
\begin{center}
\centerline{\includegraphics[width=6.5in]{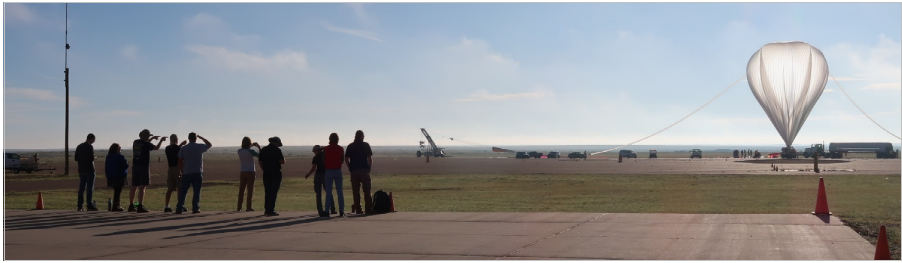}}
\caption{University students watching the launch of their experiments carried by the High Altitude Student Platform (HASP) in September 2018.
\label{fig:epo10}}
\end{center}
\end{figure}

Workforce development platforms such as piggy-back payloads and systems like HASP provide advanced degree students and early career scientists and engineers with the training and experience necessary for them to take their next step (Fig.~\ref{fig:epo10}). 
Of the 57 piggy-back payloads flown by the BPO between 2015 and 2020 close to 70\% of the payloads were focused on testing preliminary scientific instruments or engineering prototypes. 
Some of the devices tested include an autonomous paragliding lander, a neutron spectrometer, infrasound instruments, inflatable spherical reflectors, geomagnetic investigations, and air sampling instruments.  
For the HASP program specifically, flight payload outcomes include 16 peer-reviewed journal publications and 7 advanced degree theses or dissertations.

Prototype test flights on balloon platforms provide the data necessary to improve the Technical Readiness Level (TRL) of the device or experiment methodology, while at the same time early career scientists and engineers gain a unique hands-on training opportunity with ``space flight missions.'' 
In their white paper to the Balloon Roadmap PAG~\cite{EPO15}, Julie Hoover Program Coordinator for the NASA Goddard Office of the Chief Knowledge Officer, and Jimmy Acevedo Senior Educational Outreach and STEM Engagement Specialist for NASA Goddard’s Exploration and Space Communications Projects, emphasize the importance of balloon programs in addressing NASA’s future workforce needs:
\begin{quotation}
High altitude ballooning projects expose the students to a variety of nuanced and cutting-edge technologies and bridge the gaps between science, computer science, and engineering. 
…Students learn from combining their skill sets in their unique fields to reach a common goal. 
This benefits NASA and the greater engineering community by teaching the students the listening and communication skills needed to achieve a multifaceted endeavor and increases the involvement of minority students in STEM education. 
Students become even more versed in the culture of engineering work and are better prepared to merge into the workforce after graduation. 
Structured ballooning projects create a realistic NASA trial by fire production environment that will prepare students for the rigor, adventure, bureaucracy and stress of a career in engineering.
\end{quotation}

Recently, however, there is growing concern that safety and quality assurance requirements are increasing in the direction toward spacecraft mission requirements. 
\textbf{If this trend is allowed to continue it is likely that the project time of development and cost will increase while access to the instrument by untrained students or early career faculty will decrease.}
These trends would have an adverse impact on training the next generation of scientist and engineers. 
Safety and performance standards should be developed specifically for balloon projects, acknowledging that such projects are intrinsically high risk and, therefore, are reviewed at a level significantly below spacecraft standards. 
Balloon project standards should be measured relevant to the project, venue, and audience, while avoiding duplication of effort. 
The standards should also be consistent with the need to provide university students, as well as inexperienced faculty, the flexibility and opportunity to have direct involvement and even more direct responsibility for the success of balloon payloads. 
Finally, there should be investment in relevant, simple, concise, and straight forward best practices and safety training materials that will enable students as well as novice ballooning scientists and engineers to learn and retain critical information concerning ballooning.

\subsection{Are we training a diverse set of scientists and engineers?}
\label{sec:diversity}

Improving diversity and inclusion in its workforce and programs is a major goal at NASA and 
%In fact, the NASA SMD Science Plan~\cite{EPO7} directly states that ``… diversity is a key driver of innovation and more diverse organizations are more innovative.'' 
%Further, the plan specifies Strategy 4.1 to ``Increase the diversity of thought and backgrounds represented across the entire SMD portfolio through a more inclusive environment.'' 
ballooning plays an important role in developing and supporting these objectives and strategies by providing diversity in the gender/race/background of project members, science investigation topics, broad institution participation, and trained skill set.

Across participants in scientific balloon payloads and student programs, diversity in gender and traditionally under-represented members of STEM fields is encouraging but needs improvement.
%For participants in scientific balloon payloads and student programs diversity in gender and members of groups traditionally under-represented in STEM is encouraging but needs improvement. 
Astrophysics payloads, with 1/3 of such payloads reporting, indicate that the percentage of female graduate student involvement is 40 to 50\%~\cite{EPO16}. 
%For all USIP projects 79\% of participants were male and 21\% were female. 
%However, several females served as project managers for their teams~\cite{EPO17}. 
From 2012 to 2019 the HASP program includes demographic information on 969 student participants~\cite{EPO18}.
Of these students, 746 were male (77\%) and 223 (23\%) were female. 
Further, 141 (15\%) were ethnic Hispanic, 48 (5\%) were from minority groups, and 4 (0.4\%) were disabled. 
We also note that the Chief of the NASA Balloon Program Office is female but there are few female engineers or technicians at BPO and at the Columbia Scientific Balloon Facility. 
\textbf{These data indicate that gender and minority diversity in NASA balloon programs is encouraging, but still needs improvement.}

\vspace{0.2cm}

\begin{table}[h!]
\centering
\setlength{\tabcolsep}{3pt}
\begin{tabular}{|m{1.4cm}|m{1.5cm}|m{0.9cm}|m{1.2cm}|m{1.6cm}|m{0.75cm}|m{0.95cm}|m{1.2cm}|m{0.75cm}|m{1.5cm}|m{1.9cm}|}
\hline
\multirow{3}{1.4cm}{Type} & \multicolumn{1}{c|}{ARMD} & \multicolumn{3}{c|}{HEOMD} & \multicolumn{4}{c|}{SMD} & \multicolumn{1}{c|}{STMD} & \multicolumn{1}{c|}{BPO} \\\cline{2-11}
 &  \scriptsize Aeronautics & \scriptsize Space Life & \scriptsize{Physical Science} & \scriptsize{Engineering} & \scriptsize{Astro} & \scriptsize{Earth Science} & \scriptsize{Planetary Science} & \scriptsize{Helio} & \scriptsize{Technology} & \scriptsize{Programmatics \& Testing} \\
     \hline \hline
     \footnotesize{Science Payloads} & 0 & 0 & 0 & 0 & 51 & 5 & 3 & 29 & 0 & 45 \\
     \hline
     \footnotesize{HASP} & 6 & 15 & 6 & 23 & 39 & 57 & 14 & 8 & 34 & 0 \\
     \hline
\end{tabular}
\caption{General classification of science and HASP balloon payloads from 2006 through 2019. Note that some HASP payloads contain instruments that are relevant to multiple categories.}
\label{tb:payloadclass}
\end{table}

\begin{figure}[bt]
\centering
\includegraphics[width=5in]{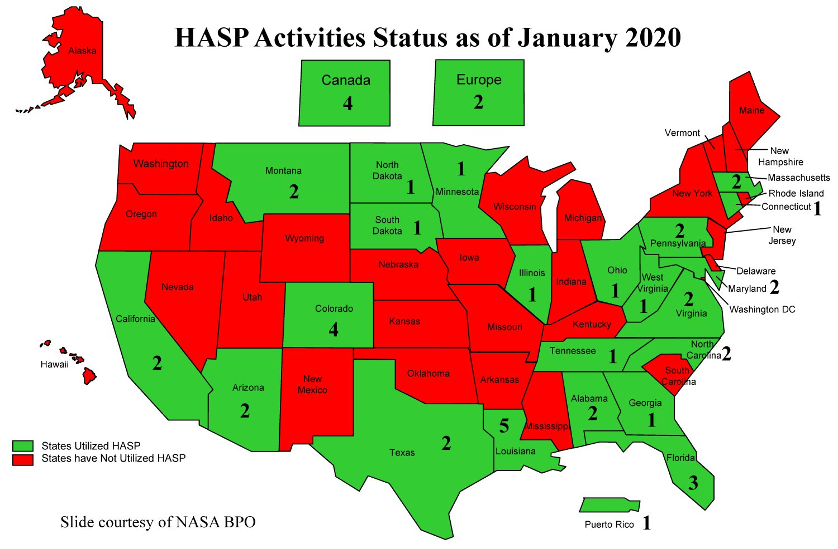}
\caption{States that have utilized HASP as a student workforce development project are indicated by the green shading.  
In addition, the number of institutions within the state that have contributed student teams is indicated by the bold number in or near the state.
\label{fig:epo11}
}
\end{figure}

In addition to the diversity of students, scientists, and engineers, balloon payload investigations also cover a diverse set of science topics that touch on all NASA mission directorates. 
Table~\ref{tb:payloadclass} shows the classification of topics across the NASA Mission Directorates for science payloads and HASP student payloads from 2006 through 2019 ~\cite{EPO16, EPO18}.
During this time period there were 88 science payloads associated with SMD topics and, in particular, there were 51 astrophysics payloads divided as 13 CMB IR/Sub mm, 8 High Energy, 21 Particle Astrophysics, and 9 UV-VIS experiment. 
Also there were 45 payloads associated with programmatic or system tests. 
The HASP student payloads touched on 6 ARMD, 44 HEOMD, 118 SMD, and 34 STMD topics. 
Further, participation in balloon payloads involves a diverse collection of institutions.  
For astrophysics science payloads, over the time period shown in Table~\ref{tb:payloadclass}, there were 22 different PI-institutions, 5 payloads from NASA centers, and 4 minority serving institutions. 
Student teams for HASP come from across the United States, Canada, and Europe (Fig.~\ref{fig:epo11}) and, in particular, there are 48 independent institutions across the world that have used HASP as part of their student training program. 
These institutions are located in 22 U.S. states plus Puerto Rico, Canada, Belgium, and the United Kingdom. 
A number of institutions have participated over multiple years with unique student teams and payloads.

Finally, scientific ballooning projects are inherently transdisciplinary in nature. 
The science mission of the student payload may be grounded in any number of disciplines including physics, biology, atmospheric science, and so on, while the technical requirements of the payload require basic knowledge in programming and electrical and mechanical engineering. 
Students also receive crucial real-world training in team building, technical communications, and project management. 
Typically students are involved in balloon payloads from conception through data analysis, and these students quickly learn that team work is critical to the success of a complex technical project.

In summary, the high altitude balloon platform provides not only excellent science results but delivers outcomes that go beyond science. 
High altitude balloon flights, particularly those associated with spectacular astronomical events, excite the imagination of the general public and are an excellent vehicle to help illustrate the NASA mission. 
High altitude balloon flights are accessible by students from middle school through graduate school and provide an inexpensive mechanism to train the next generation of scientist and engineers. 
The experiences and training these students receive in technical skills, communication, and team work are applicable and valuable in industry and academia. 
Finally, payloads associated with high altitude balloon flights are recoverable enabling innovative technology to be tested and improved with each subsequent flight. 
\textbf{In short, the high altitude balloon platform is a unique and necessary part of NASA which should be supported and expanded.}

\hyperlink{TOC}{Return to table of contents.}

%\section{Findings and Recommendations for Education \& Outreach}

%The PAG finds that the NASA Balloon Program Office has developed and supported a broad range of capabilities and platforms that engage students across the workforce development pipeline as well as providing relevant payload development experiences for early career scientists and engineers. Support for these capabilities and platforms should be expanded to provide increased workforce development opportunities in the future.

%The PAG recommends that NASA should engage with the National Space Grant College and Fellowship program to continue and expand strong support for student training ballooning programs that support the workforce development pipeline at all levels including K-12, university students, and in-service teachers.

%The PAG recommends that NASA engage with stakeholders interested in the April 8, 2024 North American total solar eclipse as early as possible to assess scientific, workforce development, and public engagement projects, as well as payload weight classes including heavy payloads, and potential launch sites.

%The PAG recommends that balloon project safety and performance standards be reviewed at a level significantly below spacecraft standards, be relevant to the project, avoid duplication of effort, be consistent with the need to provide novice personnel the flexibility to have direct involvement in the payload success, and be associated with relevant, simple, concise, and straight forward best practices and safety training materials.

\chapter{Findings \& Recommendations}
\label{ch:recommendations}

The Balloon Program analysis group provides the following findings and recommendations to NASA for the next decade of the scientific balloon program. 
We provide these in four sections below, with entries in each section enumerated in priority order.
% Figure~\ref{fig:balloonstm} provides a summary of how these recommendations connect to the different science drivers described in previous chapters.

% \begin{figure}[htb!]
%     \centering
%     \includegraphics[width = \textwidth]{figures/BalloonSTM.png}
%     \caption{The recommendations prioritized here are driven by the key science that is discussed in Chapter~\ref{ch:science}, where some examples are shown here. }
%     \label{fig:balloonstm}
% \end{figure}

%\section{General Recommendations}

%\noindent {\large \textcolor{royalblue}{\textit{Balloon Capabilities}}}

\section{Balloon Capabilities.}

% 60 M cu ft/ higher altitude/larger mass

% Balloon program should be given additional resources to support science groups with power systems

% Can NASA manage and archive the designs of common science instrumentation (i.e. star trackers, pointing systems, etc)?

\begin{enumerate}[(A)]

\item  \textbf{Super-pressure balloons} 
The PAG recommends that the NASA super-pressure balloon program continue to pursue the goal of 100-day flights including at mid-latitudes. 
NASA should strive to advance the lift capability and float altitude to the point where SPBs are commensurate with current zero-pressure balloon capabilities.

\item \textbf{Payload Telemetry} 
The PAG commends NASA for efforts to improve balloon payload downlink bandwidth.
The science return of any mission can be significantly enhanced by higher data downlink capability, avoiding the possibility of data vault loss as a mission failure mode. 
By the end of the decade, payloads will be capable of producing hundreds of Terabytes or even
Exabytes of science data in a 30-day flight. 
The PAG recommends that NASA pursue an ambitious communication 
downlink goal of 100 Mbit/s average through a balloon flight ($\sim 32$ TB in 30 days) by the end of the decade, 
enabling a higher science return at much lower mission risk. 
The PAG also recommends that NASA pursue an increase in allocations of existing telemetry links such as TDRSS and Iridium to balloon payloads to help in the near term. 

\item {\bf Lift Capabilities.} 
The PAG commends NASA for the range of qualified balloons which satisfy the lift capabilities requested by the majority of the science community. 
Atmospheric transmission increases exponentially with altitude and, in the same manner, so does scientific return. 
This motivates the development of larger balloons capable of taking payloads closer to the edge of space. 
The PAG recommends expeditious flight qualification of the 60 Mcf balloon and the development of a larger class of zero pressure balloons capable of flying payloads of >1 ton to altitudes of >160,000 ft.

\item  \textbf{Pointing Systems.}
The PAG commends the Balloon Program for the development of the Wallops Arcsecond Pointing System, and recommends building on this success to provide even better pointing knowledge and stabilization for potential observatory-class (>1m) balloon-borne telescopes. 
The PAG also recommends commencing development on a second generation WASP that is lighter, more compact, and with improved
pointing knowledge.

\item \textbf{Large-aperture telescopes.} 
The PAG finds that several areas of scientific research would be greatly augmented with the development of one or more diffraction-limited near-UV, visible, near-IR, and/or thermal IR telescopes of up to 3~m on a balloon-borne platform. 
% At IR wavelengths, there is a particular need for large aperture telescopes. 
The PAG recommends that the NASA Science Mission Directorate consider development of an observatory-class telescope of this magnitude, to be managed by the balloon program for user instruments.
% although not necessarily ones with diffraction-limited image quality. 
% The PAG recommends an active development program in three key areas to achieve this goal: thermal stabilization of optical assemblies, wavefront sensors to generate real-time in-flight wavefront errors, and a wavefront correction scheme (e.g.,deformable mirrors)  to compensate for optical assembly aberrations.

\item \textbf{Opportunities for Small Payloads.} 
The PAG finds that there is significant interest in the use of small (<75 kg) payloads for scientific research in a range of disciplines. 
The PAG recommends that the balloon program continue to develop support systems for small balloons and continue to facilitate flight opportunities for piggyback payloads. 
The PAG also recommends that the process for finding piggyback opportunities should be advertised more broadly to the balloon user community.

\item
\textbf{Commercial Opportunities: Aerostats.} The PAG finds that commercial offerings of tethered aerostats, and possibly untethered station-keeping airships, may provide scientific opportunities that could be compelling, and may also engender new ideas for investigations. The PAG finds that these vehicles appear to be developing an economy of scale for both military and non-military applications.The PAG recommends that NASA study these vehicles, 
engage the community on potential science applications, and study the practicalities of how they might be integrated with the existing Balloon Program. 

\end{enumerate}

\section{Launch Sites \& Facilities}

%\noindent {\large \textcolor{royalblue}{\textit{Launch Sites and Facilities}}}

\begin{enumerate}[(A)]

{\item \textbf{Multiple-Payload building in Wanaka.}
The PAG commends the NASA Astrophysics Division and Balloon Program for developing and maturing the 18 Mcf super-pressure balloon, and the new launch facility in Wanaka, New Zealand, which supports SPB launches.  
These developments will enable new science investigations from Wanaka with science returns comparable to significantly more costly space flight missions, and complementing to NASA flagship missions. 
The PAG recommends continued support for the growth of this facility, including a new payload integration building that could accommodate multiple payloads. 
}

{\item
\textbf{Launch crews \& Facilities.}
The PAG recommends that NASA follow through as soon as possible with the approved increases in the capacity of launch facilities and the number of ground crews that can support them.  
Ground crews and their associated facilities can only support a limited number of launches within a given launch window (as determined by prevailing weather conditions), which can result in missed launch opportunities and avoidable fatigue to all concerned.
Additionally, the ground crew capabilities must be commensurate to the multiple locations and duration of the campaigns that are currently in place.  
Therefore, The PAG recommends that completion of the approved increase in ground crews and launch facilities should be given a high priority. }

{\item \textbf{LDB Antarctic Program \& Three-payload support.}
The PAG commends the Balloon Program for the continuing success and scientific impact of the Antarctic Long Duration Balloon (LDB) program flown out of McMurdo Station. Antarctica represents a unique resource and environment for investigations across several disciplines and the opportunities afforded by the LDB merit such support. 
The PAG recommends unwavering NASA support for this flagship program and its associated facilities near Williams Field, Antarctica. 
The PAG recommends that NASA authorize a deploy a third payload building at the earliest opportunity and commit to the resources necessary to sustain a three-large-payload per season launch rate. 
The PAG recommends that NASA strive to acquire the aircraft resources necessary to ensure timely recovery of payloads.}

{\item \textbf{Mission Safety Protocols.} 
The PAG recommends that NASA consider appointing a panel consisting of both scientific balloon community and NASA to engage each other in a review of current mission safely protocols within the Balloon Program, to address the issue of burgeoning and confusing standards, which have reduced efficiency and productivity in recent years, without necessarily improving the safety of the program.}

{\item \textbf{North American launch sites and infrastructure investment.} 
The PAG finds that a North American launch site that can provide reliable night-time launch opportunities is a high priority not only for Astrophysics balloon payloads, but other disciplines as well. 
The PAG recommends that NASA identify and develop alternative launch sites  in addition to Palestine, TX, and Ft. Sumner, NM, and possible Pacific Northwest sites under consideration. 
The PAG finds also that current infrastructure in Palestine and Ft. Sumner has aged to the point that is having a negative impact on the productivity and safety of the program, and the PAG recommends that NASA invest substantially in repairs, maintenance, and upgrades on these important facilities.}

{\item \textbf{Diversity of Launch sites and scope.}
The PAG finds abundant evidence that expansion, not contraction, of the present portfolio of balloon flight options for both launch location and duration is important to the continued health of NASA astrophysics research, and the training of new investigators at every level. 
The PAG recommends that NASA give high priority to continuing operations from locations that support research in auroral and radiation belt physics, and high latitude magnetosphere-, ionosphere-, and thermosphere-coupling, which are compelling  scientific drivers in heliophysics  and require flights at magnetic latitudes ranging from 55-70 degrees (e.g., Kiruna, Sweden). }

\end{enumerate}

\section{Funding Opportunities}
%\noindent {\large \textcolor{royalblue}{\textit{Funding Opportunities}}}

% payload and tech development funding should be evaluated separately from science flights (this is being done now in Heliophysics with HFORT and HTIDeS)

% funding opportunities for balloon missions are needed in Planetary and Earth sciences

% funding opportunities that allow commercial partners where appropriate (for example piggybacks on Google Loon flights and station keeping technologies)

\begin{enumerate}[(A)]

{\item \textbf{Earth \& Planetary Science.} 
The PAG notes that currently NASA scientific ballooning offers no funding opportunities in Planetary science, and very limited opportunities in Earth Science.
As we have detailed in previous chapters, there are many different scientifically compelling investigations in both Earth and Planetary Sciences, and significant opportunities for workforce
development that are missed due to the lack of funding. 
The PAG strongly recommends that the NASA Science Mission Directorate implement consistent funding
opportunities for Earth and Planetary Science balloon payloads.}

{\item \textbf{Pioneers, and Explorers Missions of Opportunity.}
The PAG commends NASA for including balloon investigations in the new Pioneers mission class in the Astrophysics Research and Analysis Program (APRA) within Research Opportunities in Space and Earth Sciences (ROSES). 
This class of investigation (up to \$20M over 5 years) for highly meritorious and high-impact missions is of particular importance for long-duration and ultra-long duration payloads that launch from Antarctica, Sweden, and Wanaka, NZ. 
The PAG recommends that NASA continue this program as a regular part of the APRA investigations with the ROSES program.
The PAG also recommends that NASA continue to periodically include balloon-borne payloads within its Missions of Opportunity for the Explorer class investigations, to provide the possibility for exceptionally compelling science investigations that may require a level of commitment beyond the standard APRA and Pioneers opportunities.}

{\item {\bf Guest Investigators programs.} 
As balloon-borne instruments continue to increase in scientific capability in the coming decade, especially in light of the new Pioneers mission class and ultra-long duration flights achieved through the superpressure balloon, there is an expected trend towards balloon payloads following an observatory-class model. %where guest observations can be made after the primary mission is complete.
The PAG recommends that NASA provide a funding opportunity for Guest Observers/Investigators for balloon missions, in addition to accommodating data analysis from balloon-borne instruments in relevant solicitations, such as the Astrophysics Data Analysis Program (ADAP).
}

% {\item \textbf{APRA funding.} The PAG welcomes recent increases in the NASA Astrophysics Research and Analysis funding and recommends that this increased funding for balloon payloads be advanced through the next decade given the need for increased payload complexity to address the scientific challenges summarized here.}

\end{enumerate}

\section{Workforce Development, Education \& Outreach.}

%\noindent {\large \textcolor{royalblue}{\textit{Education \& Outreach}}}

% Recommend that NASA better advertise piggyback and hand launch opportunities that are accessible to groups with little to no experience

\begin{enumerate}[(A)]

{\item {\bf Workforce development} 
The PAG recommends that the community and Balloon Program Office work to foster high altitude ballooning as a key element in NASA’s workforce development pipeline from pre-college to new scientists. 

The PAG recommends specifically that NASA:
\begin{enumerate}[(i)]
\item Engage with nationwide entities that are already supporting transdisciplinary learning. 
\item  Improve accessibility to flight options for groups involved with experiential projects. 
\item Develop safety and performance standards for balloon projects at a level significantly below spacecraft standards.
\item Support technical workshops for entry level scientists to provide a venue instrument design sharing, lessons learned, and best practices.
\end{enumerate}
}

{\item {\bf Diversity.}
The PAG commends NASA Balloon Program efforts to balloon to support diversity and inclusion,  but recommends that additional effort is warranted to engage more female and minority scientists and engineers in the program at all levels: from students, postdoctoral researchers, and investigators within the scientific community, up to the Balloon Program Office and the Columbia Scientific Balloon Facility. The PAG also recommends that NASA, the BPO, and CSBF develop quantitative  assessments of their progress in addressing these issues.}

{\item \textbf{National Space Grant College and Fellowship program engagement.} 
The PAG recommends that NASA should engage with the National Space Grant College and Fellowship program to continue and expand strong support for student training ballooning programs that support the workforce development pipeline at all levels including K-12, university students, and in-service teachers.}

{\item {\bf 2024 Solar eclipse.} 
The PAG recommends that NASA engage with stakeholders interested in the April 8, 2024  North American total solar eclipse as early as possible to assess scientific, workforce development, and public engagement projects, as well as payload weight classes including heavy payloads, and potential launch sites. }

\end{enumerate}

\clearpage

\bibliographystyle{IEEEtran}
\bibliography{References.bib}

\end{document}